\documentclass[]{aa}
\usepackage[varg]{txfonts}

\usepackage{graphicx,rotating,hyperref}
\usepackage{amssymb,amsmath}
\usepackage{natbib}
\usepackage{empheq}
\usepackage{subeqnarray}
\usepackage{mathtools}
\usepackage{upgreek}
\usepackage{cases,soul}
\usepackage[normalem]{ulem}
\usepackage{xcolor}
\newcommand{\lhs}{LHS~1140~b }

\newcommand{\tml}{T_{\rm Ml}}
\newcommand{\etaq}{\eta_q}
\newcommand{\etarad}{\eta_{\rm r}}
\newcommand{\etamass}{\eta_{\rm m}}
\newcommand{\ysph}{y_{\rm sph}}
\newcommand{\rsph}{R_{\rm sph}}

\newcommand{\refd}{d_{\rm ref}}

\newcommand{\refpres}{P_{\rm ref}}

\newcommand{\trho}{\tilde{\rho}}
\newcommand{\thent}{\tilde{H}}

\newcommand{\const}{{\rm const.}}
\newcommand{\rpol}{{R_{\rm pol}}}
\newcommand{\req}{R_{\rm eq}}
\newcommand{\qpol}{{q}}
\newcommand{\qpolmin}{q_{\rm min}}
\newcommand{\qpolmax}{q_{\rm max}}
\newcommand{\qeq}{Q}
\newcommand{\neff}{n_{\rm eff}}
\newcommand{\rhoc}{\rho_{\rm c}}

\newcommand{\biga}{A}
\newcommand{\rhonp}{{\rho_{\rm s}}}
\newcommand{\interp}{P_{\rm i}}
\newcommand{\pc}{P_{\rm c}}

\newcommand{\uc}{u_{\rm 0}}
\newcommand{\mearth}{M_{\oplus}}
\newcommand{\rearth}{{R_{\oplus}}}
\newcommand{\psj}{The Planetary Science Journal}

\begin{document}

\title{Impact of rotation on synthetic mass-radius relationships of two-layer rocky planets and water worlds}

\titlerunning{}

\author{J.-M. Hur\'e\inst{1} \and P. No\'e\inst{1,2}  \and C. Staelen\inst{1} \and E. Di Folco\inst{1}}
\institute{University of Bordeaux, CNRS, LAB, UMR 5804, F-33600 Pessac, France \\\email{jean-marc.hure@u-bordeaux.fr}
 \and LPC2E, OSUC, Univ Orleans, CNRS, CNES, Observatoire de Paris, F-45071 Orleans, France}
\date{Received ??? / Accepted ???}

\abstract{We have analyzed the effects of rotation on mass-radius relationships for single-layer and two-layer planets having a core and an envelope made of pure materials among iron, perovskite and water in solid phase. The numerical surveys use the {\tt DROP} code updated with a modified polytropic equation-of-state (EOS) and investigate flattening parameters $f$ up to $0.2$. In the mass range $0.1 \, \mearth \lesssim M \lesssim 10 \, \mearth$, we find that rotation systematically shifts the curves of composition towards larger radii and/or smaller masses. Relative to the spherical case, the equatorial radius $\req$ is increased by about $0.36 f$ for single-layer planets, and by $0.30 f$ to $0.55 f$ for two-layer planets (depending on the core size fraction $q$ and planet mass $M$). Rotation is an additional source of confusion in deriving planetary structures, as the radius alterations are of the same order as i) current observational uncertainties for super-Earths, and ii) EOS variations. We have established a multivariate fit of the form $\req(M,f,q)$, which enables a fast characterisation of the core size and rotational state of rocky planets and ocean worlds. We discuss how the observational data must be shifted in the diagrams to self-consistently account for an eventual planet spin, depending on the geometry of the transit (circular/oblate). A simple application to the recently characterized super-Earth candidate \lhs is discussed.}


\maketitle

\section{Introduction}
\label{sec:mr}

The great diversity of exoplanets --- more than $5700$ objects detected over the last three decades --- represents a remarkable challenge for models of internal structure, scenarios of planetary formation and mechanisms at work, and detection methods \citep{haswell10,nielsen22}. The opportunity to find planets resembling Earth, and possibly hosting forms of life, has become an attractive reality \citep{leger04,sreview18,Wandel2018,rm24,lm25}. When transit and Doppler spectroscopy can be combined together, the radius $R$ and mass $M$ of the planet are measured with relatively good precision \citep[e.g.][]{sfs14,aplus21}. The resulting mean mass-density $3 M/4\pi R^3$ and the inspection of the planet position in pre-computed mass-radius diagrams bring informations in terms of chemical/mineralogical composition. Unfortunately, the knowledge of these data is far from sufficient to infer a unique internal structure, because the inverse problem is highly degenerate. There is a certain diversity of equation-of-states due to the variety of materials present at the period of formation, and the thermodynamical conditions are complex. The number of internal layers is another important unknown. Key informations linked to the planet environment and age of the system can also be considered to discriminate between plausible configurations and unlikely ones. In particular, the proximity of the parent star sets severe constraints in terms of chemical composition \citep{detal17,bmdd17}, stellar wind, instellation and heat transport \citep{vos06,wtsrs12,ts13,mda20,amdm21,hdaw24}, tides towards synchronization \citep{vso07,nfkr11,kbr18}. Realistic models including all these ingredients together are a formidable technical challenge.

To our knowledge, mass-radius relationships for planets (solid or even gaseous) are in general computed in the spherical/static limit. But planets rotate, more or less. It is hard to conceive that all exoplanets are either perfect spheres with no proper spin, or completely tidally locked with their parent star. Planets in our Solar System are good examples of objects exhibiting slow to moderate rotations. When the orbital separation is short, subtanstial three-dimensional deformations are expected \citep[e.g.][]{ch33bsp}. For planets moving on large orbits, the spin-orbit resonnance may not be reached, depending on the age of the system. The time scale actually varies with the separation $a$ and stellar mass $M_\star$ as $a^6/M_\star^2$ \citep[e.g.][]{gp68,zahn89,lwmm15,barnes17}. Accretion of gas and dust in a circumstellar disc is efficient in spinning up planetary embryos \citep[e.g.][]{mitra77}. Young planets are expected to rotate very fast, in particular due to accumulation of angular momentum from pebble accretion and collisions. Small size bodies like asteroids are highly affected by stochastic collisions and may end up with small rotation periods, of the order of a few hours \citep{fdpcz92}. In the giant impact theory, the rotation period of the proto-Earth is estimated to $2.7 - 5.8$ h (pre- and post-impacts), and the polar-to-equatorial axis ratio $\rpol/\req$ of the spheroidal Earth should have been well below unity, close to the mass-shedding limit/break-up velocity \citep{DEMEIJER201340,canup12}. 

Another motivation for considering rotation comes from the increasing interest for the detection of oblatness in transit light-curves \citep{Akinsanmi2020,CARADO20,liu25}. The photometric signal from a transiting planet depends on the planet size relative to the star size, and to the orbital separation \citep{haswell10}. Again, the planet is generally assumed spherical, which means that area blocking the stellar flux towards the observer is a circular disc. The existence of oblatness introduces a bias when retrieving the planets parameters, in particular the radius \citep{bdw22}. With present-day photometric sensitivity and time-sampling, the systematic detection of oblatness during the ingress/egress phases of transits is marginally accessible for gaseous planets, which represent the best targets \citep{zhzl14,liual24}. In contrast, rocky planets and water worlds that have not reached synchronisation yet orbit at large separation are out of the range of detection for the transit method. In the perspective of finer measurements (in terms of timing and relative flux) and future instruments like the Plato mission, there is a great interest in anticipating and understanding the role of rotation in mass-radius relationships, and in quantifying the additionnal level of confusion it brings in interpreting data.

Estimating the change in the radius and/or mass of a self-gravitating fluid due to rotation is an old problem. For a homogenous spheroid with flattening $f \equiv 1 - \rpol/\req \ll 1$, the equatorial radius is larger than for a sphere having the same mass by the quantity $\Delta \req \approx +\frac{1}{3}\req f$. In similar conditions, for a given equatorial radius, the mass of the rotating spheroid is smaller than for the sphere by $\Delta M \approx - M f$. It means that for an uncompressible body with flattening $f = 3\%$ (a little bit higher than for Neptune and Uranus), the relative deviation in the radius reaches $\Delta \req/\req \approx 1 \%$, which value is comparable to the precision obtained by the transit method. Solid planets are not homogeneous spheroids. The modification in the structure due to rotation has been treated in the limit of weak distorsion by \cite{ch33} for a polytropic EOS where the pressure $P$ and the mass-density $\rho$ scale as $P \propto \rho^\gamma$. The result depends on the properties on Emden's function at the equator (in particular, the non-analytical derivative), and especially on the polytropic exponent $\gamma$ of the gas. As rotation increases, the equatorial elongation changes faster than the pole contracts. For a given mass, the central density is slightly reduced;see also \cite{w88}, \cite{vd20} and \cite{rc24} for $\gamma=2$. How $\Delta \req$ and $\Delta M$ are modified for rocky planets and water worlds ? And how these changes caused by rotation compare with observational error bars  $\delta R$ and $\delta M$ ? In this article, we investigate the impact of rotation on synthetic, single-layer and two-layer planets with internal compostion typical of the Earth (iron/silicate) and water-rich planets. The case of planets supporting an atmosphere is planned for another paper. 

Authorizing rotation renders the problem technically much more complex than in the static/spherical case, as there are two spatial coordinates to manage simultaneously, instead of a unique, radial variable. For this reason mainly, we deliberately abandon some ``ingredients'' that are undoubtedly decisive in any realistic approach. We assume single-layer and two-layer solid planets in rigid rotation, without any atmosphere. The radius is greatly changed in the presence of a relatively massive atmosphere \citep{ase08,howe14}. Regarding the EOS, there are several degrees of sophistication and realism, especially when the temperature is an explicit parameter \citep{s12,hakim18,bdf20}. For simplicity, we use a barotropic EOS appropriate to rocky/icy planets, in the form of the modified polytrope, according to \cite{skhm07}. This is a reliable approximation to more complex ansatz. In this context, different mineralogical compositions can be considered, from pure to impure materials. We have selected only three kinds of materials, namely iron, silicate and water (all in solid phase). We also assume that the parent star is far enough, and therefore neglect the effects of tides (gravitational attraction and internal energy release) and irradiation that produce surface heating and trigger circulation caused by temperature imbalance at the surface \citep[e.g.][]{afbdl24}. We assume rigid body rotation, which is a common hypothesis. 

In Sect. \ref{sec:equilibriumSCF}, we briefly summarize the assumptions and equation set relevant to model a multi-layer planet in rotation, which are mostly described in \cite{bh21}, and the main properties of the DROP code used to numerically solve the equilibrium on a cylindrical $(R,Z)$-grid \citep[see also][]{hh17,hhn18}. Attention is paid to the specific link between the enthalpy $H$ and the mass-density $\rho$ associated to the EOS. Sections \ref{sec:surveys_sl} and \ref{sec:surveys_tl} are devoted to an exploration of the parameter space in the single-layer case and two-layer case, respectively. Note that there are two key parameters, namely the core fractional size $q \in [0,1]$ and the flattening $f=1-\rpol/\req$. In particular (see Sect. \ref{sec:mr}), we take the liberty to investigate the influence of rotation on mass-radius relationships well beyond the standard limit of slow rotations, by considering flattening parameters $f$ up to $0.2$ (simulations can go much above). A global fitting procedure is perfomed, leading to a general, fully reversible formula of the form $\req=\req(M,\qpol,f)$, mixing quadratic laws and a sigmoid. In Sect. \ref{sec:errorbars}, we discuss the diversity of structures in terms of $\delta q$ and $\delta f$ compatible with observational error bars $\delta M$ in the mass and $\delta \req$ in the radius. Although this is not the nominal target regarding our purpose, we consider LHS~1140~b \citep{dittman17,ment19,ll20,cad24}, a potential ocean world that has been recently modelled by \cite{rm24} and \cite{dal24}. The main results are summarized in the conclusion, including an example of bias correction appropriate for both circular and oblate transits. A few perpectives and possible improvements of the present work are listed.

\section{Single-layer and two-layer equilibrium configurations from the Self-Consistent-Field method and the {\tt DROP} code}
\label{sec:equilibriumSCF}

\subsection{Equation set}

The structure of an isolated, spherical planet in steady state is obtained, like for stars, by integrating the one-dimensional equation for hydrostatic equilibrium, in which local gravity force balancing the pressure gradient is directly available from Gauss's theorem. The integration proceeds from the center/surface towards the opposite boundary, and shots are repeated until the boundary conditions are fulfilled, optionnally with interior matching points \citep{cox_1968,hdaw24}. With this technique, it is relatively easy to incorporate various details in the physical and chemical processes, depending on the local conditions \citep{slattery77,gui99,vplus05,skhm07,rm24}. In the presence of rotation, the equation of hydrostatic equilibrium and the Poisson equation involve two spatial coordinates, and the shooting methods are less direct. If the rotation rate $\Omega$ inside the planet depends only on the cylindrical distance $R$ from the axis of rotation and the fluid is a barotrope \cite[e.g.][]{amendt1989}, then the two components of the Euler equation are the gradient of the following scalar field
\begin{equation}
  H + \Psi + \Phi = \const,
  \label{eq:bernoulli}
  \end{equation}
where $\Phi = - \int{\Omega^2 R dR}$ is the centrifugal potential, $H=\int{dP/\rho}$ is the enthalpy, $\rho$ is the mass density, $P$ is the pressure, and $\Psi$ is the gravitational potential, which obeys the Poisson equation
\begin{equation}
  \Delta \Psi = 4 \pi G \rho.
\label{eq:poisson}
\end{equation}
The problem then consists in finding $H$ satisfying these two equations at each point in the fluid, and this is usually performed via an iterative procedure, known as the Self-Consistent-Field (SCF) method. The enthalpy $H$ is guessed, then $\Psi$ is computed from Eq.\eqref{eq:poisson} by using $\rho(H)$, and it is injected in Eq.\eqref{eq:bernoulli} to get the enthalpy $H$. This new field is compared to the previous one, and the cycle is repeated until convergence. A wide litterature has been devoted to the determination of the solutions for Eq.\eqref{eq:bernoulli}, in various contexts, in particular assuming a polytropic EOS for the material where $P \propto \rho^\gamma$ \citep[e.g.][]{hachisu86,horedttextbook2004}. A new difficulty arises when combining rotation and layering, due to the presence of mass-density jumps \cite[e.g.][]{kiu10,schu11,ka16,bh21,hr23}. Actually, when the body is made of ${\cal L}$ well-differenciated layers in total, and provided the two conditions above are still satisfied, then there is one Bernoulli equation per layer to consider, because each layer has its own EOS. The gravitational potential is still defined by Eq.\eqref{eq:poisson}, but the second member becomes a piece-wise function. Pressure balance at each interface $S_l$ bounding layers $l \in [1,{\cal L}]$ and $l+1$ writes
\begin{equation}
\left. P_l \right|_{S_l} = \left.  P_{l+1} \right|_{S_l},
\label{eq:pbalance}
\end{equation}
means that the knowledge of all interfaces is mandatory (layer ${\cal L}+1$ denotes the free space). In total, there are $2{\cal L}+1$ equations to solve simultanesouly. We proceed according to \cite{bh21}, by solving one Poisson equation per layer on a dedicated computional grid with appropriate boundary conditions, namely $\Delta \Psi_l = 4 \pi G \rho_l$, and the total potential $\Psi$ is obtained by superposition. This is well adapted to treat correctly mass-density jumps at the interface between layers.

\begin{table*}[h]
  \caption{Data$^\star$ for the modified EOS from \cite{skhm07}, according to Eq.\eqref{eq:mpeos}.}
  \label{tab:eosdata}
  \begin{tabular}{ccccccc} \hline \hline
         & $\rhonp$ (g/cm$^3$) & $K^{-m}$ & $m$ & index $n$ & exponent $\gamma$ & $\frac{1-n}{3-n} \equiv \beta$ \\ \hline
   iron  &  $8.30$ & $1.034726 \times 10^{-6}$ & $0.528$ & $1.11864$ & $1.89393$ & $-0.06306$ \\
perovskite   & $4.10$  & $0.463261 \times 10^{-6}$ & $0.541$ & $1.17864$ & $1.84842$ & $-0.09809$ \\
water  & $1.46$ &  $0.954465 \times 10^{-6}$ & $0.513$ & $1.05338$ & $1.94931$ & $-0.02743$ \\ \hline
  \end{tabular}\\
  $^\star$The coefficient $K$ is given in g/cm$^3 \times (\text{dyn/cm}^2)^{-m}$, and assumes that $P$ is in dyn/cm$^2$ (see note \ref{note:keos}). Coefficients $n$ and $\gamma=1+\frac{1}{n}$ are computed in the limit of high pressures, i.e. $\gamma=\frac{1}{m}$.
\end{table*}

\begin{figure}[h]
  \centering
\includegraphics[width=7.5cm,bb=360 270 1000 791,clip==]{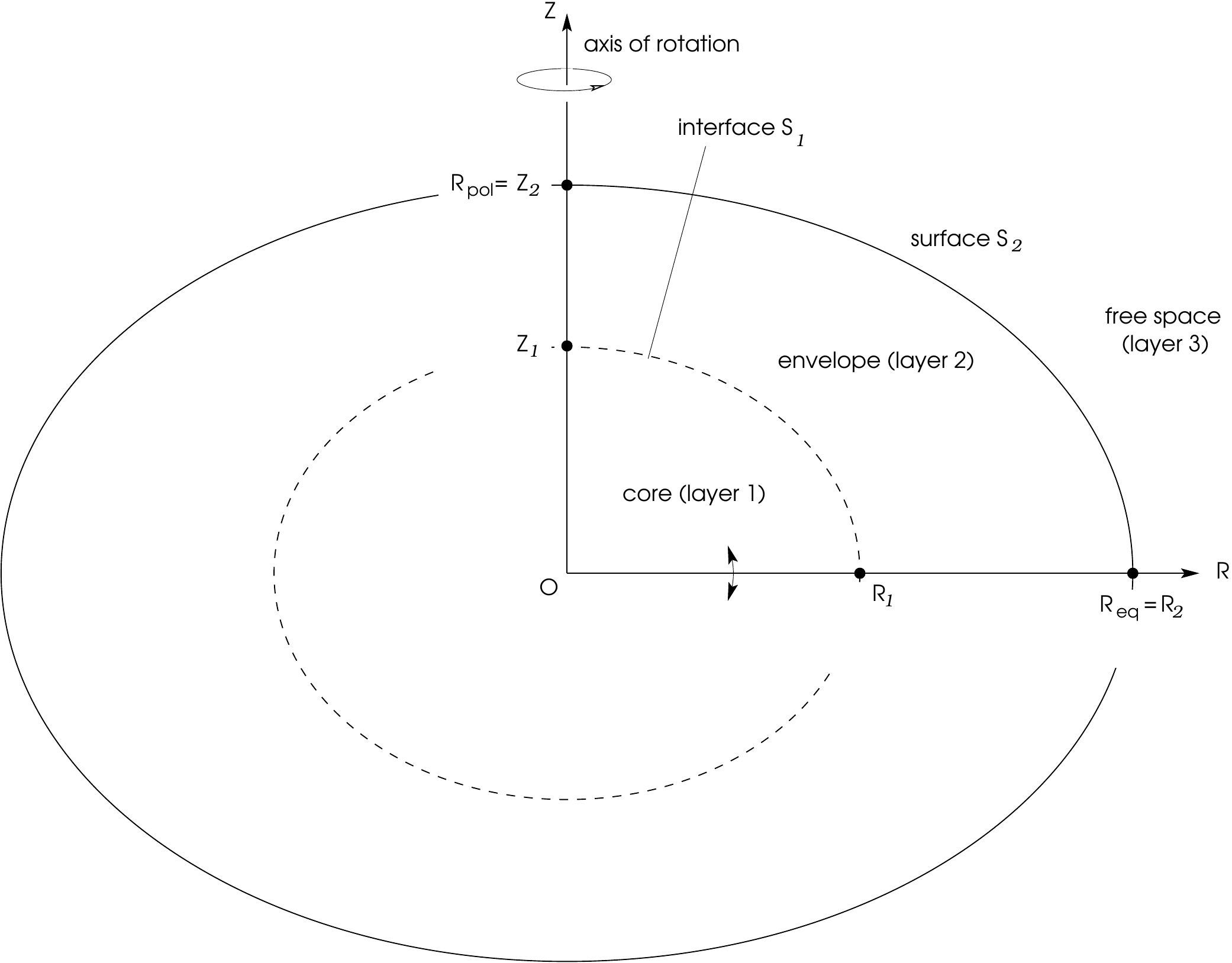}
\caption{Typical configuration and notations for a rotating system made of ${\cal L}=2$ differenciated layers: the core (C) as layer 1, and the envelope (E) as layer 2. The polar radius is $\rpol \equiv Z_2$ and the equatorial radius is $\req \equiv R_2$. All layers rotate at the same rate $\Omega$ around the $Z$-axis.}
\label{fig:llayers2.eps}
\end{figure}

As mentionned in the introduction, we consider single-layer and two-layer planets. For  ${\cal L}=2$, the core (C) is layer 1, and the envelope (E) is layer 2 as depicted in Fig. \ref{fig:llayers2.eps}. The equatorial and polar radius of the core are $R_1$ and $Z_1$ respectively, and we have $R_2 \equiv \req \ge R_1$ and $Z_2 \equiv \rpol \ge Z_1$ for the envelope. Usually, rotation flattens the structure so that we expect $Z_l \le R_l$. The two layers are in rigid rotation at the same rate $\Omega$.

\subsection{The modified polytropic EOS and the selection of mineralogical compositions for solid planets}
\label{subsec:meos}
  
As the diversity of planetary interiors is wide, the coverage in mass-density, pressure and temperature must be sufficient \citep[e.g.][]{ts13,hakim18,bdf20,halde20}. For telluric and ocean planets with masses in the range $0.1 \mearth$ to $10 \, \mearth$, different EOS can be selected \citep[e.g.][]{gss09,sfs14,rc24}; see Sect. \ref{sec:mr}. These are generally combinations of experimental data and theoretical calculations, and consist of piece-wise functions and tables of data specific for various phases. Our primary objective is to test the role of rotation and to simplify the problem as much as possible. In this context, we work with the barotropic EOS given by \cite{skhm07},  namely
\begin{equation}
  P=K(\rho-\rhonp)^{1/m}
  \label{eq:mpeos}
\end{equation}
where $\rhonp$ is the mass-density at null pressure, $K$ and $m$ are constants\footnote{The notation differs slightly from \cite{skhm07} where Eq. \eqref{eq:mpeos} is written as $\rho = \rho_0 + c P^n$. The connection between the two formula requires $\rho_0 \rightarrow \rhonp$, $n \rightarrow m$ and $c \rightarrow K^{-m}$. As a consequence, if the pressure $P$ is measured in dyn/cm$^2$, then the values of $K^{-m}$ reported in Tab. \ref{tab:eosdata} are in g/cm$^3 \times$ (dyn/cm$^2$)$^{-m}$. \label{note:keos}}. This analytical model works for several types of solid materials, in particular for water, silicates and iron, and it includes quantum effects important at very high pressures \citep{zs69}. It is well suited to bodies with central pressures from $10^{11}$ to $10^{13.5}$ dyn/cm$^2$, expected for the mass range of interest here \citep{cox_1968}; see also Sect. \ref{subsec:pc}.

The parameters $m$, $\rhonp$ and $K$ needed in Eq.\eqref{eq:mpeos} are listed in Tab. \ref{tab:eosdata} for solid iron (I), perovskite (P) and water (W), following \cite{skhm07}. The table also contains the index $n$ and the exponent $\gamma=1+\frac{1}{n}=\frac{1}{m}$ corresponding to the high-pressure limit. Regarding water, we are aware that $\rhonp = 1.46$ g/cm$^3$ is larger than the standard values for ice and liquid phase in standard conditions, but we follow the authors' prescription for the sake of consistency.

If we consider only configurations where $\rhonp(E) < \rhonp(C)$, including the unrealistic case of an iron core and a water envelope, there are six different possibilities:
\begin{itemize}
  \setlength{\itemsep}{+2pt}
\item C $\in\{$I,\,P,\,W$\}$ for single-layer planets,
\item CE $\in \{$IP,\,IW,\,PW$\}$ for two-layer planets. 
 \end{itemize}

\subsection{The associated enthalpy}
\label{subsec:h}

With respect to a polytropic EOS where $P \propto \rho^\gamma$ and $\gamma=1+\frac{1}{n}$, the bimodal prescription advantageously incorporates the uncompressible case $n=0$ (i.e. $\gamma \rightarrow \infty$) at vanishing/low pressure and a compressible case at high pressure, with $\gamma \rightarrow\frac{1}{m}$. With Eq.\eqref{eq:mpeos}, this change is especially significant in terms of the associated enthalpy $H$ required in Eq.\eqref{eq:bernoulli}, as we have 
\begin{equation}
H(\rho) = \gamma K\int{(\rho-\rhonp)^{\gamma-1}\frac{d\rho}{\rho}}.
\end{equation}
Unfortunately, a close form for $H$ exists only for certain values of $m$. For instance, for $m=\frac{1}{2}$, the enthalpy is of the form $H \sim \rho- \const \times \ln \rho$. In the general case, $H$ is given by
\begin{equation}
  H = K\rhonp^{\gamma-1} u^{\gamma} {}_2F_1(\gamma,\gamma;\gamma+1;u) + \const,
  \label{eq:h2F1}
\end{equation}
where ${}_2F_1$ denotes the Gauss hypergeometric function, and
\begin{equation}
  u=1-\frac{\rhonp}{\rho} \le 1.
\end{equation}
The non-linear change of variable (a Pfaff transformation) ensures that the hypergeometric series unconditionnaly converges since $\rho \ge \rhonp$ in the structure. It can be verified that, in these conditions, $H(\rho)$ remains bijective, and it can therefore be inverted to get $\rho(H)$. This inversion, which is accomplished by numerical means, is unavoidable since $\rho$ is directly required to get the potential $\Psi$ from the Poisson equation. The integration constant in Eq.\eqref{eq:h2F1} is of fundamental importance (there is one such constant per layer). It represents the enthalpy of the material at zero pressure, i.e. for $u=0$. If we define
\begin{equation}
\biga=\frac{\rhonp}{\rho_0} \le 1,
\label{eq:biga}
\end{equation}
where $\rho_0$ is, for instance, the largest value in the actual layer, then $\uc=1-\biga$ and so the integration constant in Eq.\eqref{eq:h2F1} is
\begin{equation}
\const =H_0-K\rho_0^{\gamma-1}\biga^{\gamma-1} \uc^{\gamma} {}_2F_1(\gamma,\gamma;\gamma+1;\uc).
\end{equation}

\subsection{Simulations of rotating two-layer planets at equilibrium with the updated {\tt DROP} code}
\label{subsec:drop}

We use the {\tt DROP}-code to determine numerically the equilibrium state of a rotating body made of ${\cal L}$ layers \citep{bh21} from the Self-Consistent-Field (SCF) method. The code is scale-free. In particular, we set $\trho=\rho/\rho_0$ for the mass density, and $\thent=H/H_0$ for the enthalpy, where $\rho_0$ and $H_0$ is specific to each layer. In the new version of the code, the modified polytropic EOS defined by Eq.\eqref{eq:mpeos} is implemented. The critical point is the automatic detection, at each step of the SCF-cycle, of the core-envelope interface $S_1$ where Eq.\eqref{eq:pbalance} has to be fulfilled; see Fig. \ref{fig:llayers2.eps}. Running a simulation with the {\tt DROP} code requires $3$ input parameters for ${\cal L}=2$:
\begin{itemize}
\setlength{\itemsep}{+1pt}
\item the flattening parameter $f=1- \rpol/\req \ge0$,
\item the parameter $\biga=\rho_{s,{\cal L}}/\rho_{\cal L}^{\rm bottom} \le 1$ which sets the top-to-bottom mass-density ratio in the surface layer,
\item the relative size of the core $\qpol=Z_1/\rpol$ along the $Z$-axis.
\end{itemize}
For a single-layer planet, $\biga$ is basically the surface-to-center mass density ratio and the last parameter is useless. The value of this parameter mostly determines the mass of the planet. For $A \simeq 0.3 - 0.85$, we typically scan two decades in mass around one Earth mass, depending on the composition.

At convergence, the code delivers the enthalpy at equilibrium $H_l(R,Z)$ for each layer $l=\{1,2\}$, the interface $S_1$ and $S_2$, in the form of discrete values $\{R_i,Z_i\}$, the rotation rate $\Omega$, and the equatorial radius $\req$. From these data, we can compute the mass $M_l$ of each layer, the total mass $M$, the core mass-fraction $M_1/M$, the central pressure $\pc$, the rotation period $T$, etc. (see below). The accuracy of simulations is mainly controlled by the numerical resolution. We work with $N+1=129$ grid nodes per direction, with uniform spacing. This is a moderate resolution. As the quadrature and finite difference schemes are second-order accurate in the unit-square grid, the relative precision in output quantities (mass, volume, Virial parameter, etc.) is $1/N^2 \approx 10^{-4}$, typically. Most outputs are given with $5$ digits.

\subsection{An example of simulation}

\begin{figure}
       \centering
       \includegraphics[trim={1.cm 7.7cm 14.6cm 0cm},clip,width=1.\linewidth]{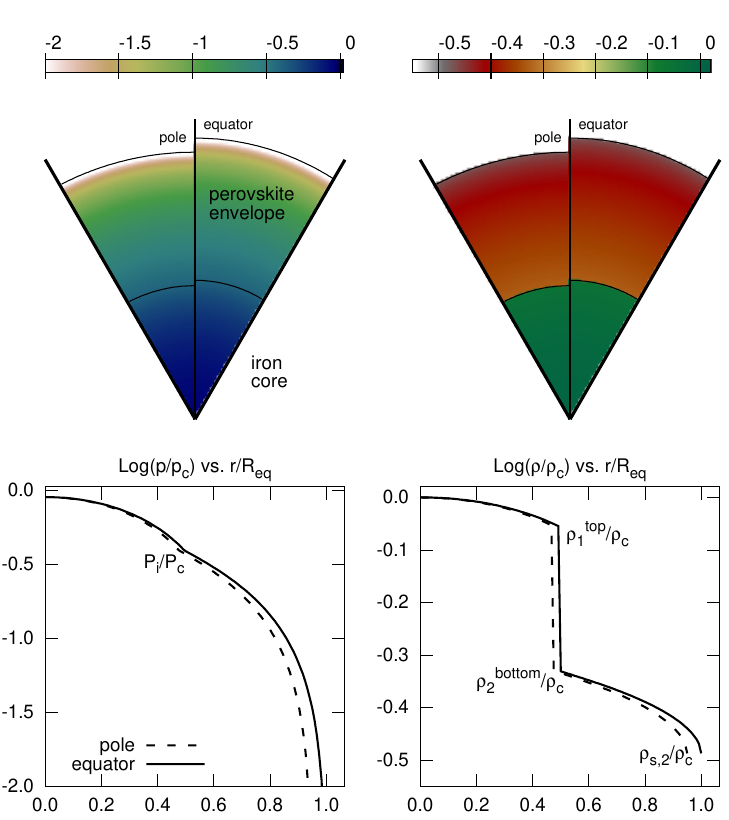}
       \caption{Normalized pressure ({\it left}) and mass-density ({\it right}) in color code ({\it top panels}) close to the pole and to the equator for a two-layer, IP planet computed with the {\tt DROP} code. The input/ouput parameters are in Tab. \ref{tab:ref}. Profiles along the rotation axis (spherical radius $r \equiv Z$) and along the equator ($r \equiv R$) are also shown  ({\it bottom panels}). }
       \label{fig:graph1refbis_IP.pdf}
\end{figure}

We show in Fig. \ref{fig:graph1refbis_IP.pdf} the pressure and mass-density profiles obtained for a two-layer planet made of an iron core and a perovskite envelope, i.e. CE=IP. For this run, we take $\qpol=0.5$ and $f=0.05$. The last parameter $\biga \approx 0.697$ has been selected such that we get a planet with $M \approx 1 \mearth$. This example therefore represents a very simplified Earth, but with flattening parameter $f$ comparable to Jupiter's value (instead of $\simeq 0.003$). We see the smooth connection of the pressure at the interface, while the mass-density undergoes a jump, as expected. The ouput data are gathered in Tab. \ref{tab:ref}. The resulting equatorial radius is about $6346$ km, and this is $6029$ km at the pole. The rotation period is about $6$ h $22$ min.

\begin{table}
   \caption{Results for the two-layer, IP planet displayed in Fig. \ref{fig:graph1refbis_IP.pdf}.}
   \label{tab:ref}
  \begin{tabular}{lr} \hline \hline
         $\blacktriangledown$ input\\
         \quad flattening parameter $f$ & $0.05$\\
         \quad $\biga=\rho_{s,2}/\rho_2^{\rm bottom}$ & $0.69689$\\
         \quad relative core size (pole) $\qpol$ & $0.5$\\ \hline
          $\blacktriangledown$ output\\
         \quad equatorial radius $\req$ in $\rearth$ &  $0.99621$\\
         \quad total mass $M$ in $\mearth$ & $1.00000$\\
        \quad relative mass of the core $M_1/M$ & $0.24305$\\
   \quad central-to-mean mass-density ratio $\rhoc/\langle \rho \rangle$ & $2.14162$\\
   \quad top-to-central mass density ratio (core) $\rho_1^{\rm top}/\rhoc$ & $0.88047$\\
   \quad rotation period $T$ (h) & $6.36260$\\
   \quad  relative size of the core (equator) $\qeq$  & $0.49404$\\
   \quad interface mass-density jump $\rho_1^{\rm top}/\rho_2^{\rm bottom} \equiv \alpha$ & $1.88106$\\
   \quad central-to-surface mass density ratio $\rhoc/\rho_{s,2}$ & $3.06567$\\
   \quad central pressure $\pc$ ($10^9$ dyn/cm$^2)$ & $2696.31$\\
   \quad interface pressure $\interp \equiv P(Z_1)$ ($10^9$ dyn/cm$^2)$ & $1488.09$\\\hline
  \end{tabular}
\end{table}

\begin{figure*}[h]
       \centering
       \includegraphics[trim={1.cm 0.5cm 1.cm 0.5cm},width=17cm]{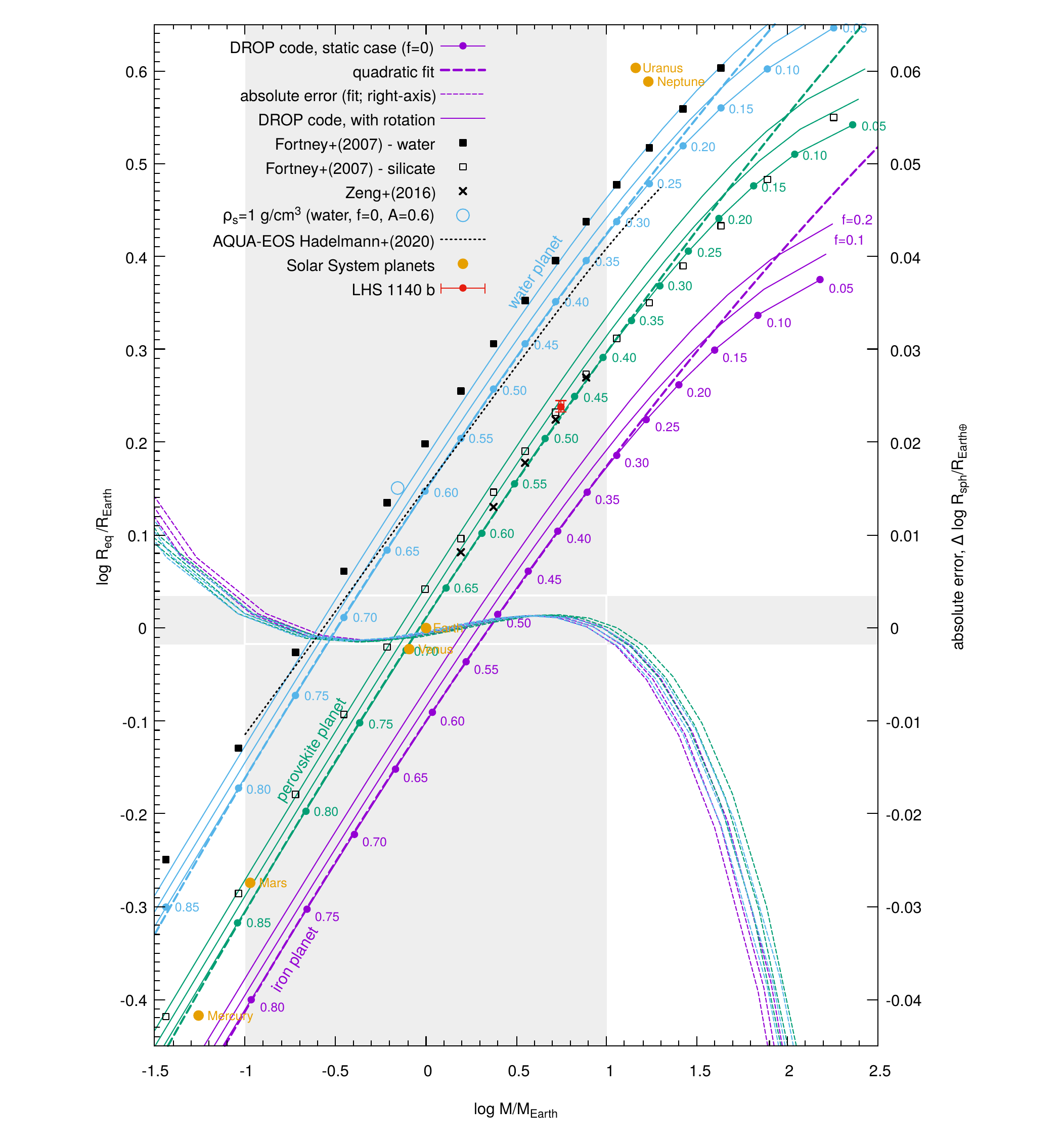}
       \caption{Mass-radius relationships ({\it left axis}) computed with the {\tt DROP} code in the single layer case, for iron ({\it purple}), perovskite ({\it green}), and water ({\it cyan}), and for three flattening parameters $f \in \{0,0.1,0.2\}$. Values of the input parameter $\biga$ are indicated along the curves for $f=0$; see Eq. \eqref{eq:biga}. The domain where the fits are performed is indicated ({\it grey, shaded zone}) as well as the absolute error ({\it right axis}). The result obtained with $\rhonp=1$ g/cm$^3$, $f=0$ and $A=0.6$ for water in Eq. \ref{eq:mpeos} is shown for comparison ({\it cyan, open circle}). Also plotted are the empirical relationship valid in the range $1-8\, \mearth$ and built from the Earth's model by \cite{zsj16} ({\it black crosses}), the data from Tab. B.1 of \cite{halde20} obtained from the AQUA-EOS for water assuming a $300$ K surface temperature ({\it thin, dotted lines}), and the results by \cite{fmb07} for pure-water and pure-silicate planets ({\it open and filled squares}), all obtained in the static case. See Sect. \ref{sec:lhs1140b} for a discussion about LHS~1140~b ({\it red dot}).}
       \label{fig:mr_static_singlelayer}
\end{figure*}

\subsection{Sampling for the numerical surveys. Tables}

We have run the {\tt DROP} code in the same conditions as in the previous example, by varying $f$, $\biga$, and $\qpol$ independently to each other. Given the moderate resolution (see above), a good compromise between the computing time and the quality of the data on output is obtained with the following samplings:
\begin{itemize}
 \setlength{\itemsep}{+1pt}
\item $f \in [0,0.2]$, with $\Delta f=0.025$ ($f=0$ is for the static case),
\item $\biga \in [0.05,0.95]$ with $\Delta \biga = 0.025$,
\item $\qpol \in ]0,1[$ for two-layer planets, with $\Delta \qpol = 0.1$.
\end{itemize}
This represents $171$ runs for a single-layer planet, and $1536$ runs in a two-layer case. As we have five different configurations (see Sect. \ref{subsec:drop}), the number of runs amounts to $5130$ in total. The list of input/output is given in the Appendix \ref{app:mrf_table} together with a sample of data obtained for a water planet and for a IW-planet and $q=0.5$, with $f=0.1$ in both cases. The full tables obtained for the six types of planets are available in electronic form on Zenodo\footnote{DOI:{\tt10.5281/zenodo.15397897}\label{fn:zenodo}. Specific runs can be provided upon request. The presence of ${}_2F_1(u)$ in Eq.\eqref{eq:h2F1} largely increases the computing time with respect to a standard polytropic EOS (see Sec. \ref{sec:hyperg} for a discussion about convergence acceleration and the computing time). The production of each cube of data requires about $6$ h of computation on a standard computer for a two-layer configuration. }

\section{Results in the single-layer case}
\label{sec:surveys_sl}

Before considering configurations with ${\cal L}=2$, it is instructive to examine first the case where the planet is a single-layer body made of pure material, i.e. C $\in \{$I,P,W$\}$. The mass-radius relationships obtained for $f \in \{0,0.1,0.2\}$ are displayed in Fig. \ref{fig:mr_static_singlelayer}. In the static case, we have $\req($W$)>\req($P$)>\req($I$)$ for a given mass, which is expected from the mass density of the species.  We find that:
\begin{itemize}
\item decreasing $\rhonp$ shifts the pure-composition lines quasi-horizontally, towards low masses (i.e. a left shift in the plane),
\item decreasing the constant $K^{-m}$ shifts the pure composition lines towards large radii and high masses (a rightward and upward shift)
\item decreasing the power-law index $m$ has a similar effect as decreasing the gaz constant.
\end{itemize}

In the graph, we have also reported the semi-empirical mass-radius relationship computed in \cite{zsj16} based on the Preliminary reference Earth model \citep{prem81}, the data computed by \cite{halde20} from the AQUA-EOS for water (assuming a $300$ K surface temperature), and the results by \cite{fmb07} for pure water and pure silicate planets. Given the differences in the EOS and models of internal structure with these authors (explicit treatment of temperature and heat transport mainly), the comparisons with these authors are indeed satisfactory in the spherical case. Besides, we have checked that the agreement with \cite{fmb07} is much better for the water line by setting $\rhonp \simeq 1$ g/cm$^3$ in the code.

\subsection{Trends and fits}

We observe that changing $f$ at constant mass (radius) shifts the equatorial radius (mass, respectively) towards larger (lower, respectively) values, which is the consequence of the deficiency of matter along the rotation axis, i.e. $\rpol < \req$. The mass-radius relationships obtained for different values of $f$ are parallel to each other, with a relatively linear/regular spacing (in log. scale). The slopes are clearly not very sensitive to rotation. However, the immediate consequence is that rotation introduces a confusion for interpretating observations, in the sense that a ``high'' position in the plane can correspond to a static planet with light elements, or to a rotating planet with heavy compounds. The curves are monotonic and smooth, and can clearly be fitted by simple functions. A significant curvature in the mass-radius relationship appears beyond about $10$ Earth masses (see below). This is the reason why, in the following, fits are constructed by using only data falling in the reference mass-range $0.1 \, \mearth \lesssim M \lesssim 10 \, \mearth$. In terms of radius, this roughly corresponds to $0.4 \, \rearth \lesssim \req \lesssim 3. \, \mearth$. This domain is much smaller than in \cite{skhm07}, and slightly extends into the domain of gaseous planets \citep{ed23}. In these conditions, the mass-radius relationship can be well fitted by a quadratic law 
\begin{equation}
y(x) = c_0 +c_1x + c_2 x^2,
\label{eq:quadfit}
\end{equation}
where $x \equiv \log M/\mearth$ and $y \equiv \log\req/\rearth$. Much better functions can indeed be found, in particular a sum of power laws in $y$, but a second-order polynomial is easily reversible \citep[e.g., see][]{fmb07}. The coefficients $c_0$, $c_1$ and $c_2$ are reported in Appendix \ref{sec:coef_singlelayer} for each value of the flattening parameter $f$. As shown in Fig. \ref{fig:mr_static_singlelayer}, the fit reproduces $\log \req$ with an absolute precision better than $0.003$ for each compound, for any $f$. Clearly, $c_0$ exhibits the largest variations. In log. scale, the local slope $\beta$ of the mass-radius relationship is deduced from Eq.\eqref{eq:quadfit}, namely
\begin{flalign}
  \left.\frac{\partial y}{\partial x}\right|_{f=0} = c_1+ 2c_2x \equiv \beta,
  \label{eq:beta_single}
\end{flalign}
which is about $0.29$ for iron and water, and $0.30$ for perovskite. This is slightly lower than $\frac{1}{3}$, typical of a homogeneous body. The leading term is coefficient $c_1$, but the slope slightly decreases with increasing mass ($c_2$ are negative values). We see from Tabs. \ref{tab:ci_iron} to \ref{tab:ci_water} that the $c_i$'s depend monotonically on $f$. The slope $\beta$ marginally increases. It happens that the $c_i$'s can be very well fitted with a quadratic form in $f$. We therefore get a bi-quadratic formula for the mass-radius-flattening relationship of the form
\begin{equation}
y \approx  \sum_{i=0,2} x^i \times \sum_{j=0,2} c_{ij} f^j \equiv y(x,f),
\label{eq:mro_relationship}
\end{equation}
where the $c_{ij}$'s are given in the Appendix \ref{sec:coef_singlelayer_general}  for iron, perovskite and water planets.

\subsection{Relative changes in the mass and radius due to rotation}
\label{subsec:eorot}

By differenciating Eq.\eqref{eq:mro_relationship} at $f=0$, we find
\begin{flalign}
  d y \approx & \underbrace{\left(\sum_{i=0,2}{c_{i1} x^i}\right)}_{\rm rotation}  df+\underbrace{\left(c_{10}+2c_{20}x\right)}_{\rm static} dx.
  \label{eq:mr1storder}
\end{flalign}
It follows that, in the limit of slow rotations, the mass-radius relationship is linearly shifted upwards with respect to the static case by the quantity $\Delta \req = \etarad f \req$ where
\begin{equation}
\etarad \equiv \frac{1}{f}\frac{\Delta \req}{\req}  \approx \ln 10 \times \left.\frac{\partial y}{\partial f}\right|_{f}\approx  \ln 10 \sum_{i=0,2}{c_{i1} x^i},
\label{eq:etarad}
\end{equation}
which depends on the mass. In the range of reference, $\etarad$ varies from $0.33$ at low mass to $0.39$ at high mass. For $x=y=0$, we have
\begin{flalign}
\etarad  \approx  \ln 10 \, c_{01} \approx
  \begin{cases}
 0.3611 \quad \text{for iron planets},\\
 0.3551 \quad \text{for perovskite planets},\\
 0.3614 \quad \text{for water planets},
 \label{eq:dreq}
  \end{cases}
  \end{flalign}
which values are close to $\frac{1}{3}$ (see Sect. \ref{sec:mr}). We see from Tab. \ref{tab:cij} that $c_{01}$ is the largest among the $c_{i1}$'s, whatever the material. Besides, $c_{11}$ being positive, $\etarad$ slightly increases as the mass increases.

 We can make a similar analysis at constant radius. From Eq.\eqref{eq:mr1storder}, the change in the mass is $\Delta M = \etamass f M$ in the limit of slow rotations, with
\begin{equation}
\etamass \equiv \frac{1}{f}\frac{\Delta M}{M} \approx - \ln 10 \times \frac{\left.\frac{\partial y}{\partial f}\right|_{f}}{ \left.\frac{\partial y}{\partial x}\right|_{f} } \approx - \ln 10  \frac{\sum_{i=0,2}{c_{i1} x^i}}{c_{10}+2c_{20}x},
\label{eq:etamass}
\end{equation}
which depends on the mass.  The mass-radius relationship is therefore shifted leftwards with respect to the static case. We see from Tab. \ref{tab:cij} that $c_{11}$ and $c_{21}$ are positive, while $c_{20}$ is always negative. It means that $\etamass <0$ monotonically decreases with the mass (it varies from $-1$ at to about $-1.5$). For $x=y=0$, we get 
\begin{flalign}
  \etamass  \approx -\frac{c_{01}}{c_{10}} \ln 10  \approx
  \begin{cases}
 -1.2308 \quad \text{for iron planets},\\
 -1.1797 \quad \text{for perovskite planets},\\
 -1.2338 \quad \text{for water planets}.
  \end{cases}
  \label{eq:dm}
\end{flalign}

\subsection{Inverse formula and error bars}

We can determine $f$ as a function of $M$ and $\req$ from Eq.\eqref{eq:mro_relationship}. If we define $\mu_j = \sum_{i=0,2}{c_{ij}x^i}$, then $y=\mu_0+\mu_1f+\mu_2f^2$, which is easily solved for $f$. The root is
\begin{equation}
  f=\frac{1}{2\mu_2}\left[-\mu_1+\sqrt{\mu_1^2+4\mu_2(y-\mu_0)} \right],
  \label{eq:fsingle}
\end{equation}
which becomes $f \approx (y-\mu_0)/\mu_1$ for $f \rightarrow 0$. This is therefore the flattening required to match a given mass and radius (see Sects. \ref{subsec:slm} and \ref{subsec:mrfsummary}). Any uncertainty $\delta \req$ in the radius (or  $\delta M$ in the mass) produces an error $\delta f$ in the flattening, which is also the meaning of $\etarad$ and $\etamass$ (see below).

\begin{figure}
       \centering
       \includegraphics[trim={3cm 0.5cm 2.9cm 0.5cm},clip,width=0.98\linewidth]{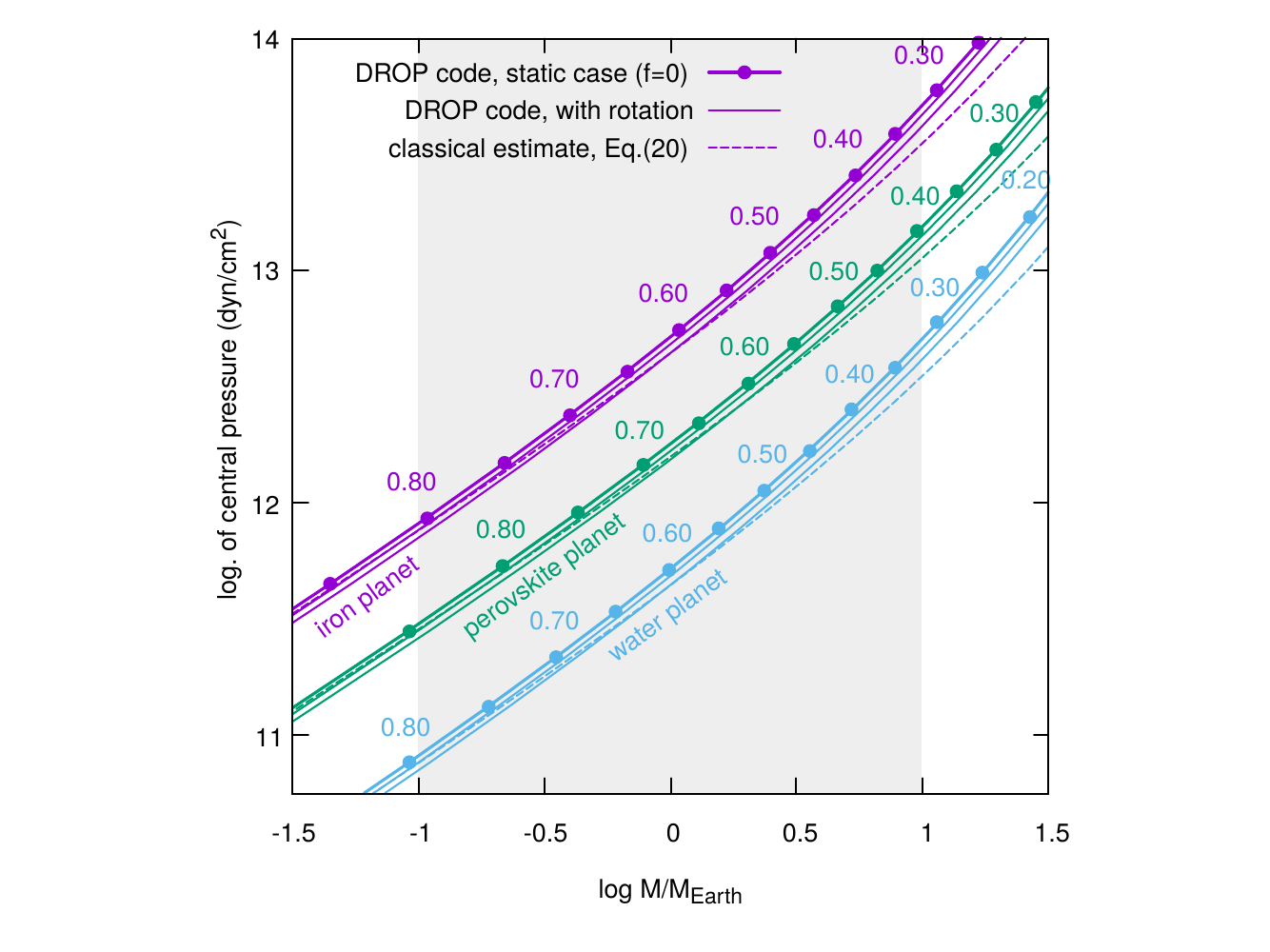}
       \caption{The central pressure $P_c$ in the same conditions as for Fig. \ref{fig:mr_static_singlelayer} for $f \in \{0,0.1,0.2\}$, and the classical estimate $P_0$ for spherical systems ({\it dashed line}) according to Eq.\eqref{eq:p0}.}
       \label{fig:mr_pc_singlelayer}
\end{figure}

\subsection{Central pressure}
\label{subsec:pc}

The central pressure $\pc$ is displayed in Fig. \ref{fig:mr_pc_singlelayer} in the same conditions as above, together with the classical estimate for spherical stars \citep{cox_1968}
\begin{equation}
P_0 = \frac{3 GM^2}{8\pi\req^4},
\label{eq:p0}
\end{equation}
which quantity is, in fact, a lower limit. We first notice the remarkable correlation between $\pc$ and $P_0$ in the static case, which both increase with the mass. This is expected because $P_0 \propto  M^{2(1-2\beta)}$, where the power-law exponent of the mass-radius relationship is, as seen before, less than $\frac{1}{2}$. For a given total mass, $\pc$ is slightly larger in the static case than for a rotating configurations. This result, which concerns a two-layer planet with modified EOS, is therefore similar with the case of a (single-layer) polytropic gas sphere, as established in \cite{ch33}. 

\subsection{Rotation period}

\begin{figure}
       \centering
       \includegraphics[trim={3cm 0.5cm 2.9cm 0.5cm},clip,width=0.98\linewidth]{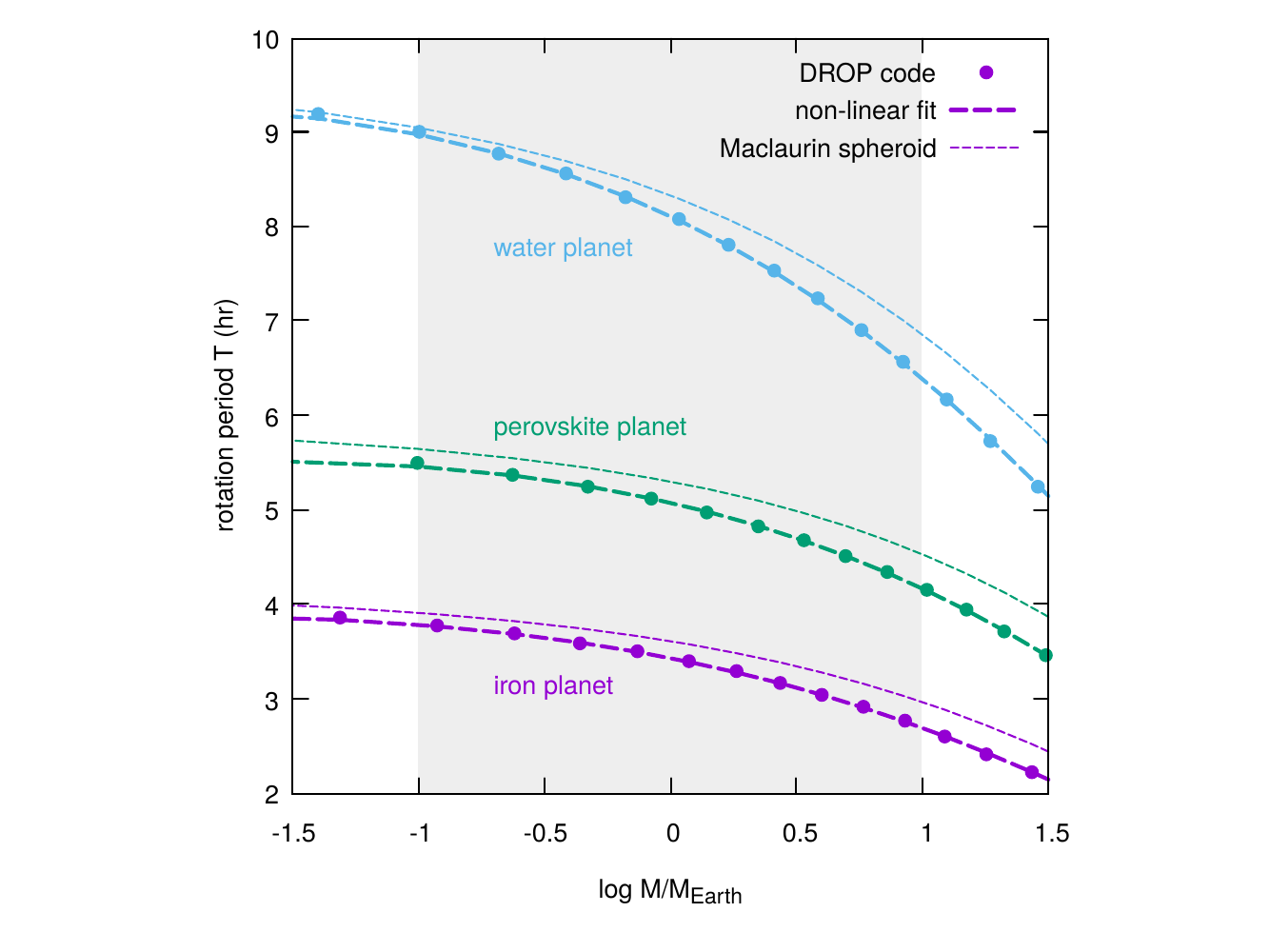}
       \caption{The rotation period $T$ of single-layer planets in the same conditions as for Fig. \ref{fig:mr_static_singlelayer} for $f=0.1$, together with the fit with Eq. \eqref{eq:periodfit}. The period of the Maclaurin spheroid ({\it dotted lines}) having with the same flattening and same mean mass density according to Eq.\eqref{eq:maclaurin0} is also given.}
       \label{fig:mr_period_singlelayer}
\end{figure}

The rotation period $T$ of the planet is displayed versus the mass in Fig. \ref{fig:mr_period_singlelayer} for $f=0.1$. As expected, $T$ increases as $f$ decreases, and is larger for smaller masses. A reliable fit can be obtained with the help of Maclaurin's formula for homogeneous spheroids, which gives the rotation rate $\Omega$ as function of the mean-mass density $\langle \rho \rangle$ and flattening $f$. In the limit of slow rotations, the period of the Maclaurin spheroid $\tml=2 \pi /\Omega$ is given by \citep[e.g.][]{ch33}
\begin{equation}
  \tml^2 \approx \frac{15\pi}{2 G \langle \rho \rangle \left(1- \frac{\rpol^2}{\req^2}\right)} \approx \frac{15\pi}{4 G \langle \rho \rangle f\left(1-\frac{f}{2}\right)}.
  \label{eq:maclaurin0}
\end{equation}
This formula works remarquably well with the modified EOS once the mean-mass density is known, as Fig. \ref{fig:mr_period_singlelayer} shows, and the scaling with $f$ is reliable. By expliciting the mean mass density, we can see that the quantity $T^2 M f/\req^3$ should roughly be a constant (of the order of $0.3$ in the present case). A quadratic fit in $x$ and a linear fit in $f$ appears sufficient, as follows
\begin{equation}
  2\log T - 3y+x+\log f=\sum_{i=0,2}x^i\sum_{j=0,1}{t_{ij}f^j},
  \label{eq:periodfit}
\end{equation}
where the coeficients $t_{ij}$ are given on Appendix \ref{sec:coef_singlelayer_period}. The quality of the fit is visible in Fig. \ref{fig:mr_period_singlelayer}.

\begin{figure*}
  \centering
       \includegraphics[trim={2.5cm 0.5cm 1.5cm 0.5cm},width=17cm]{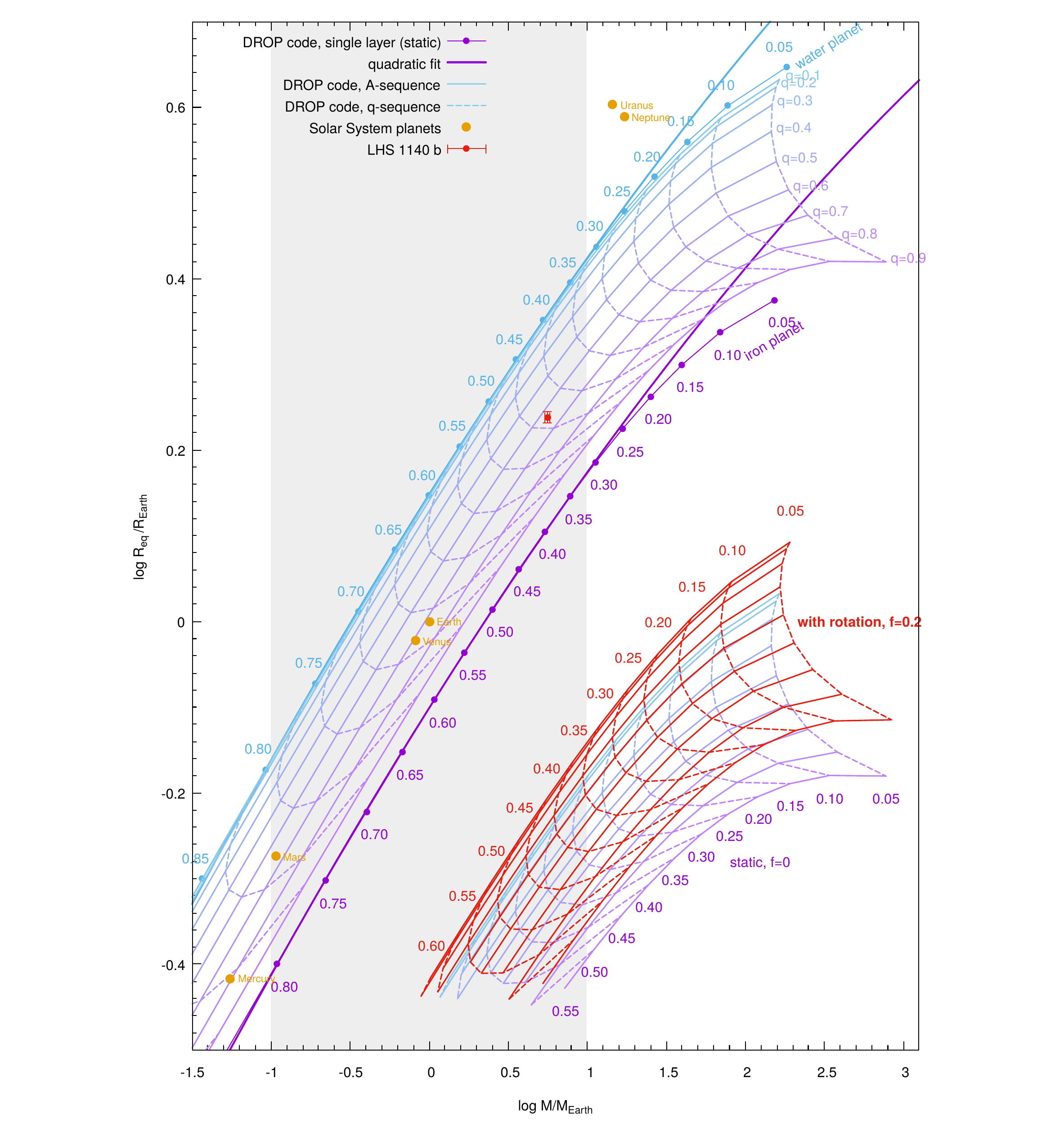}
       \caption{Same caption as for Fig. \ref{fig:mr_static_singlelayer} but in a two-layer IW-planet, in the static case. Labels refer to values of top-to-bottom mass density ratio $\biga$ (single-layer), and the relative size of the core $\qpol$ along the sequences.  The effect of rotation is also shown for $f=0.2$ ({\it bottom right, red lines}; a vertical shift of $-0.6$ applied for clarity).}
       \label{fig:mr_static_IW_full}
\end{figure*}

\subsection{Note about the very-high mass part of the diagram}

At very high mass, the radius goes through a maximum and then decreases with the mass. This behavior is the consequence of the bimodal EOS used here, and is fully expected \citep{zs69,sfs14}. Actually, according to the theory of polytropic spheres, the mass $M$ and radius $\req$ are linked as follows \citep[e.g.][]{horedttextbook2004}
\begin{equation}
  \req^{3-n} \times  M^{n-1} = \const,
  \label{eq:mrpoly}
\end{equation}
and we have $\beta = \frac{1-n}{3-n}$, with the consequence that the radius remains a decreasing function of the mass (i.e. $\beta < 0$) as soon as $n > 1$, which is effective at very high pressure; see Tab. \ref{tab:eosdata}. The maximum occurs at about 550 $\mearth$ for iron and silicate and much beyond for water. It is therefore not visible in the figure, and out of the range of interest here.

\section{Results for two-layer planets}
\label{sec:surveys_tl}

We have run the code with ${\cal L}=2$ for the same parameters $\biga$ and $f$ as above, and with $\qpol \in \{0.1,0.2,\dots,0.9\}$. The mass-radius relationships obtained  for CE=IW in the static case are displayed in Fig. \ref{fig:mr_static_IW_full} as a function of $\qpol$ for a given $\biga$ (hereafter, a $q$-sequence), and as a function of $\biga$ for a given $\qpol$ (hereafter, a $\biga$-sequence). A sub-figure shows the effect of rotation for $f=0.2$. The data obtained for $f=0.1$ and $q=0.5$, which corresponds to Saturn's oblateness, are gathered in Tab. \ref{tab:mr_water_IW} (bottom). We see that the mass-radius relationship stands between the two curves obtained in the single-layer cases C=$\{$I,W$\}$, and we have similar trends as for single-layer cases (i.e. the radius increases as $f$ increases). The $\biga$-sequences are roughly linear in the domain of interest, parallel to each other, but with a marked curvature at high mass. All $\qpol$-sequences are also roughly parallel to each other but are non-bijective, for any $\biga$. As $\qpol$ increases, the mass and radius first decrease until $\qpol \approx 0.6$, then the variation is fully reversed. The results obtained for iron/perovskite and for perovskite/water are displayed in Figs. \ref{fig:mr_static_IP_full} and \ref{fig:mr_static_PW_full}, respectively. We make similar observations as for the IW planet. The diagrams are a little bit more compact because $\rhonp(C)/\rhonp(E)$ is smaller than for the IW case.

\subsection{Fitting procedure}

Whatever the mineralogical composition, we can fit the data in the reference mass range (see Sec. \ref{sec:surveys_sl}) by a quadratic law, i.e. with Eq.\eqref{eq:quadfit}.  The process is a little bit more complex than for the single-layer case, because there is an additionnal parameter. The coefficients obtained for each $\biga$-sequence where $\qpol$ and $f$ are held fixed are now denoted $c_i'$, with $i=\{0,1,2\}$. These are listed in Tab. \ref{tab:ciprim_static_ofqpol} for the static IW-planet, together with mean values $\langle c_i' \rangle$ and half-amplitudes of variation $\Delta_i$ defined by
\begin{equation}
  \begin{cases}
  \langle c'_i\rangle = \frac{1}{9} \sum_{\qpol}{c'_i},\\
  2\Delta_i \equiv \max\{c'_i\}-\min\{c'_i\},
  \end{cases}
  \label{eq:meancprim}
\end{equation}
where the single-layer data are excluded. This procedure is repeated for each value of the flattening parameter.

\subsubsection{A sigmoid for the coefficient $c_0'$}

\begin{figure}
       \centering
       \includegraphics[trim={1.5cm 0.1cm 1.5cm 0cm},clip,width=1.02\linewidth]{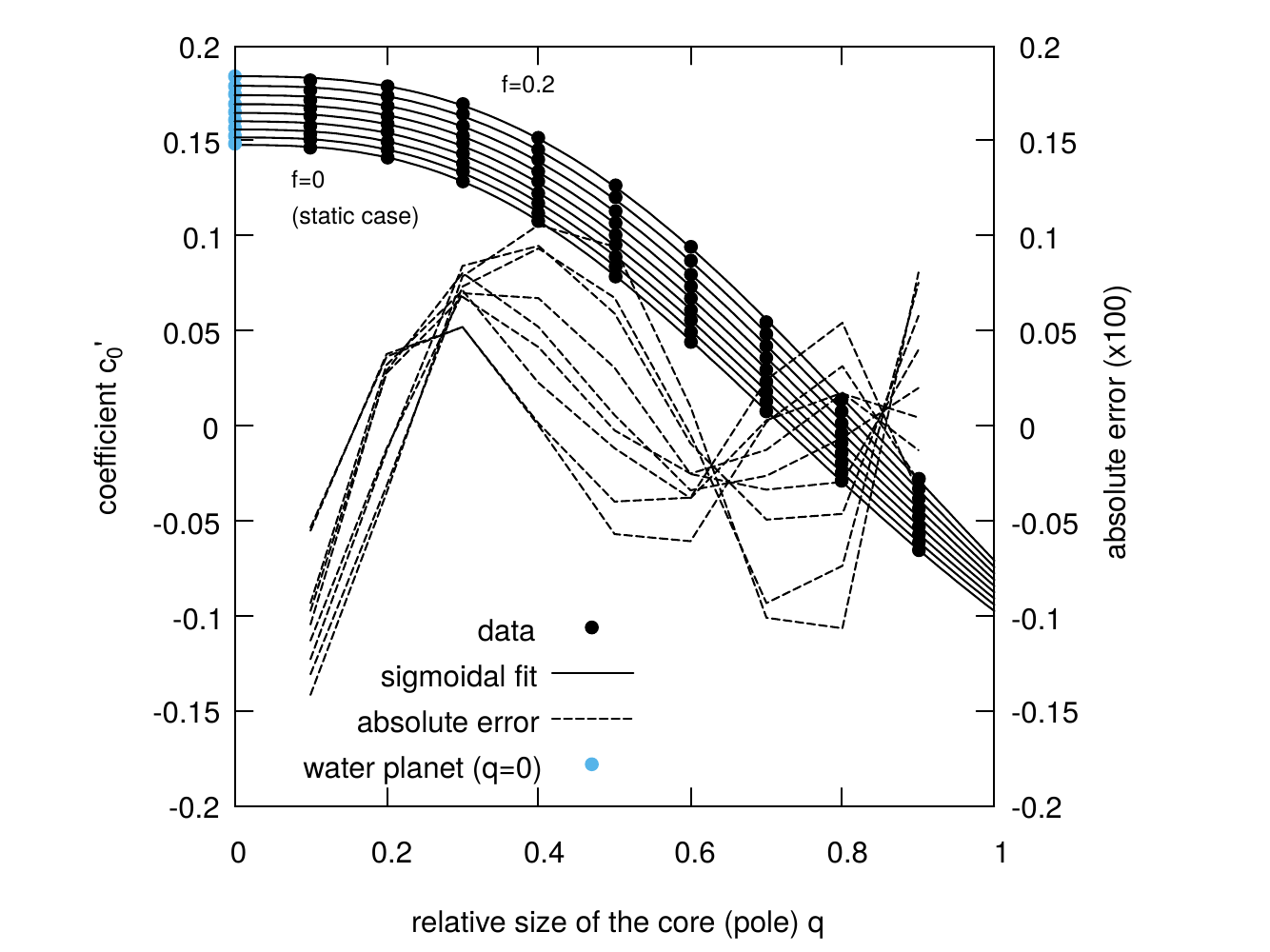}
       \caption{Coefficient $c_0'$ obtained in the IW-case fitting the mass-radius relationship for a given value of the core fractional size $\qpol$, for each flattening parameter $f$ ({\it dots}). The fits by a sigmoid are superimposed ({\it plain lines}), and the absolute error is also shown ({\it dashes lines, right axis}).}
       \label{fig:mr_cprim0}
\end{figure}

We find that $c_0'$ exhibits marked variations with the core size. It is plotted versus $\qpol$ in the IW-case in Fig. \ref{fig:mr_cprim0} for all the flattening parameters considered, including the static case. In fact, $c_0'(\qpol)$ can be approximated by a smooth function, but a quadratic law is not really appropriate. A better choice is a symmetrical sigmoid, namely
\begin{equation}
c_0'(q) = \frac{d_0+d_1 q^\sigma}{1+ d_2 q^\sigma},
\label{eq:sigmoidalfit}
\end{equation}
where the coefficient $d_0=y(0)$ corresponds in principle to the single layer case; see the coefficients $c_{0j}$ in Tabs. \ref{tab:ci_iron} to \ref{tab:ci_water}. The analysis reveals that $d_0$, $d_1$, $d_2$ and the exponent $\sigma$ vary very weakly with $f$, and can all be fitted by a parabola. However, for Eq.\eqref{eq:sigmoidalfit} to be easily/analytically reversible in the variable $f$, we deliberately use a unique value for $\sigma$ regardless of the flattening parameter $f$. We then decide to use the value obtained in the static case, namely $\sigma \approx +2.7$. The parameters $d_0$, $d_1$ and d$_2$ of the sigmoidal fit are given in Tab. \ref{tab:d1d2sigmoid_IW} for the IW-planet. It happens that these quantities vary monotonically with $f$, and can be very well approximated by a parabola, again, i.e. $d_i=\sum_{j=0,2}{d_{ij}f^j}$. The coefficients $d_{ij}$ of these three fits are listed in Tab. \ref{tab:cprimij_dijsigmoid_IWIPPW} for CE=\{IW,IP,PW\}. Figure \ref{fig:mr_cprim0} also displays the absolute error in fitting $c_0'$ by Eq. \eqref{eq:sigmoidalfit}, which turns out to be less than about $0.001$ for $\qpol \le 0.9$ and $f \le 0.2$. We get a similar magnitude for the IP and PW-cases.

\subsubsection{A parabola for the next two, averaged coefficients}

Figure \ref{fig:mr_cprim12mean} shows the mean values $\langle c_1' \rangle$ and $\langle c_2' \rangle$ and the associated deviations $\Delta_i$ defined by Eq. \eqref{eq:meancprim} versus $f$, still in the IW-case. We see that, in contrast with $c_0'$, the coefficients $c'_1$ and $c_2'$ do not vary much with the core size $\qpol$. The size of the error bars is about $0.002$ at most for $c_1'$ and about $0.003$ for $c_2'$, in absolute. As $c_2'$ is lower than $c_1'$ by a factor $10$ for any $f \le 0.2$, working with mean values introduces an absolute error in $y$ of the order of $0.002$, mainly caused by the substitution $c_1' \rightarrow \langle c_1' \rangle$. As $|x| \le 1$ for the reference mass range, the impact is an error of the order of $0.002$ in $y$. Next, we use a parabola to fit $\langle c_1' \rangle$, i.e.
\begin{equation}
  \langle c_1' \rangle = \sum_{j=0,2}{c_{1j}'f^j},
  \label{eq:meancprimij}
\end{equation}
while a linear function appears sufficient for $\langle c_2' \rangle$. The coefficients $c_{1j}'$ and $c_{2j}'$ are listed in Tab. \ref{tab:cprimij_dijsigmoid_IWIPPW} of the Appendix for the three types of composition. The quality of these two fits is visible in Fig. \ref{fig:mr_cprim12mean} (right axis) for CE=IW. As the absolute error is less than $10^{-4}$, the error is dominated by the averaging procedure, i.e. $0.002$ in absolute, and not by the fit of mean coefficients.

\subsection{The parametric mass-radius relationship}
\label{subsec:empriicallaw}

It follows that we get a complete parameteric expression of the mass-radius relationship for two-layer planets, namely
\begin{flalign}
  y & \approx  \underbrace{\frac{d_0+d_1 \qpol^\sigma }{1+ d_2 \qpol^\sigma }}_{\text{coefficient } c_0'} + \langle c_1' \rangle x + \langle c_2' \rangle x^2 \equiv y(x,q,f).
\label{eq:mroq_relationship}
\end{flalign}
The fractional size of the core $\qpol$ is uniquely contained in the first term, while the effect of rotation is hidden in all the coefficients. Given the discussion above, the absolute error in fitting $y$ is estimated to
\begin{flalign}
  \Delta y \approx \underbrace{\Delta c_0'}_{0.01} +\underbrace{\Delta \langle c_1' \rangle}_{0.002} |x|+  \underbrace{\Delta \langle c_2' \rangle}_{0.003} |x|^2 \lesssim 0.015,
\end{flalign}
which corresponds, at most, to a relative error of $1.5\%$ in the radius for the whole reference mass range. Depending on the application and context, Eq.\eqref{eq:mroq_relationship} can be adapted or degraded, for instance by omitting quadratic terms in $x$ and/or $f$. The advantage of this formula is that it is easily inversible in any of the two variables $M$ and $\qpol$ (see below). We can test Eq.\eqref{eq:mroq_relationship}, for instance in the IP-case, by setting $M=1 \, \mearth$ (i.e. $x=0$), $\qpol=0.5$ and $f=0.05$, which corresponds to the example of Fig. \ref{fig:graph1refbis_IP.pdf}. The formula gives $y \approx -0.00196$, leading to $\req \approx 0.99551 \, \rearth$. This is in excellent agreement with the value determined by the {\tt DROP} code and reported in Tab. \ref{tab:ref}.

\begin{figure}
       \centering
       \includegraphics[trim={1.9cm 0.cm 1.7cm 0cm},clip,width=0.98\linewidth]{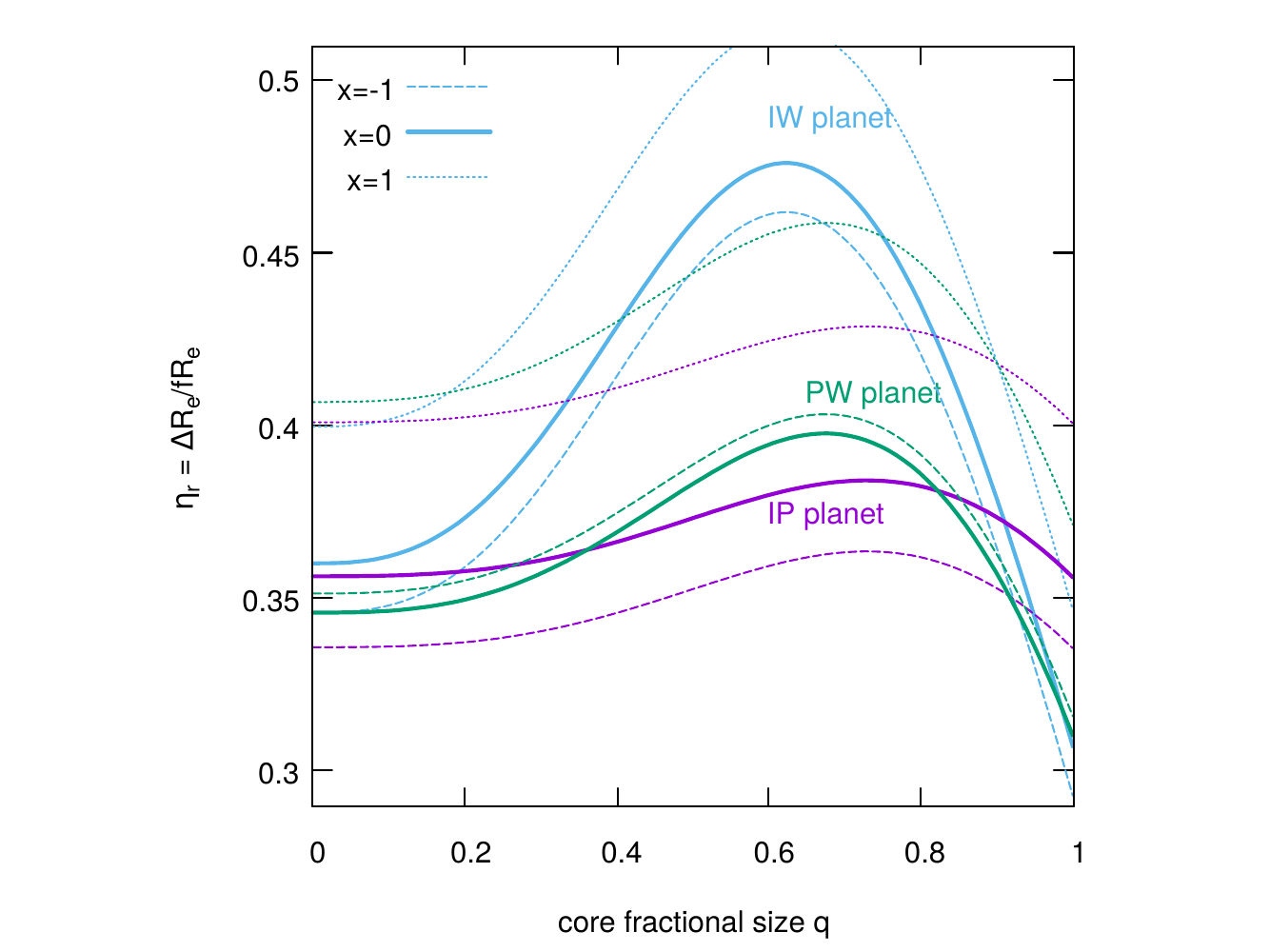}\\
       \includegraphics[trim={1.9cm 0.cm 1.7cm 0.5cm},clip,width=0.98\linewidth]{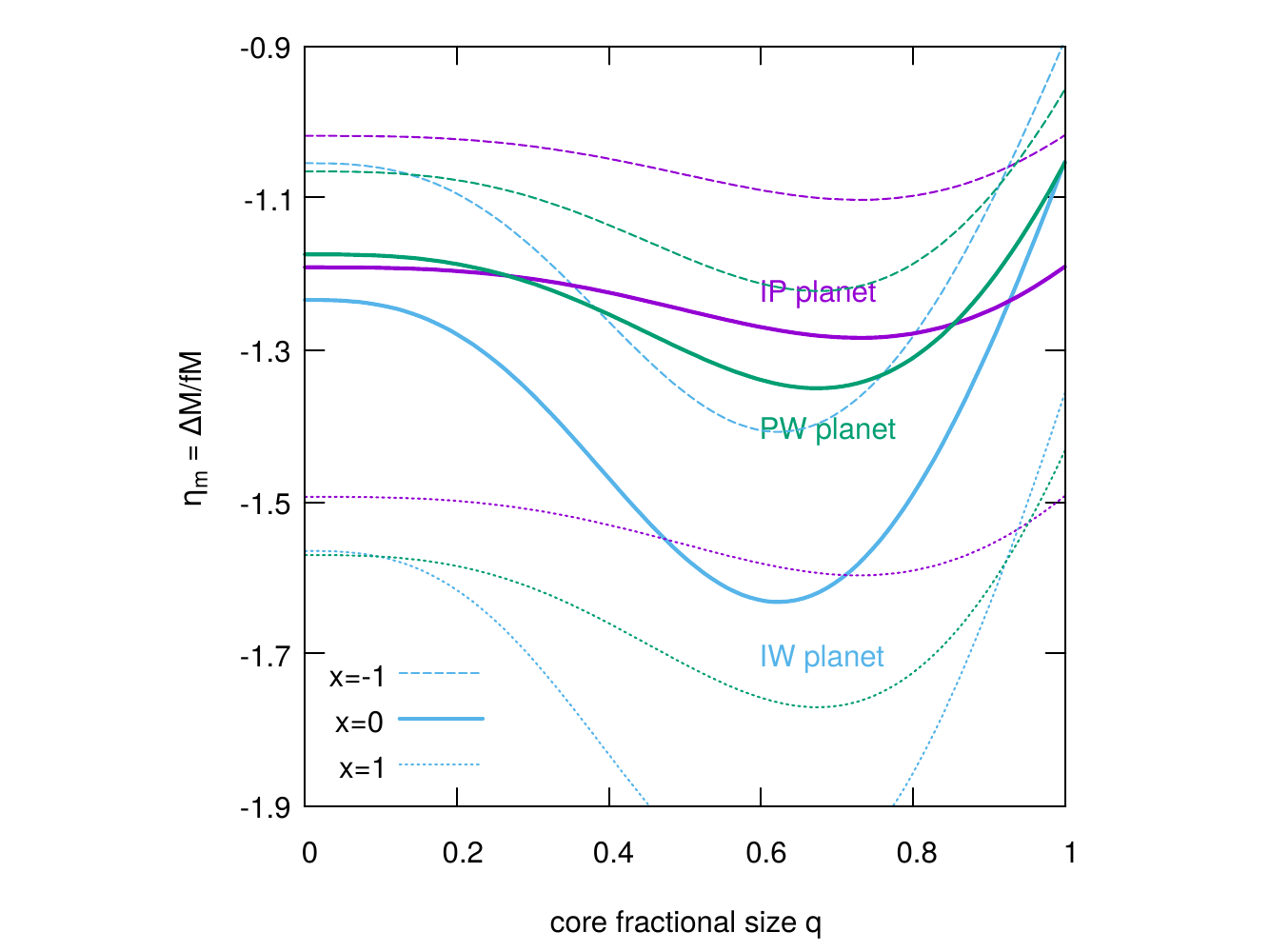}\\
       \includegraphics[trim={1.9cm 0.cm 1.7cm 0.5cm},clip,width=0.98\linewidth]{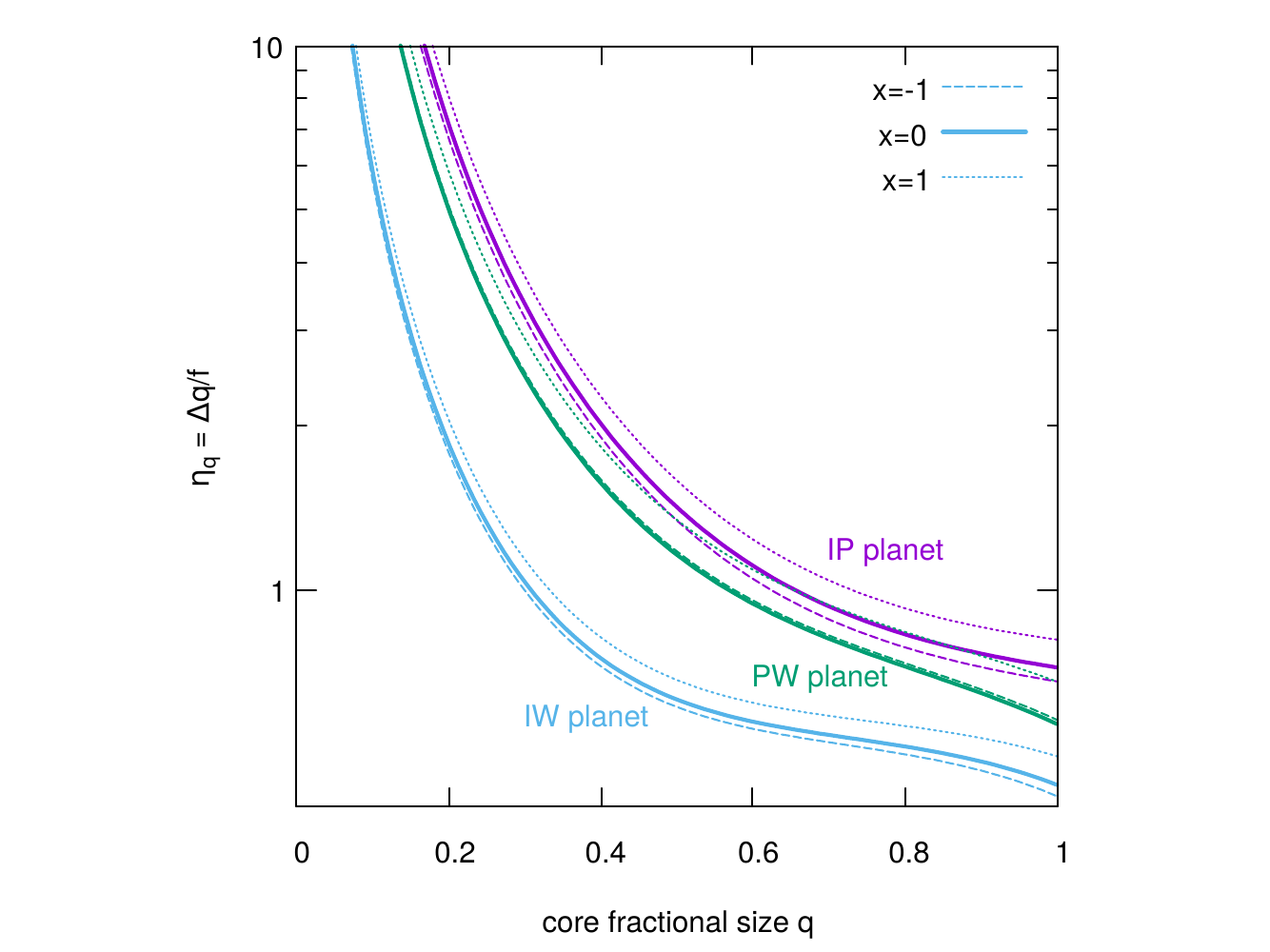}
       \caption{Parameters $\etarad$ ({\it top}), $\etamass$ ({\it bottom}) and  $\etaq$ ({\it bottom}) measuring the relative change in the equatorial radius, mass and core fraction respectively with rotation (in units of $f$) as a function of the core size $\qpol$ for $0.1 \, \mearth$, $1 \, \mearth$ and $10 \, \mearth$.}
       \label{fig:etar_double}
\end{figure}

\subsection{Effect of rotation detailed}
\label{subsec:rot_twolayer}

As done for the single-layer case, we can differenciate Eq.\eqref{eq:mroq_relationship} in the vicinity of $f=0$. The partial derivatives are
\begin{flalign}
  \left.\frac{\partial y}{\partial x}\right|_{q,f} \approx c_{10}'+ 2c_{20}'x \equiv \beta,
  \label{eq:derivx_twolayer}
\end{flalign}
and
\begin{flalign}
  \nonumber
  \left.\frac{\partial y}{\partial f}\right|_{q,f} \approx & \left[\frac{d_{01}+d_{11}q^\sigma }{1+d_{20}q^\sigma}-\frac{(d_{00}+d_{10}q^\sigma)d_{21}q^\sigma }{(1+d_{20}q^\sigma)^2}\right]\\
  & \qquad\qquad\qquad\qquad\qquad+c_{11}'x+ c_{21}'x^2,
  \label{eq:derivf_twolayer}
\end{flalign}
and these depend on the mass and especially on the fractional size. Note that the power-law index $\beta$ of the mass-radius relationship is given by $\langle c_1' \rangle$, which is mainly represented by the coefficient $c'_{10}$. This coefficient is close to $0.3$, as in the single-layer case. We can then deduce $\etarad$ by varying the radius at constant mass, and $\etamass$ by varying the mass at constant radius, namely
\begin{flalign}
  \etarad = \ln 10 \times \left.\frac{\partial y}{\partial f}\right|_{q,f},
  \label{eq:etarad_twolayer}
\end{flalign}
and
\begin{flalign}
  \etamass = - \ln 10 \times \frac{\left.\frac{\partial y}{\partial f}\right|_{q,f}}{ \left.\frac{\partial y}{\partial x}\right|_{q,f} }.
  \label{eq:etamass_twolayer}
\end{flalign}

\begin{figure}
       \centering
       \includegraphics[trim={1.8cm 0.5cm 2.9cm 0.5cm},clip,width=0.96\linewidth]{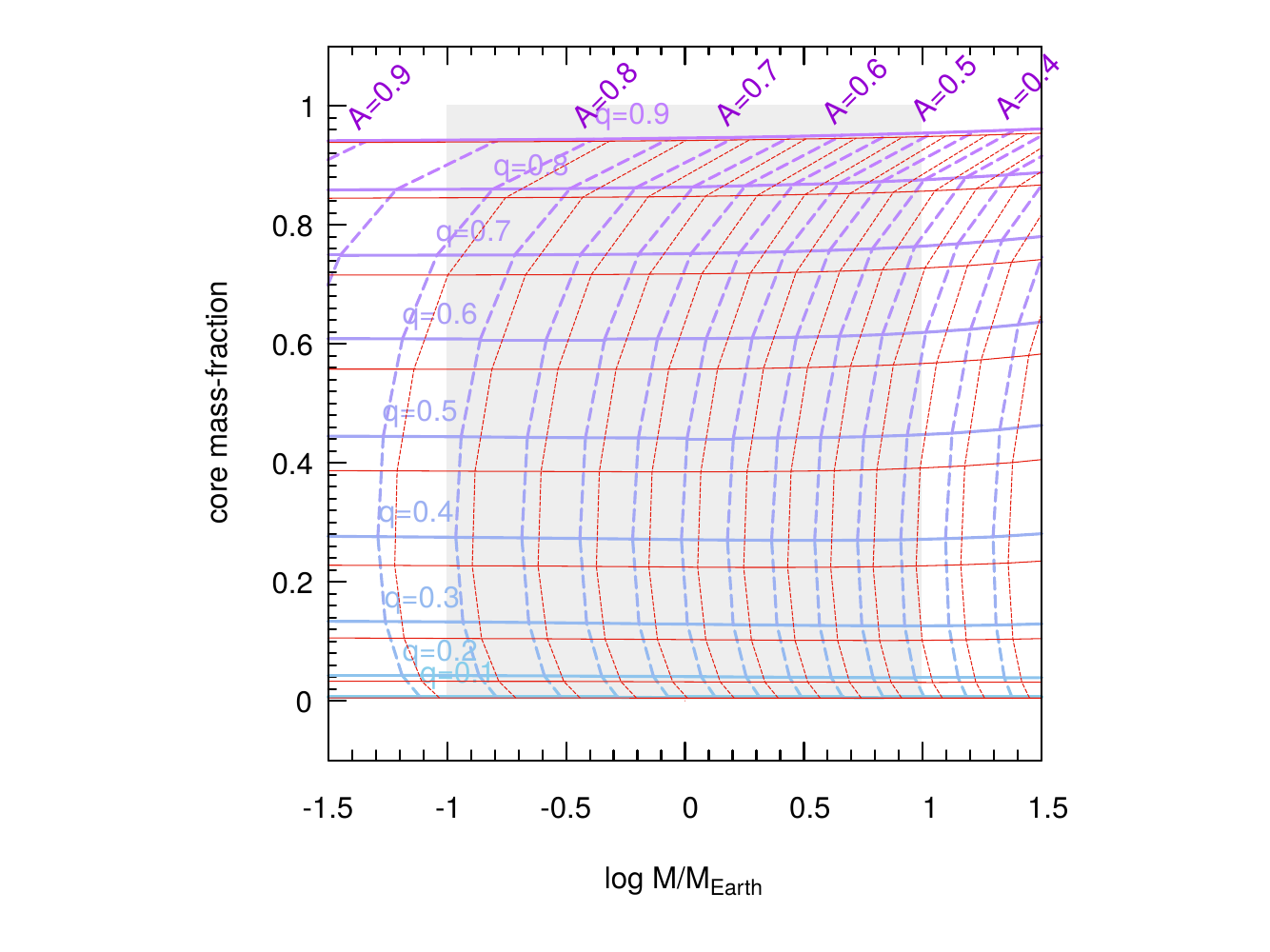}
        \caption{Core mass-fraction versus the mass for the IW planet in the static case ({\it color gradient}) and for $f=0.2$ ({\it red}); see the Appendix \ref{app:mr_cmfIPPW} for the IP and PW-planets.}
       \label{fig:mr_cmf}
\end{figure}

Figure \ref{fig:etar_double} show $\etarad$ and $\etamass$ versus $q$ for the three types of compositions. We notice that response to rotation goes through an extremum, which is fully calculable from Eq.\eqref{eq:derivf_twolayer}. In the static case,  the extremum stands at $\qpol \approx 0.62$ for an IW-planet, which corresponds to a core mass fraction of $0.65$ from Fig. \ref{fig:mr_cmf}. This is $\qpol \approx 0.73$ and $M_1/M \approx 0.60$ for the IP-planet, and $\qpol \approx 0.67$, and $M_1/M \approx 0.56$ for the PW-planet.

We can continue this analysis by considering the third derivative ${\partial y}/{\partial q}$, and define
\begin{flalign}
  \eta_q = \frac{1}{f} \Delta q \approx - \frac{\left.\frac{\partial y}{\partial f}\right|_{x,y}}{\left.\frac{\partial y}{\partial q}\right|_{x,y}},
  \label{eq:etaq_twolayer}
\end{flalign}
which gives the link between $f$ and $q$ keeping simultaneously the mass and the radius of the planet fixed. For instance, a PW-configuration with $f=0$ and $q=0.5$ has the same mass and same radius as another PW-confguration with $f=0.1$ and $q =0.5+0.1\times 1.16 \approx 0.616$ (since we get $\etaq \approx 1.16$ from the graph). This formula must, however, be manipulated with caution because $f$ is intended to be much smaller than unity as well as $\Delta q$. Figure \ref{fig:etar_double} shows $\eta_q$ versus $\qpol$. As expected from Figs. \ref{fig:mr_static_IW_full}, \ref{fig:mr_static_IP_full} and \ref{fig:mr_static_PW_full}, the changes are relatively small for core sizes larger than about $0.6$, and very large at small $q$.

\subsection{Core/envelope discrimination}
\label{subse:cediscrimination}

Following the fitting procedure, $\qpol$ appears only in the sigmoidal representation of coefficient $c_0'$. As a consequence, we can express the relative size of the core as function of the mass, radius and flattening parameter $f$. From Eq.\eqref{eq:mroq_relationship}, we find
\begin{flalign}
  \qpol = \left[\frac{y - c_1' x -c_2'x^2-d_0}{d_1-d_2\left(y - c_1' x -c_2'x^2\right)} \right]^{\frac{1}{\sigma}} \equiv q(x,y,f),
  \label{eq:qpol}
\end{flalign}
which assumes that the term inside the brakets is positive (and less than unity). It is of particular interest because it gives a first estimate of the relative size of the core for a given mass and radius \citep{brinkman24}. This avoids a full scan of the parameter space. Note that any uncertainty in the radius and/or mass induces an error $\delta \qpol$ in the fractional core size $\qpol$ (see Sect. \ref{sec:errorbars}).

\begin{figure}
       \centering
       \includegraphics[trim={1.8cm 0.5cm 2.9cm 0.5cm},clip,width=0.96\linewidth]{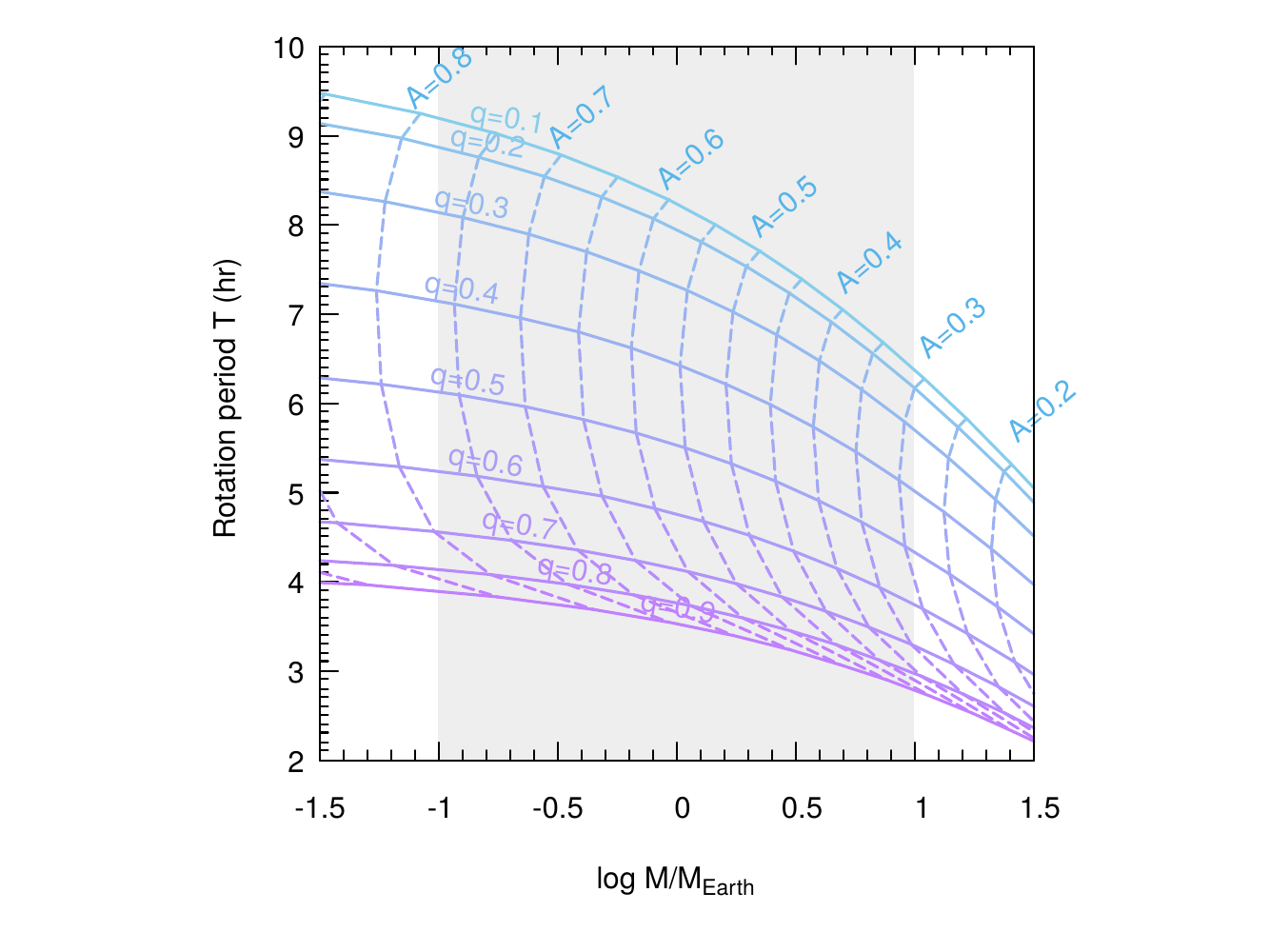}
       \caption{Same caption as for Fig. \ref{fig:mr_cmf}, but for the period $T$ computed for $f=0.1$; see the Appendix \ref{app:mr_cmfIPPW} for the IP and PW-planets.}
       \label{fig:mr_period}
\end{figure}

\begin{figure*}[h]
       \centering
       \includegraphics[trim={8.2cm 1.5cm 3.5cm 4.5cm},clip,width=1.15\linewidth]{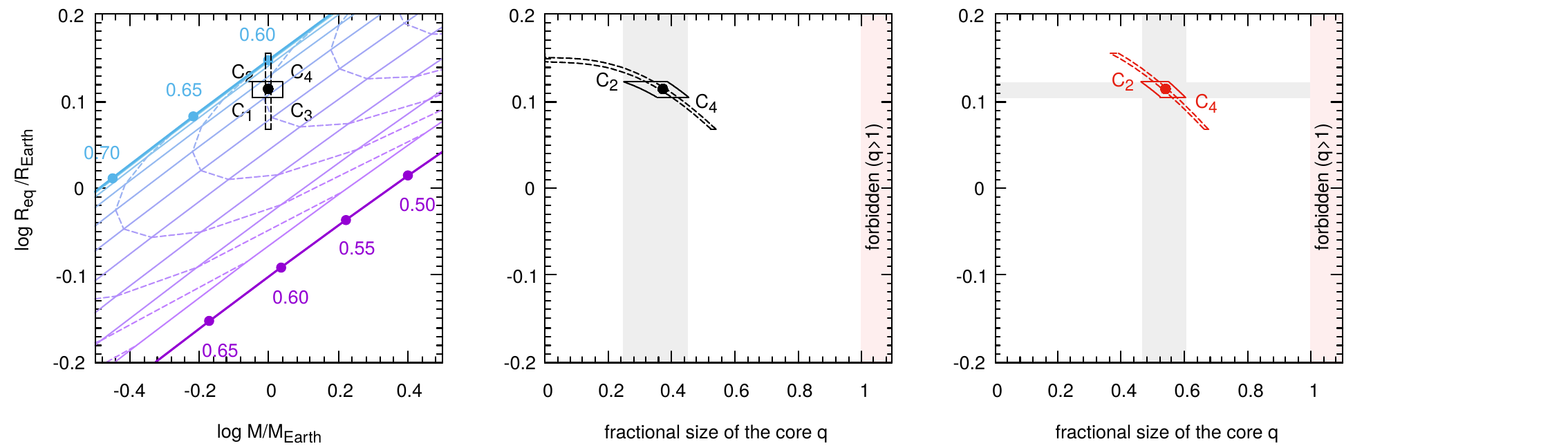}\\
       \includegraphics[trim={8.2cm 1.5cm 3.5cm 4.5cm},clip,width=1.15\linewidth]{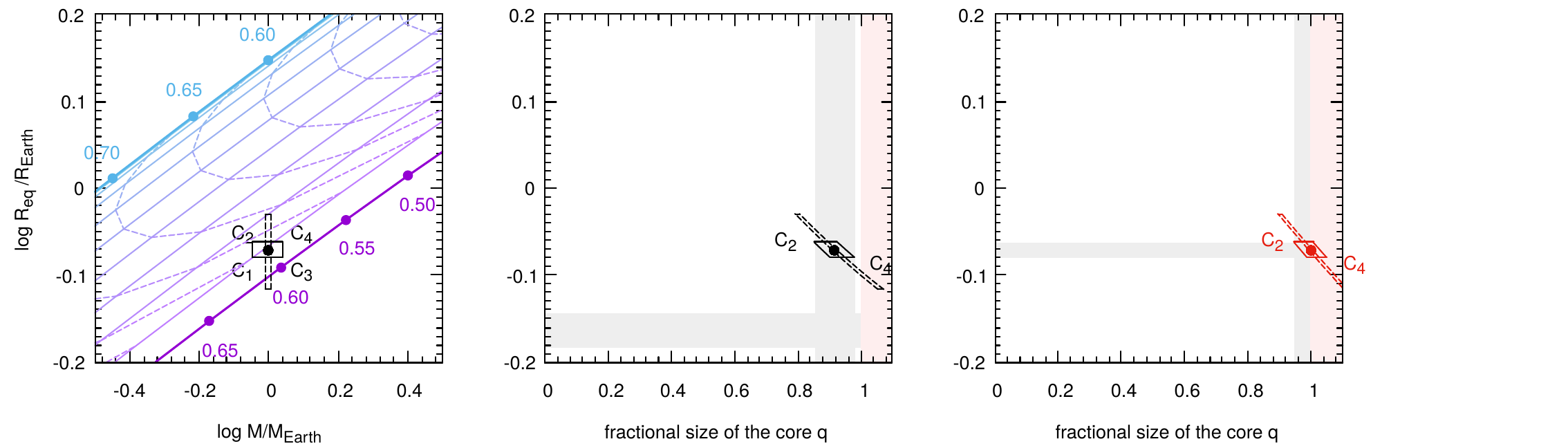}
       \caption{Four sets of fiducial data $(M \pm \delta M,\req \pm \delta \req)$ placed in the mass-radius relationship for IW-planets ({\it left panels}) for $M=1 \, \mearth$, and $\req=1.3 \, \rearth$ ({\it top}) and  $\req=0.85 \, \rearth$  ({\it bottom}). Error bars are  $\delta M/M=0.1$ and $\delta \req/\req=0.02$ ({\it black, plain lines}) and $\delta M/M=0.02$ and $\delta \req/\req=0.1$ ({\it black, dotted lines}). The image of the error bars in terms of fractional size $\qpol$ is shown in the static case ({\it middle panels}) and with fast rotation $f=0.2$ ({\it right panels}). See also Fig. \ref{fig:mr_static_IW_full}.}
       \label{fig:mr_DELTAimage}
\end{figure*}
mass=1.;dmass=0.1;radius=1.3;dradius=0.13
mass=1;dmass=0.1;radius=0.85;dradius=0.085

\subsection{Core mass fraction and rotation period}

The core mass-fraction $M_1/M$ is unfortunately not an input in the simulations, but an output. It is shown in Fig. \ref{fig:mr_cmf} for the IW configurations, and for $f \in \{0,0.2\}$; the figures obtained in the IP and PW-cases are reported in the Appendix \ref{app:mr_cmfIPPW} . We see that the core mass-fraction is very well correlated with the core size $\qpol$. This is true whatever the rotation state of the planet. The effect of rotation is weak. For a given total mass and fractional size of the core $\qpol$, the core mass-fraction is the largest when the planet is static, and it decreases with rotation. Figure \ref{fig:mr_period} shows the rotation period $T$ versus the mass for $f=0.1$. As for the single-layer case, the scaling with $f$ given by Eq. \eqref{eq:maclaurin0} is reliable. The rotation period decreases with increasing mass, which is the consequence of the rise of the mean mass density $\langle \rho \rangle$ with $M$. As we have $\req(IW) > \req(IP) \req(PW)$ for a given mass, $T$ is the largest for IW-planet, and the lowest in the PW-case.

\section{Error bars}
\label{sec:errorbars}

Positionning a planet,  in precomputed mass-radius relationships for single layers enables to determine how far/close to a pure composition the body is, and therefore to deduce plausible combinations of compositions, either in the form of a mixture of materials or in the form of distinct layers (see below). In terms of number of layers, composition and rotational state, the level of degeneracy depends not only in the position but also on the size of error bars on the radius $\delta \req$ and on the mass $\delta M$.
        
\subsection{Single-layer models}
\label{subsec:slm}

As shown in Sect. \ref{sec:surveys_sl}, including rotation moves the mass-radius relationship upwards (i.e. towards larger radii). So, if observational error bars permit, a position in the diagram can be interpreted in terms of a deviation to a pure composition or in terms of rotation, or both. This can easily be checked by running the {\tt DROP} code by selecting all values of $f$ that lead to a given pair $(M \pm \delta M, \req \pm \delta \req)$, but a scan of the parameter space is costly. The advantage of the simulation is that we can impose high values of $f$, up to the break-up velocity. In the theory of ellipsoidal figures of equilibrium \cite{chandra69}, the homogeneous spheroids in rigid rotation are stable for values of the flatenning parameter up to $f \simeq 0.42$, which corresponds to the transition towards states of lower energies in the form of triaxial bodies. However, if $f \le 0.2$, we can use Eq.\eqref{eq:fsingle} as an approximation, and we can include the corresponding error bars $\delta x$ and $\delta y$, and even use the $\etamass$ and $\etarad$ parameters for small $f$ values. An illustration is given in Sect. \ref{sec:lhs1140b}.

\subsection{Two-layer models}

With two-layer diagrams, the degeneracy increases and the interpretation is less clear. For instance, we see from Figs. \ref{fig:mr_static_IW_full}, \ref{fig:mr_static_IP_full} and \ref{fig:mr_static_PW_full} that the data $(x,y)=(1,0.3)$ is compatible with static IW, IP and PW-planets, differing by the size of their core. Again, in the spherical case, we can scan the region in the diagram numerically with the {\tt DROP} and select the right pairs $(\qpol, \biga)$ on input compatible with error bars. We then get a range $\delta \qpol$ of fractional sizes as the image of $(x_0 \pm \delta \req,y_0 \pm \delta M)$. As shown, rotation represents another degree of freedom (and a source of confusion in the inverse problem). We can therefore repeat this treatment by varying $f$, and it follows that $\delta \qpol$ becomes a function of $f$. In fact, we can directly access the image of error bars in terms of $\qpol$ and $f$ by using Eq. \eqref{eq:qpol}.

As an illustration, we show in Fig. \ref{fig:mr_DELTAimage} two positions $(x_0 \pm \delta x,y_0 \pm \delta y)$ in the IW diagram, and the image of error bars in terms of core size $\qpol$ obtained from Eq.\eqref{eq:qpol} in the static case $f=0$ and for $f=0.2$. The extreme points C$_i$ obtained by considering $\pm \delta x$ and $\pm \delta y$ are also reported. We see that:
\begin{itemize}
 \setlength{\itemsep}{+0.5pt}
\item the image of the error bar in the radius $\req \pm \delta \req$ is an error bar in the core size around $\qpol_0$,
\item the image of the error bar in the mass $M \pm \delta M$ is another error bar in the core size around $\qpol_0$,
\item the two error bars in the core size are not equal, not symmetrical around $\qpol_0$, non-linear in $\delta \req$ and $\delta M$, and sensitive to rotation.
\end{itemize}

{\bf Error bars in terms of fractional size}. We see from the figures that $\qpol \in [\qpolmin,\qpolmax]$, where the lower and upper bounds correspond to points $C_2(x_0 - \delta x,y_0 + \delta y)$ and $C_4(x_0 + \delta x,y_0 - \delta y)$, respectively. We can therefore retain
\begin{equation}
  \begin{cases}
    \qpolmin \equiv  q(x_0-\delta x,y_0+\delta y,f),\\
    \qpolmax \equiv q(x_0+\delta x,y_0-\delta y,f),
  \end{cases}
\end{equation}
where $\delta x>0$ and $\delta y>0$. Note that $\qpolmin$ and $\qpolmax$ are not symmetrical with respect to $q_0=q(x_0,y_0,f)$, unles the uncertainties are small. In these conditions, we have
\begin{equation}
  (x_0-\delta x,y_0+\delta y) \rightarrow 2\delta q \approx \qpolmax - \qpolmin
\end{equation}
in the general case. In the limit of small uncertainties, we can deduce an expression for $\delta q$ by differenciation of Eq.\eqref{eq:qpol}. This approach uses the derivatives established in Sect. \ref{subsec:rot_twolayer}.\\

{\bf Error bars in terms of flattening}. The flattening parameters compatible with a given observational data can be estimated from $\etarad$ and $\etamass$, by using Eqs. \eqref{eq:etarad_twolayer} and \eqref{eq:etamass_twolayer} where $\Delta \req$ and $\Delta M$ are to be replaced by $\delta \req$ and $\delta M$, respectively. Assuming that the spherical configuration $f=0$ is reachable by the model and $q$ is fixed, we have
\begin{equation}
  \delta f \approx \max \left( \frac{1}{\etarad}\frac{\delta \req}{\req} , -\frac{1}{\etamass}\frac{\delta M}{M} \right),
  \label{eq:fmax}
\end{equation}
which differs from one composition to another. So, for a given fractional size $\qpol$, we have
\begin{equation}
  (x_0-\delta x,y_0+\delta y) \rightarrow \delta f
\end{equation}
where $\delta f$ can be directly read  from  Fig. \ref{fig:etar_double}.

\section{Illustration with the parameters of LHS~1140~b}
\label{sec:lhs1140b}

For a suitable application, we searched for a transiting super-Earth with a confirmed status and a good precision on the mass and on the radius. We have selected the targets from the Nasa Exoplanet Archive \citep{akeson13} and European exoplanet database with a maximum radius of $1.8\rearth$. From the semi-major axis $a$ and mass $M_\star$ of the host star, the candidates have then been ordered with increasing parameter $a^6/M_\star^2$ which scales with their tidal locking time \citep{zahn89,lwmm15,barnes17}. Unfortunately, all robust super-Earth candidates with a radius measurement currently discovered are expected tidally locked, because of their short orbital periods and presumed age. The comparison with our models can therefore not account for the  final spin state of the planet, and therefore remains purely illustrative. We have selected LHS~1140~b, which represents a fair trade off, because it exhibits among the largest synchronisation time scale. This planet has been detected by both photometric transit and radial-velocity measurements by different groups \citep{dittman17,ment19,ll20,cad24}. It orbits a M-dwarf star located at about $12$ pc away from the Sun. It has an averaged mass-density compatible with an iron/silicate content. In addition, the instellation conditions seem to permit water, if present, to be in liquid phase, making this object of particular interest in several respects \citep{Wandel2018}. This planet has recently been modelled by \cite{rm24} and \cite{dal24}, with up to four differenciated layers (including a thin atmosphere). The water layer could be $2000$ km deep, and the ocean might represent up to $7 \%$ of it in the nominal cases. The internal structure of LHS~1140~b is still under debate. It is commonly interpreted either as water world or a mini-Neptune, but the recent spectroscopic observation with JWST disfavor a body with dihydrogen- and methane-rich atmosphere, hence the second option \citep{cadieux24b}.

In this section, we use $\rsph = 1.730 \pm 0.025 \, \rearth$ as the spherical radius and the mass is $M = 5.60 \pm 0.19 \, \mearth$, according to \cite{cad24}. In decimal logarithm, these data correspond to $x_0 \pm \delta x \approx 0.7482^{+0.0145}_{-0.0150}$ and $y_0 \pm \delta y \approx 0.2380^{+0.0062}_{-0.0063}$. In the absence of any information about the orientation of the rotation axis with respect to the line of sight, we assume $\rsph = \req$ for simplicity. This setting is correct only in the case of a circular transit, i.e. the planet is exceptionnally seen pole-on. We can then directly place the planet in the mass-radius diagrams, without any bias. Otherwise, a correction must be applied (see Sect. \ref{subsec:correction}).

\begin{figure}
       \centering
       \includegraphics[trim={3cm .7cm 1.8cm 0.5cm},clip,width=8.4cm]{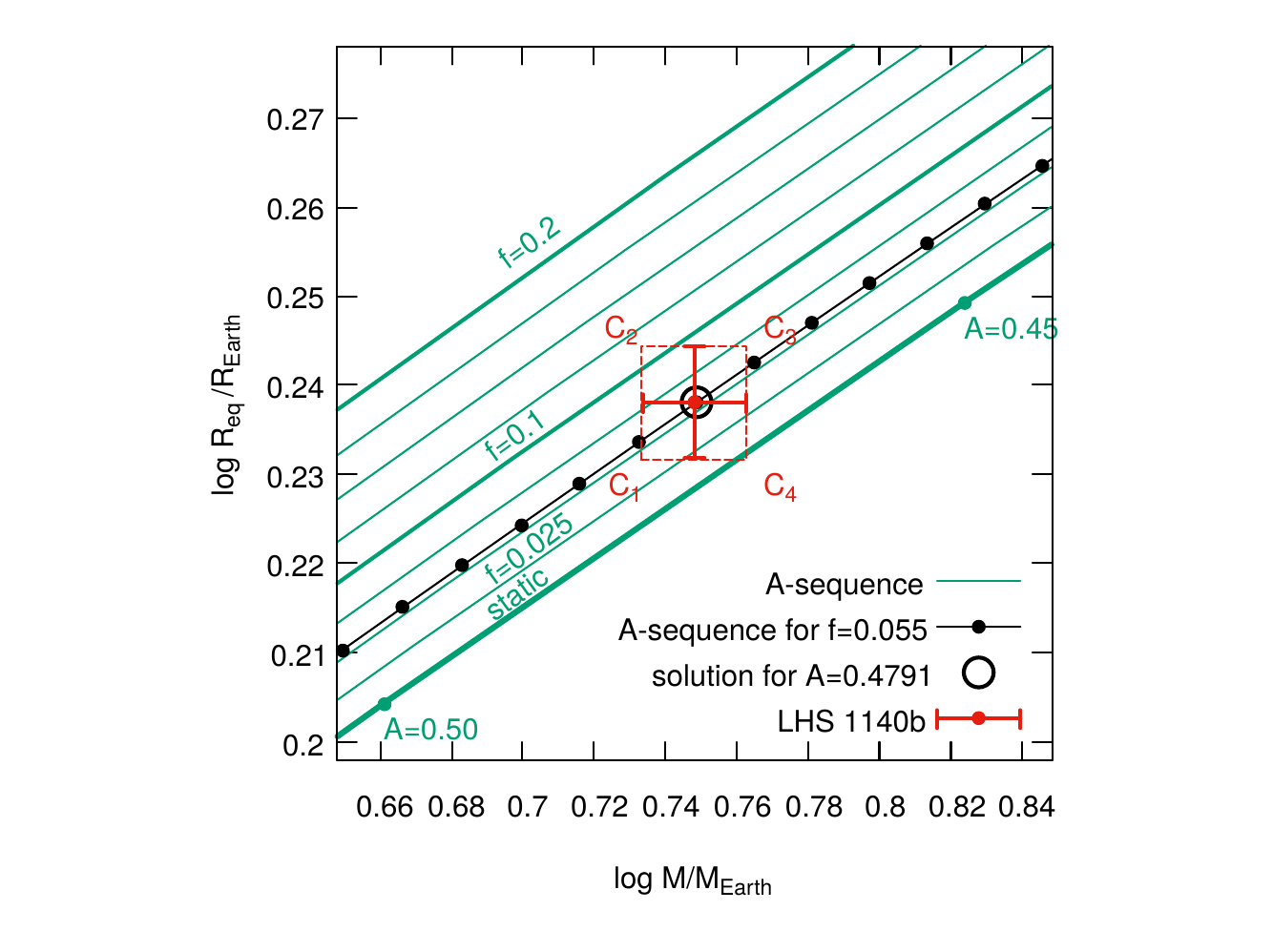}
       \caption{Position of LHS~1140~b ({\it red}) in the mass-radius diagrams for single-layer, pure-perovskite planets obtained for different flattening parameters. The $A$-sequence obtained wit the {\tt DROP} code for $f=0.055$ is also plotted ({\it black}); see Fig.\ref{fig:mr_static_singlelayer}.}
       \label{fig:lhs1140b_perovskitediagram}
\end{figure}

\subsection{Single-layer models}

We see from Fig. \ref{fig:mr_static_singlelayer} that a pure water planet is not possible, and that the iron hypothesis is not sustainable, unless a very large rotation rate is invoked. Actually, as shown before, the mass-radius relationships are shifted upwards when $f>0$, and the faster the rotation, the larger the shift. The only plausible solution is obtained with perovskite, but some rotation is necessary. According to the figure, $f < 0.1$ at a one sigma confidence level. A zoom is presented in Fig. \ref{fig:lhs1140b_perovskitediagram}. By using the {\tt DROP}-code, we find $f = 0.055$ for the central value of the observational interval, while we get $f \approx 0.064$ from the fit by Eq.\eqref{eq:fsingle}. The deviation between these two values is due to the quality of the fit. By considering errors bars and especially points C$_2$ and C$_4$ (see Sect. \ref{sec:errorbars}), we get the values of Tab. \ref{tab:lhs1140b_pplanet}, where the associated rotation period is also reported.

\begin{table}
  \caption{Flattening from Eq.\eqref{eq:fsingle} and rotation period from Eq.\eqref{eq:periodfit} for the parameters of LHS~1140~b assuming a pure perovskite planet, compared to the Earth.}
  \label{tab:lhs1140b_pplanet}
  \begin{tabular}{lllrr}\hline\hline
    point   &  $x$     & $y$       &  flattening $f$     & period $T$ \\\hline
$(x_0,y_0)$ & $0.74822$ & $0.23804$ & $0.0636$ & $5.51$\\
C$_2$       & $0.73320$ & $0.24428$ & $0.1204$ & $4.08$\\
C$_4$       & $0.76268$ & $0.23172$ & $0.0026$ & $26.78$\\\hline
Earth       & $0$       & $0$       & $0.0034$ & $24 \quad \,$\\\hline
  \end{tabular}
\end{table}

\begin{figure}
       \centering
       \includegraphics[trim={3cm .7cm 2.1cm 0.5cm},clip,width=8.4cm]{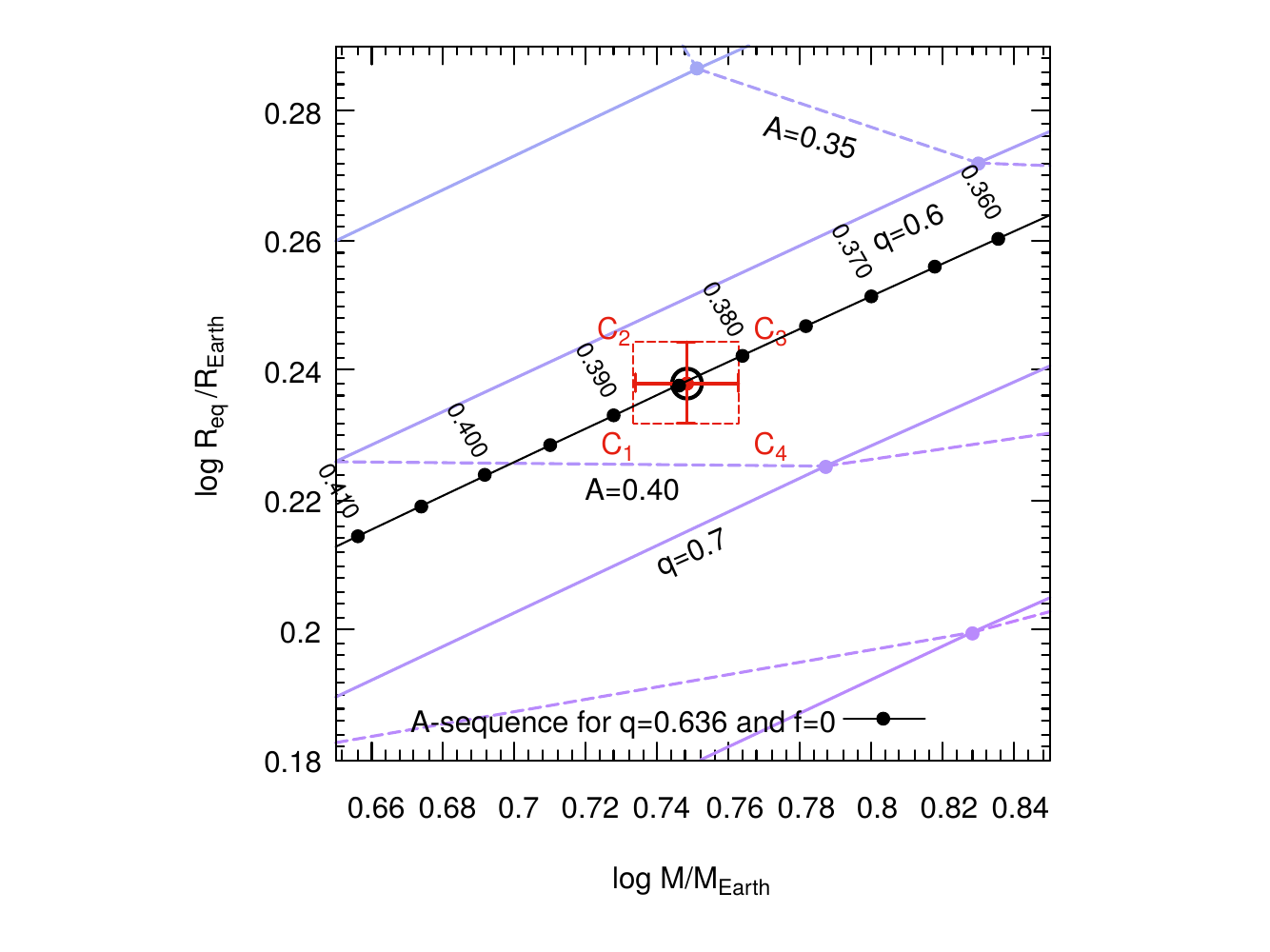}\\
       \includegraphics[trim={3cm .7cm 2.1cm 0.5cm},clip,width=8.4cm]{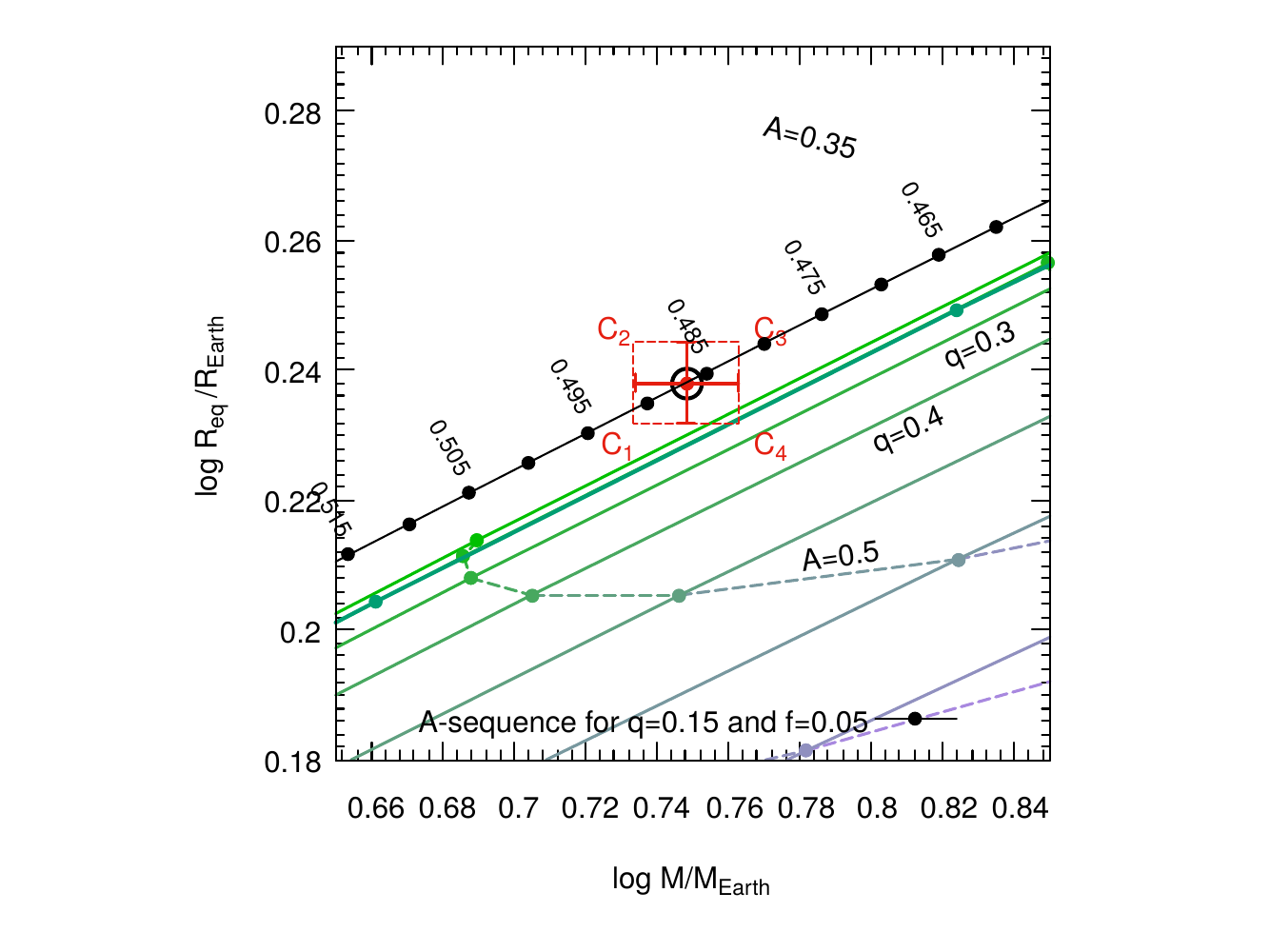}\\
     \includegraphics[trim={3cm 0.5cm 2.1cm 0.5cm},clip,width=8.4cm]{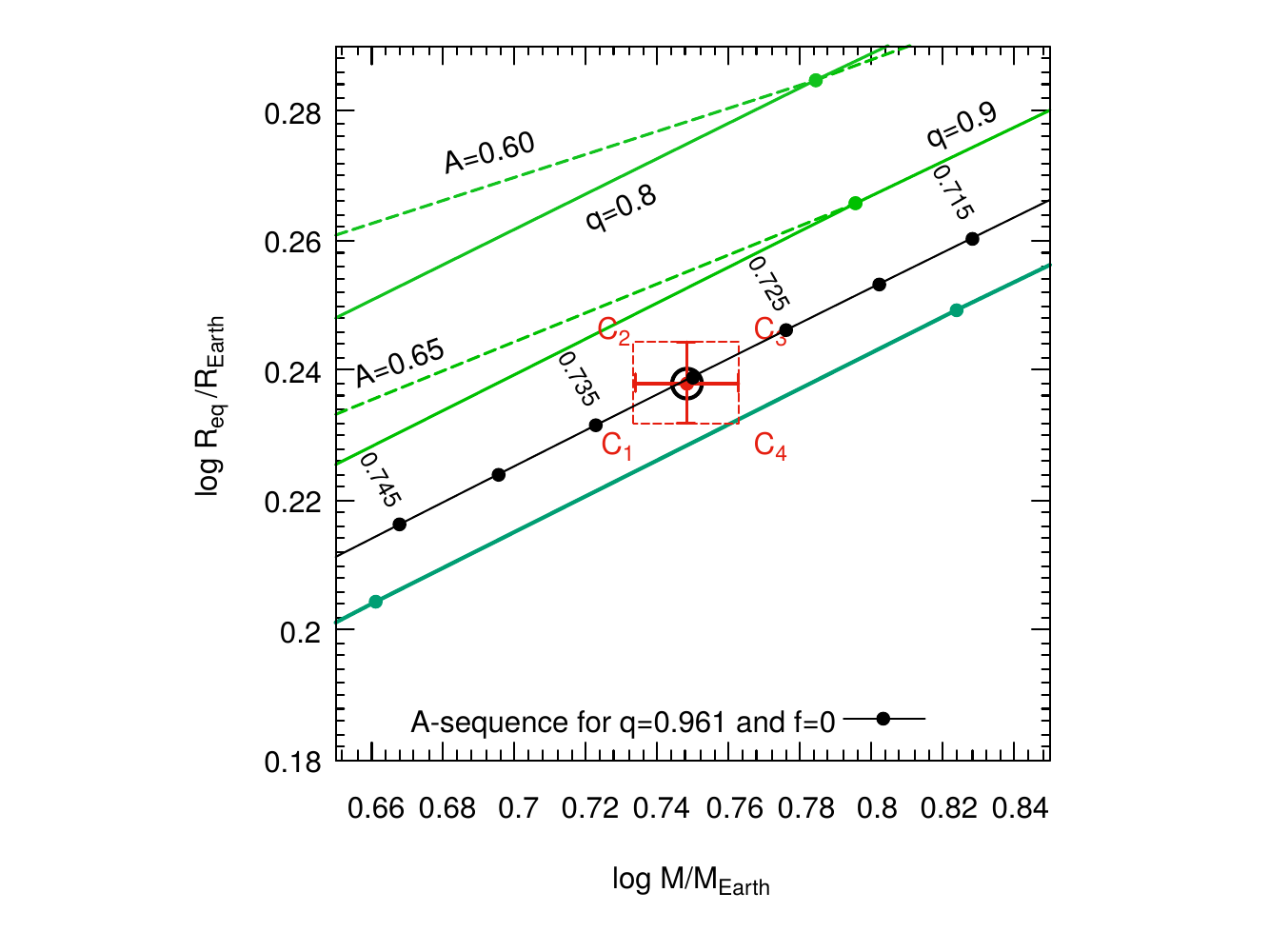}
       \caption{Position of LHS~1140~b ({\it red}) in the mass-radius diagram for IW-planets ({\it top panel}), for IP-planets ({\it middle planel}) and for PW-planets ({\it bottom planel}). The one-sigma error bars are taken from \citep{cad24}. The best static $A$-sequence is also plotted for IW- and PW-configurations ({\it black}); see also Fig. \ref{fig:graph1_lhs1140b.pdf} and Tab. \ref{tab:ref_lhs1140b}.}
       \label{fig:lhs1140b.pdf}
\end{figure}

\subsection{Two-layer models} Figure \ref{fig:lhs1140b.pdf} shows the position of the planet in the mass-radius diagrams for an IW-, an IP- and a PW-planet, together with the error bars through the points C$_i$'s. We can feed Eq.\eqref{eq:mroq_relationship} with the observational data, and in particular with the error bars, as done in the single-layer case. The results obtained for three rotational states are gathered in Tab. \ref{tab:lhs1140b_twolayer}. In the static case, LHS~1140~b is compatible with an iron core/water envelope planet for $\qpol \approx 0.64 \pm 0.03$, and with a perovskite core/water envelope planet for $\qpol \approx 0.94 \pm 0.04$, while the IP solution is ruled out. New solutions are permitted with rotation. While we can consider relatively large values for $f$, it is interesting to use Eq.\eqref{eq:fmax} to get the range of rotation states compatible with the error bars. For LHS~1140~b, we have $\delta \rsph/\rsph \approx 0.01445$, and $\delta M/M \approx 0.03393$. As a consequence, from the fractional size obtained in the spherical case, we find for the range of flattening parameters compatible with the error bars
\begin{itemize}
  \setlength{\itemsep}{+1pt}
\item $\delta f \approx 0.029$ for the IW-planet,
\item $\delta f \approx 0.038$ for the PW-planet,
\end{itemize}
and the corresponding period is $T \approx 7.5$ h in both cases.\\

\begin{figure}
       \centering
       \includegraphics[trim={1.4cm 1.cm 1.3cm 2cm},clip,width=9cm]{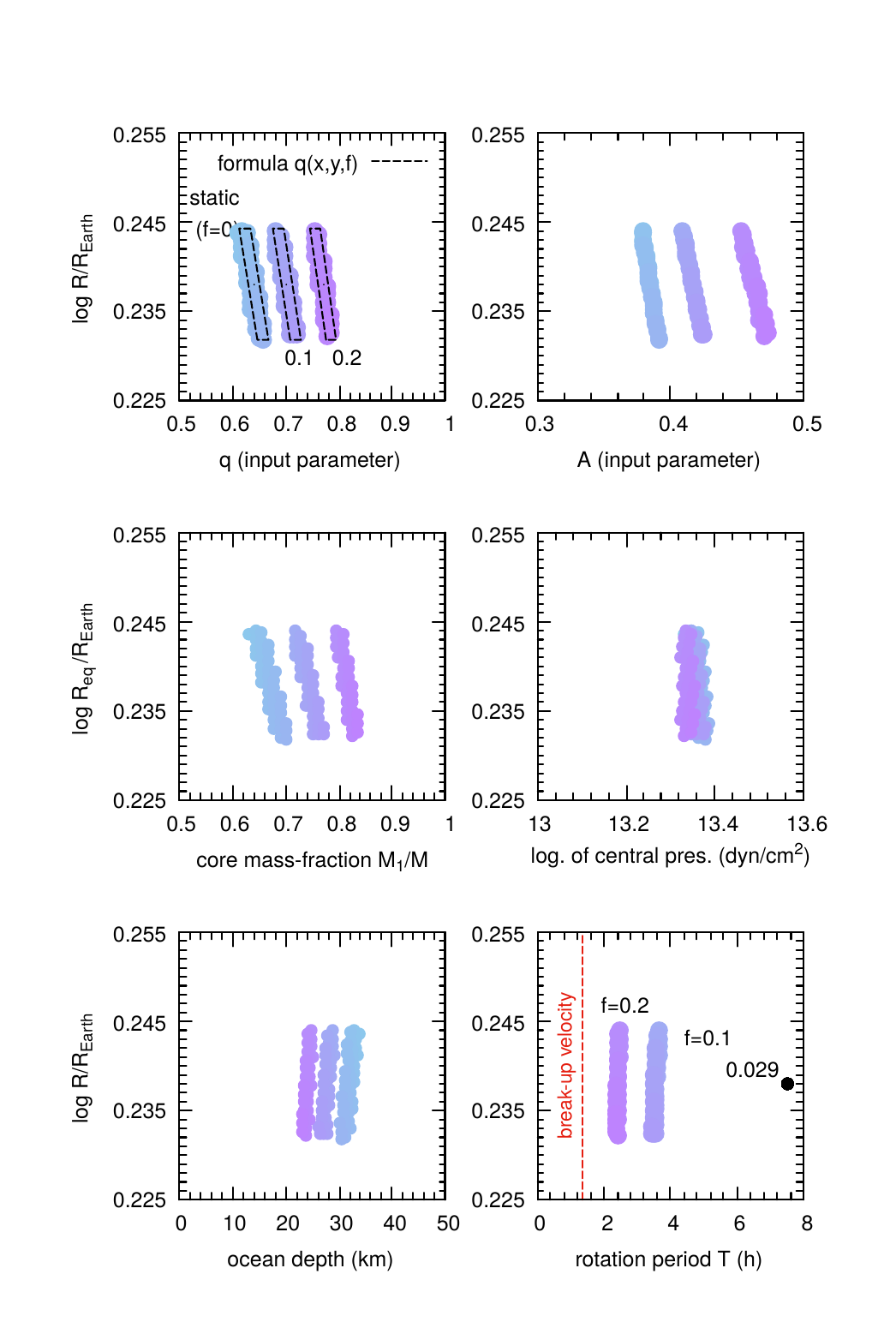}
        \caption{Structure parameters of LHS~1140~b compatible with the error bars (see Fig. \ref{fig:lhs1140b.pdf}, top panel) computed with the {\tt DROP} code, assuming a two-layer, IW-planet in three rotational states. The color code refers to values of the core fractional size, which is smaller in the static case than with rotation; see also Fig. \ref{fig:mr_static_IW_full}. Regarding the ocean depth, see Sect. \ref{sec:odepth}. As an indication of the break-up velocity (which depends on $\qpol$), we have reported in the last panel the Keplerian period at the planet equator ({\it red}) as well as the period associated to $\delta f = 0.029$, which amplitude corresponds to the error bars in the static case; see Eq. \eqref{eq:fmax}.}
       \label{fig:lhs1140b_details_IW}
\end{figure}

\begin{table}
  \caption{Relative core size $\qpol$ calculated from Eq.\eqref{eq:qpol} for LHS~1140~b in different spin states $f \in \{0,0.1,0.2\}$, assuming a two-layer planet.}
  \label{tab:lhs1140b_twolayer}
  \begin{tabular}{llrrrrr}\hline\hline
     & $f$  & $(x_0,y_0)$   & C$_1$    & C$_2$    & C$_3$     & C$_4$ \\\hline
   IW & $0$   & $0.641$ & $0.647$ & $0.613$ & $0.634$ & $0.668$ \\
& $0.10$ & $0.703$ & $0.708$ & $0.676$ & $0.696$ & $0.729$ \\
& $0.20$ & $0.771$ & $0.776$ & $0.746$ & $0.765$ & $0.795$ \\\hline
 IP  & $0$   & $q^\sigma <0$ & $q^\sigma <0$  & $q^\sigma <0$ & $q^\sigma <0$ & $q^\sigma <0^\ddagger$ \\
    & $0.0681^\star$ & $0$ \\
    & $0.0636^\dagger$ & $0$ \\
& $0.10$ & $0.3277$ & $0.364$ & $q^\sigma <0$ & $0.280$ & $0.466$ \\
& $0.20$ & $0.5638$ & $0.580$ & $0.470$ & $0.547$ & $0.637$ \\\hline
 PW & $0$   & $0.937$ & $0.946$ & $0.894$ & $0.927$ & $0.979$ \\
    & $0.0962^\star$ & $1$ \\
    & $0.10$ & $q^\sigma >1$ & $q^\sigma >1$ & $0.964$ & $0.994$ & $q^\sigma >1$ \\
& $0.20$ & $q^\sigma >1$ & $q^\sigma >1$  & $q^\sigma >1$ & $q^\sigma >1$ & $q^\sigma >1$  \\\hline
  \end{tabular}
$^\star$single-layer limit of Eq.\eqref{eq:qpol}\\
  $^\dagger$from Eq. \eqref{eq:fsingle}\\
  $\ddagger$tiny core solution with the {\tt DROP} code 
\end{table}

\noindent{\bf The IW hypothesis}. If rotation is accounted for, type IW solutions remains fully available. The iron core is larger and more massive. By using the {\tt DROP} code, we can scan the parameter space and extract the structures compatible with the observational data.  In this task, Eq.\eqref{eq:qpol} is of great help as the right range of fractional sizes $\qpol$ is fastly identified. Figure \ref{fig:lhs1140b_details_IW} shows the fractional size $\qpol$, the parameter $A$, the core mass fraction $M_1/M$, the central pressure $\pc$, an estimate of the ocean depth (see Sect. \ref{sec:odepth}) and the rotation period $T$ obtained for $f \in \{0,0.1,0.2\}$. We see that the range of fractional sizes as given by Eq.\eqref{eq:qpol} is in very good agreement with the numerical approach. We find $\qpol \approx 0.635 \pm 0.025$, and the core-mass fraction is $0.66 \pm 0.04$ in the non-rotating case. These ranges are pushed towards higher values with rotation. The ocean-depth is estimated to be at least about $30$ km, and it decreases as $f$ increases. The rotation period, which roughly varies like $1/\sqrt{f}$, is already as low as about $4$ h for $f=0.1$.\\

\begin{figure}
       \centering
       \includegraphics[trim={1.cm 7.7cm 14.6cm 0cm},clip,width=1.\linewidth]{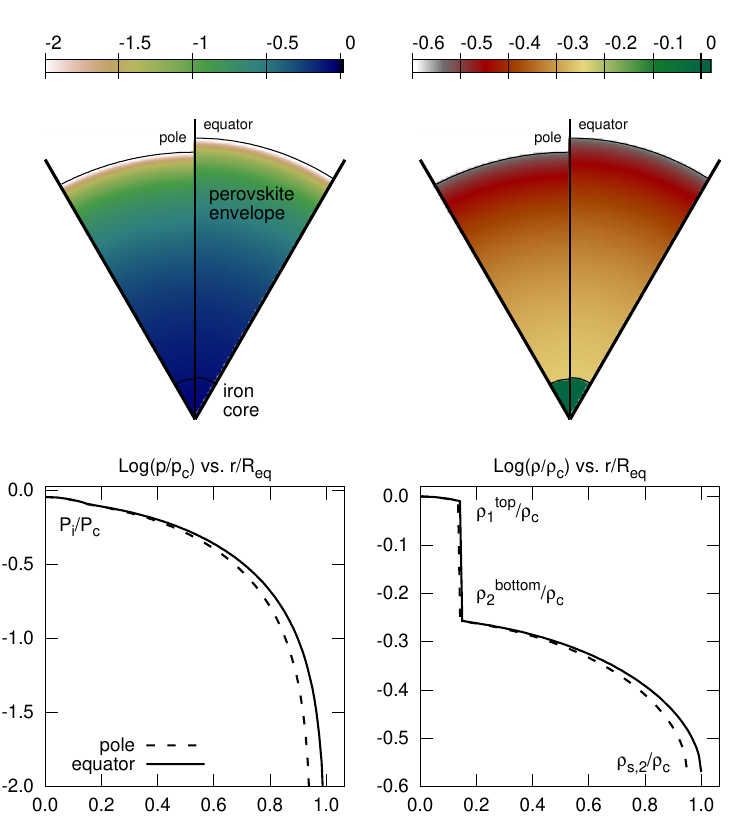}
       \caption{Same caption as for Fig. \ref{fig:graph1refbis_IP.pdf} but for LHS~1140~b, assuming a rotating IP planet with $q=0.15$ and $f=0.05$; see also Tab. \ref{tab:ref_lhs1140b} and Fig. \ref{fig:lhs1140b.pdf} (middle panel).}
       \label{fig:graph1_lhs1140b.pdf}
\end{figure}

\noindent{\bf The IP hypothesis}. As Tab. \ref{tab:lhs1140b_twolayer} shows, an iron core/perovskite envelope planet is a plausible option for LHS~1140~b, but this requires some rotation, again. The core has a small size and a low mass in relative. For $f=0.1$ for instance, we have $\qpol \approx 0.33^{+0.13}_{-0.04}$. We have plotted in Fig. \ref{fig:lhs1140b.pdf} (middle panel) the $A$-sequence obtained with the {\tt DROP}-code for $f=0.05$ and $q=0.15$. The solution obtained for $A \approx 0.4866$ matches the central value $(x_0,y_0)$ very well. Output data are gathered in Tab. \ref{tab:ref_lhs1140b} and the pressure and mass-density maps are displayred in Fig. \ref{fig:graph1_lhs1140b.pdf}. In this simulation, the iron core is less than $1 \%$ of the planet's mass, the rotation period is $6$ h $24$ min.\\

\begin{table}
   \caption{Input parameters and results for the two-layer, IP planet displayed in Fig. \ref{fig:graph1_lhs1140b.pdf}.}
   \label{tab:ref_lhs1140b}
  \begin{tabular}{lr} \hline \hline
         $\blacktriangledown$ input\\
         \quad flattening parameter $f$ & $0.05$\\
         \quad $\biga=\rho_{s,2}/\rho_2^{\rm bottom}$ & $0.4866$\\
         \quad relative core size (pole) $\qpol$ & $0.15$\\ \hline
          $\blacktriangledown$ output\\
         \quad equatorial radius $\req$ in $\rearth$ &  $1.7303$\\
         \qquad estimated from the fit, i.e. Eq.\eqref{eq:mroq_relationship} & $1.7145$ \\
         \quad total mass $M$ in $\mearth$ & $5.5999$\\
        \quad relative mass of the core $M_1/M$ & $0.0079$\\
        \quad mean mass-density $\langle \rho \rangle$  (g/cm$^3$) & $  6.2717$\\
   \quad central-to-mean mass-density ratio $\rhoc/\langle \rho \rangle$ & $  2.4284$\\
   \quad top-to-central mass density ratio (core) $\rho_1^{\rm top}/\rhoc$ & $  0.9763$\\
   \quad rotation period $T$ (h) & $  6.4036$\\
   \quad relative size of the core (equator) $\qeq$  & $  0.1479$\\
   \quad interface mass-density jump $\rho_1^{\rm top}/\rho_2^{\rm bottom} \equiv \alpha$ & $  1.7649$\\
   \quad central-to-surface mass density ratio $\rhoc/\rho_{s,2}$ & $  3.7147$\\
   \quad central pressure $\pc$ ($10^9$ dyn/cm$^2)$ & $8470.46$\\
   \quad interface pressure $\interp \equiv P(Z_1)$ ($10^9$ dyn/cm$^2)$ & $7653.08$\\\hline
  \end{tabular}
 \end{table}

\noindent{\bf The PW hypothesis}. This solution is relatively constrained in the sense that it persists only if $f$ is very small according to Fig. \ref{fig:lhs1140b.pdf} and Tab. \ref{tab:lhs1140b_twolayer}. As quoted above, $f \le \delta f \simeq 0038$. It means that LHS~1140~b can be a huge, slowly rotating silicate body with a shallow water envelope. 

\subsection{Discussion}

The results presented above are in good agreement with the ternary diagrams presented in the spherical case in \cite{dal24}, based in the one-dimensional computational code by \cite{zs08} with updated EOS. These authors, actually, do not rule out a two-layer structure for this planet, and the core mass-fractions predicted for the IW and PW configurations are similar to those reported in this work.

Although the IW-configuration is not really realistic, the large $f$-values presented in Fig. \ref{fig:lhs1140b_details_IW} better illustrate the additional degeneracies for the internal structure of such terrestrial planets when rotation is accounted for. For LHS~1140~b, they correspond to very small rotation periods of a few hours only (close to the critical angular velocity), which must be regarded as illustrative cases and cannot represent the genuine, current spin rate of LHS~1140~b. Indeed, this planet orbits at a short stellocentric distance ($<0.1$\,au), where the tidal locking time is expected to be much shorter than the system age of 5\,Gyr. Its initial rotation must therefore have converged towards an equilibrium state between gravitational and thermal atmospheric tides, if any atmosphere is present \citep[e.g.][]{lwmm15}. Actually, for each composition of our double-layer model, we can estimate the maximum (‘primordial’) value of $f$ compatible with the observational error bars (for instance, $f \approx 0.029$ corresponds to a minimum period $T \simeq 7.5$ h for the IW case, see Fig.\ref{fig:lhs1140b_details_IW}).

Large spin rates are nevertheless not excluded for young, solid planets. In the current paradigm, giant impacts are expected to play a significant role at late stages of the planet formation process, and can spin-up planets close to their breakup velocity \cite[e.g.][]{ki07,mb10}. In their recent collision scenario for the Earth-Moon system, to form the Moon from the Earth’s mantle, \cite{cs12} even invoked a pre-impact spin rate of $2.3$ h only for the proto-Earth. Pebble accretion is also thought to offer a valuable mechanism to significantly spin-up solid protoplanets through angular momentum transfer from the pebbles flow to the surface of the forming planet \citep{jl10,vodi20}. Pebble accretion might be able to yield such fast (inital) rotation periods close to the critical breakup value for terrestrial planets \citep{tkk23}, although recent developments suggest that the final outcome for the planetary spin strongly depends on the details of the modeling of the angular momentum transfer through the planet’s primordial atmosphere \citep{yvd23}. Due to a methodology bias and sensitivity limitation of the transit technique, studying spin distributions and constraining planetary oblateness remains limited to the gaseous giants, as no genuine super-Earth has been detected by the transit technique beyond the tidal locking region so far; see e.g. \cite{liu25} for Kepler167e, and \cite{lw24} for Kepler51d. Within the next decade, the PLATO mission \citep{raueral24} is expected to populate this domain with dozens of Earth-like and Super-Earth planets with orbital periods long enough ($\gtrsim100-200 \,$ d) to escape tidal braking.

\section{Summary, discussion and perspectives}
\label{sec:conclusion}

\subsection{Mass-radius-flattening relationships (summary)}
\label{subsec:mrfsummary}

In this paper, we have computed mass-radius relationships for single-layer and two-layer planets, including solid-body rotation by a conservative approach, leading to ``mass-radius-flattening'' relationships. We have selected three types of materials composing the core and the envelope, namely iron, perovskite and water, and the bimodal EOS from \cite{skhm07}. Tables of data are available in electronic form (see note $\ref{fn:zenodo}$); see Appendix \ref{app:mrf_table} for a sample. Specific simulations can be performed upon request.\\

In the reference mass-range from $0.1$ to $10 \, \mearth$, the mass-radius relationships are, in log. scale, very well fitted by a quadratic law, and this is still possible when rotation is accounted for. Despite the fits are valid over only two decades in mass, these are fully reversible in any of the $3$ parameters $M$, $\qpol$ and $f$. This is very useful to get rapid insight of permitted configurations, in terms of core size and rotational states.
For single-layer planets, we can summarize the results as follows:
\begin{itemize}
\item the mass-radius relationships are given by Eq.\eqref{eq:mro_relationship} where the $9$ coefficients $c_{ij}$ are found in Tab. \ref{tab:cij},
\item the inverse formula $f(M,\req)$ is given by Eq.\eqref{eq:fsingle},
\item in the static case, we have 
\begin{equation}
  \req \propto
  \begin{cases}
 M^{0.293 \pm 0.037}\quad \text{for iron planets},\\
 M^{0.301 \pm 0.030} \quad \text{for perovskite planets},\\
 M^{0.293 \pm 0.034} \quad \text{for water planets},\\
  \end{cases}
\end{equation}
where the upper (lower) value refers to the lower (resp. upper) bound in the mass reference range,
\item the relative change in the radius (at constant mass) and in the mass (at constant radius) due to rotation are $\etarad f$ and $\etamass f$, respectively; see Eqs. \eqref{eq:etarad} and \eqref{eq:etamass},
\item the rotation period is available from Eq.\eqref{eq:periodfit}, where the coefficients are listed in Tab. \ref{tab:tij},
\item for a given mass, the central pressure is slightly decreased, which meets the result established in the single-layer, polytropic case by \cite{ch33}.
\end{itemize}

For two-layer planets, the core fractional size $\qpol$ is the additional parameter. The mass-radius-flattening relationships can still be fitted in this case, with a sigmoid for $\qpol$. It follows that:
\begin{itemize}
\item the mass-radius relationships are described by Eq.\eqref{eq:mroq_relationship}, where the coefficients $c_{ij}'$, $d_{ij}$ and $\sigma$ are given in Tab. \ref {tab:cprimij_dijsigmoid_IWIPPW},
\item in the static case and for a given fractional core size $\qpol$, we have from Eq.\eqref{eq:derivx_twolayer} and Tab. \ref{tab:cprimij_dijsigmoid_IWIPPW},
\begin{equation}
  \req \propto
  \begin{cases}
 M^{0.292 \pm 0.036}\quad \text{for IW planets},\\
 M^{0.299 \pm 0.031} \quad \text{for IP planets},\\
 M^{0.294 \pm 0.035} \quad \text{for PW planets},\\
  \end{cases}
\end{equation}
where the upper (lower) value refers to the lower (resp. upper) bound in the mass reference range.
\item the relative change in the radius (at constant mass) and in the mass (at constant radius) due to rotation are found from $\etarad f$ and $\etamass f$, respectively; see Eqs. \eqref{eq:etarad_twolayer} and \eqref{eq:etamass_twolayer} and Fig. \ref{fig:etar_double},
\item  for a given mass, radius and associated uncertainties $(x_0 \pm \delta x,y_0 \pm \delta y)$, the relative size of the core (and uncertainty) can be determined from Eq.\eqref{eq:qpol}, and it depends on rotation (see for instance Fig. \ref{fig:mr_DELTAimage}),
\item in the limit of small flattenings, $f$ can be expressed as a function of $M$, $\req$ and $\qpol$ (this part is left to the reader),
\end{itemize}

The numerical suveys reveals that the relative change in the radius (mass) at constant mass (radius respectively) due to planet spin (see Sect. \ref{sec:mr}), which is mesured by the parameter $\etarad f$ ($\etamass f$ respectively) is significantly larger than for a rotating, homogeneous body. This is marginally significant for single-layer planet, but more prononced for two-layer planets with core size $\qpol \simeq 0.70 \pm 0.05$ depending on the composition, as Fig. \ref{fig:etar_double} shows.

\begin{figure*}
       \centering
       \includegraphics[trim={-0.2cm 0.cm 0.1cm 0},clip,width=0.8\linewidth]{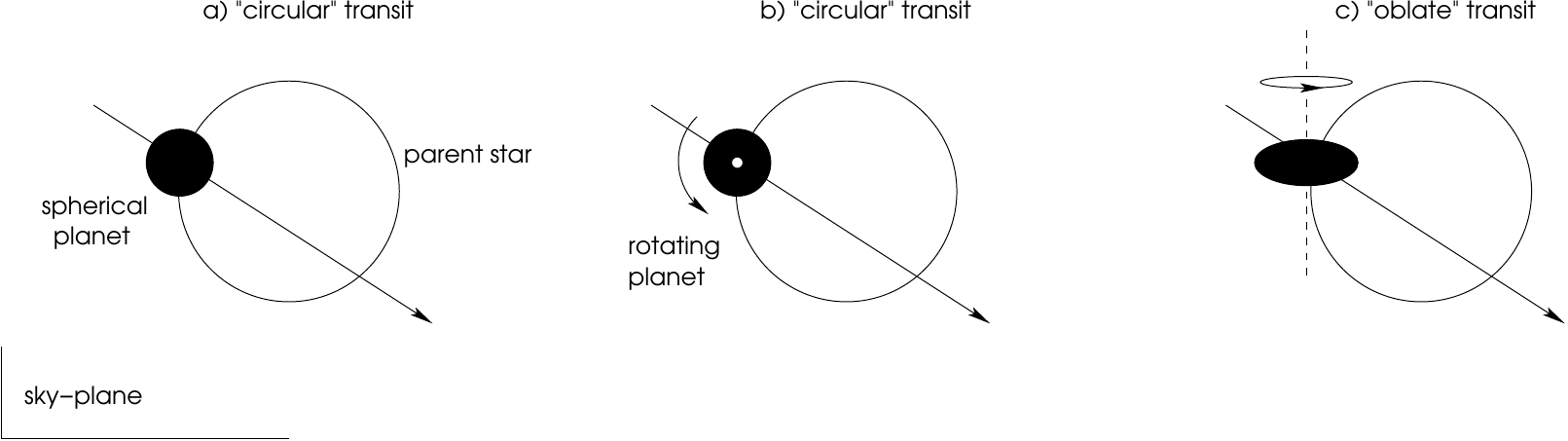}
       \caption{Two extreme orientations of the rotation axis of the planet with respect to the line-of-sight (perpendicular to the figure) during a transit, relative to the spherical case.}
       \label{fig:transits}
\end{figure*}

\subsection{Slope and ''effective'' polytropic index}

As a direct consequence of the EOS, the power-law exponent is $\beta \approx 0.3$ for single and two-layer planets, for the mass range of reference. This is slightly larger than in \cite{skhm07} who considered a wider range of reference, and in \cite{bdf20}, and more compatible with values extracted from exoplanet catalogs \citep{pbdh24}. This remains in good agreement with \cite{fmb07}, both in terms of absolute position and slopes (sligtly larger here), despite slight differences in the EOS (in particular $\rhonp$). We have seen that $\beta$ depends on $f$, through the coefficient $c_1$ and $\langle c_1' \rangle$, and is larger with rotation. It would therefore be very interesting to understand the individual role or effect of the three parameters $\rhonp$, $\biga$ (i.e. $K$) and $m$ in these laws, and more generally, to determine how the coefficients in the fits depend on these three parameters. In the static case, this can be performed directly from the Lane-Emden equation.

According to Eq.\eqref{eq:mrpoly}, a power-law of the form $R \propto M^\beta$ corresponds to a single-layer polytropic EOS with index
\begin{equation}
  n = \frac{1-3 \beta}{1-\beta} \equiv \neff.
  \label{eq:neff}
\end{equation}
 As $\beta \approx 0.3$ in all cases presented in the article, the bimodal EOS corresponds to an ``effective'' polytropic index $\neff \approx \frac{1}{7}$. Values of $\neff$ obtained for the single-layer and two-layer planet examined in the article are listed in Tab. \ref{tab:neff}. It means that the mass-radius relationships computed from Eq. \eqref{eq:mpeos} can be reproduced with a single layer polytrope with index $\neff$. The fact that $\neff$ remains close to $0$ means that the global internal structure is close to a piling of homogeneous layers, as mass density profiles show.
\begin{table}
  \caption{Effective index $\neff$ according to Eq.\eqref{eq:neff}.}
  \label{tab:neff}
  \begin{tabular}{llcc} \hline \hline
${\cal L}$   & case & $\beta$ & $\neff$ \\ \hline
$1$  & iron planet & $0.293 \pm 0.037$ & $0.170^{0.141}_{-0.156}$ \\
  & perovskite planet & $0.301 \pm 0.030$ & $0.139^{0.120}_{-0.131}$ \\
  & water planet & $0.293 \pm 0.034$ & $0.171^{0.132}_{-0.146}$ \\ \hline
$2$   & IW planet & $0.292 \pm 0.036$ & $0.177^{+0.138}_{-0.153}$ \\
  & IP planet  & $0.299 \pm 0.031$ & $0.147^{0.120}_{-0.131}$\\
  & PW planet & $0.294\pm 0.035$ & $0.166^{0.136}_{-0.150}$\\ \hline
  \end{tabular}
 \end{table}
 
\subsection{Note on the ocean depth of water-rich planets}
\label{sec:odepth}

When the outer layer is made of water, and the conditions are met at the surface for water to be in liquid phase, then we can determine the depth $d$ of the liquid ocea by considering the pressure and density at which phase transition to high pressure ice/ice VII occurs. This point in principle depends on temperature. As a typical/low value \citep{leger04}, we can take $10^{10}$ dyn/cm$^2 \equiv \refpres$ and solve $P(R,Z) = \refpres$ numerically for any structure computed with the {\tt DROP} code. In contrast to the spherical case, $d$ is expected to depend on the latitude. We can give an estimate of the ocean depth at the pole from the equation of hydrostatic equilibrium, namely
\begin{equation}
- \refpres = - \int_{\rpol-d}^\rpol{\rho d \Psi},
\label{eq:pwater}
\end{equation}
 and the impact of rotation. In the limit of small flattening values $f \rightarrow 0$, we have for any homogenous spheroid \citep[e.g.][]{binneytremaine87}
\begin{equation}
  \Psi(0,Z) \approx - \frac{2}{3}\pi G \rho \left[(3-2f) \req^2 - \left(1+ \frac{4}{5}f \right) Z^2 \right],
\end{equation}

Assuming $d^2 \ll \rpol^2$ and that the effect of rotation on the gravitational acceleration in the surface layer is not greatly altered by spheroidal stratification with respect to the Maclaurin spheroid, we have $\refpres \approx \rhonp g [\Psi(\rpol)-\Psi(\rpol-d)]$. So, we find from Eq.\eqref{eq:pwater}
\begin{equation}
\refpres \approx \frac{GM  \rhonp d}{\req^2} \left(1+ \frac{4}{5}f \right),
\label{eq:dref_lowm}
\end{equation}
and it follows that $d$ decreases with the mass like $d \propto M^{2\beta-1}$, and decreases as rotation increases. By setting $\refpres \approx 10^{10}$ dyn/cm$^2$, we get
\begin{equation}
  \refd=\frac{\refpres \rearth^2}{G\mearth \rhonp} \approx 70 \text{ km},
\end{equation}
which is for a non-rotating planet with same size and mass as the Earth (note that this value is obviously sensitive to $\rhonp$, and we would get a $30 \%$ larger depth for $\rhonp=1$ g/cm$^3$). In the general case, we find
\begin{equation}
  \log \frac{d}{\refd} \approx 2y-x-\frac{4}{5}f,
\label{eq:docean}
\end{equation}
and we can use Eq.\eqref{eq:quadfit} or \eqref{eq:mroq_relationship} to get a function of $x$, $\qpol$ and $f$ only. Note that the impact of rotation, which is weak, is not only supported by the term $\frac{4}{5}f$, but also by $y$ which depends on $f$ too. For LHS~1140~b, the ocean depth estimated from the {\tt DROP} code is about $33$ km in the static IW-case (see Fig. \ref{fig:lhs1140b_details_IW}), while Eq. \eqref{eq:docean} predicts $37$ km. We have verified that the depth at the equator is a little bit larger than at the pole, which is the consequence of the mass-density gradient, slightly weaker along the equatorial axis.
  
\subsection{Corrections in the mass-radius relatationships due to the planet oblateness and axis orientation}
\label{subsec:correction}

When the radius is deduced from transit method, the planet is genrally assumed spherical, meaning that the obstacle to the star light seen by the observer is a circular disc. This is slightly different if the planet is spinning. Following \cite{bdw22}, two extreme geometries of the primary eclipse can be considered, as sketched in Fig. \ref{fig:transits}:
\begin{itemize}
\item during a circular transit, the rotation axis is always oriented towards the observer (case b), then the blocking section is a disc, and the area is $\pi \rsph^2 = \pi \req^2$,
\item during an oblate transit, the rotation axis is always perpendicular to the line-of-sight (case c), then the relevant shape is roughly an elliptic disc, and the area is  $\pi \rsph^2 = \pi \req^2(1-f)$.
\end{itemize}

Regarding photometric signals, there is no difference between a spherical planet (case a) and a pole-on, spinning planet (case b). In the latter situation, however, the position of the planet in the mass-radius relationship is higher than it would be in the static case, leaving the impression that the planet contains a larger fraction of light compounds. For an oblate-transit (case c), we have $\rsph= \req \sqrt{1-f}$, which leads to
\begin{equation}
  \ysph=y+\frac{1}{2}\log(1-f)
  \label{eq:mr_corrected}
\end{equation}
with the same notation as for Eq.\eqref{eq:mroq_relationship}. In this case, without sufficient resolution to detect the rotation state at the ingress/egress of the transit, then the position of the planet in the mass-radius relationship based upon $\rsph$ leaves the impression that the planet contains heavier elements, because it is in fact higher in the diagram. We see that the correction for an oblate transit in the mass-radius relationships can be accounted self-consistently for  by inserting Eq.\eqref{eq:mr_corrected} into Eq.\eqref{eq:qpol}. We find
\begin{flalign}
  \qpol = \left\{\frac{\ysph - \frac{1}{2} \tau \log(1-f) - c_1' x -c_2'x^2-d_0}{d_1-d_2\left[\ysph - \frac{1}{2} \tau \log(1-f) - c_1' x -c_2'x^2\right]} \right\}^{\frac{1}{\sigma}},
  \label{eq:qpo_correctedl}
\end{flalign}
where we have introduced the parameter $\tau \in [0,1]$ (i.e. it is zero for a circular transit and unity for an oblate transit).

For LHS~1140~b, an oblate transit means a larger equatorial radius. With $f=0.05$ as considered in the figures, the ``true'' equatorial radius becomes $\req \approx 1.77$.

\begin{figure}
       \centering
     \includegraphics[trim={3cm 0.5cm 2.1cm 0},clip,width=8.5cm]{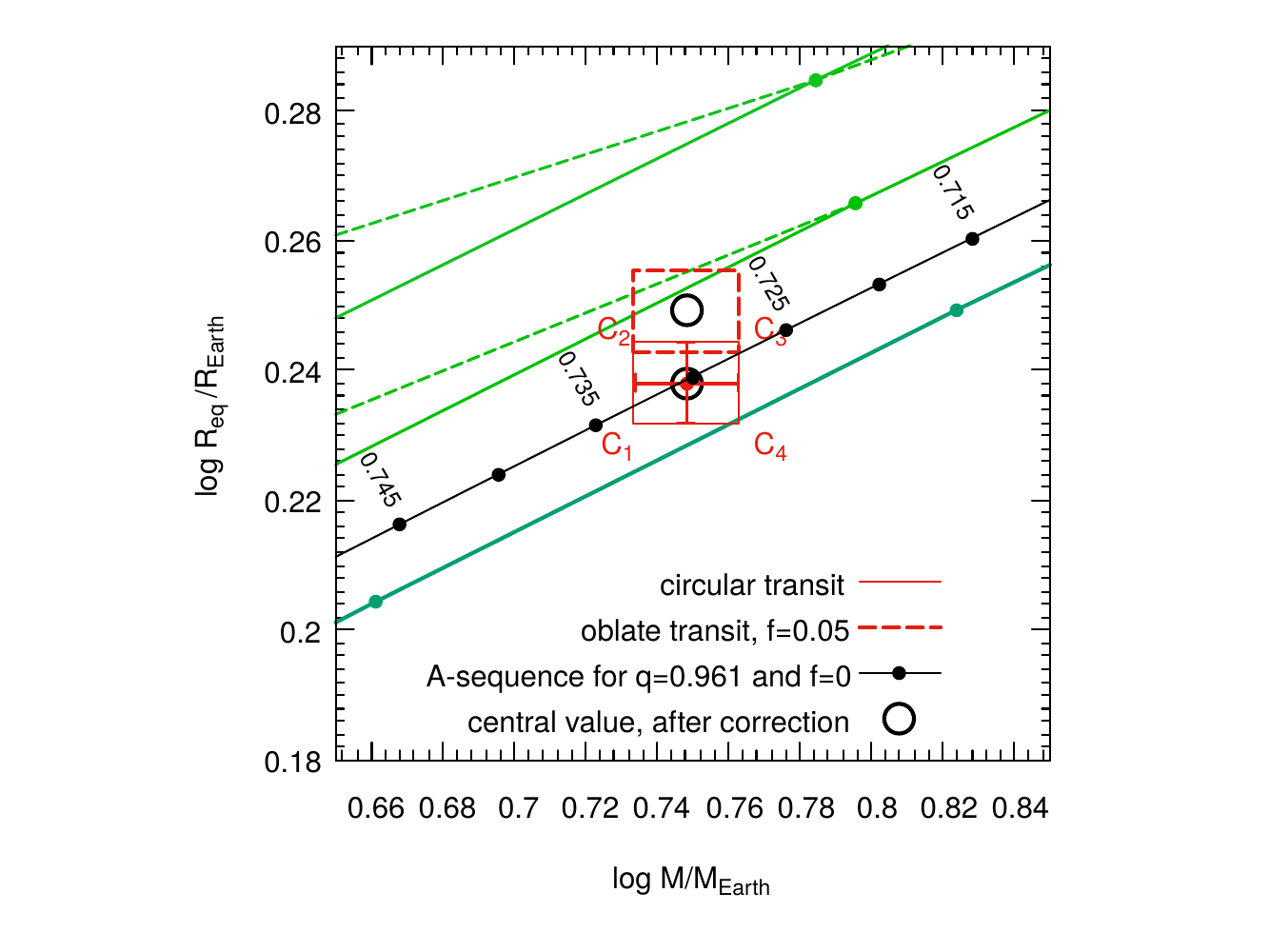}
       \caption{Effective position of LHS~1140~b in the PW-diagram if the rotation axis of the planet is perpendicular to the line-of sight, corresponding to an oblate transit for $f = 0.05$; see also Figs.  \ref{fig:lhs1140b.pdf} (bottom) and \ref{fig:transits}.}
       \label{fig:lhs1140b_fcorrected.pdf}
\end{figure}

\begin{figure}
       \centering
       \includegraphics[trim={1.cm 7.7cm 14.6cm 0cm},clip,width=1.\linewidth]{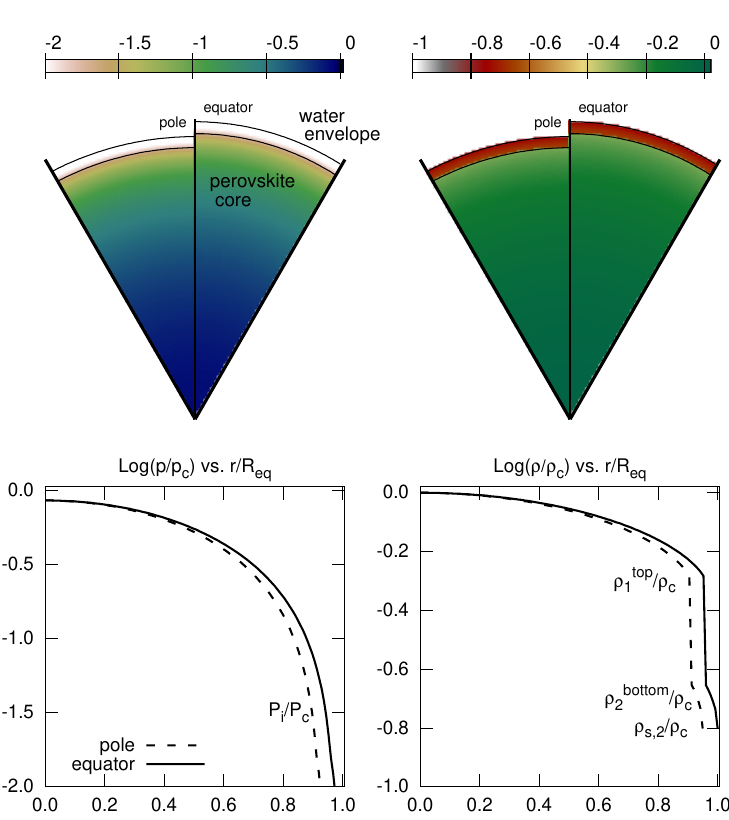}
       \caption{Same caption as for Fig. \ref{fig:graph1_lhs1140b.pdf} but assuming a PW planet and an oblate transit with $f = 0.05$ (period $T \approx 6$ h); see also Tab. \ref{tab:ref_lhs1140b_oblate005} and Fig. \ref{fig:lhs1140b_fcorrected.pdf}.}
       \label{fig:graph1_lhs1140b_oblate005.pdf}
\end{figure}

We show in Fig. \ref{fig:lhs1140b_fcorrected.pdf} the effective position of LHS~1140~b in the PW-diagram with flattening $f=0.05$, assuming an oblate transit, together with the $\biga$-sequence that crosses the new ''effective'' central value, i.e. $(x_0,\ysph-\frac{1}{2}\log(1-f))$. In this case, the ``true'' equatorial radius becomes $\req \approx 1.77$. Figure \ref{fig:graph1_lhs1140b_oblate005.pdf} displays the structure corresponding to this new point, and Table \ref{tab:ref_lhs1140b_oblate005} lists the new data for this equilibrium.

\begin{table}
   \caption{Input parameters and results (column 3) for the PW planet shown in Fig. \ref{fig:graph1_lhs1140b_oblate005.pdf}, assuming an oblate transit and $f=0.05$.}
   \label{tab:ref_lhs1140b_oblate005}
   \begin{tabular}{lrr} \hline \hline
         $\blacktriangledown$ input \\
         \quad flattening parameter $f$ & $0^\star$& $0.05$\\
         \quad $\biga=\rho_{s,2}/\rho_2^{\rm bottom}$ & $0.7300$ & $0.7108$\\
         \quad relative core size (pole) $\qpol$ & $0.9606$ & $0.9524$\\ \hline
          $\blacktriangledown$ output\\
         \quad spherical radius $\req$ in $\rearth$ & $1.7317$ &  $1.7303$\\
         \quad equatorial radius $\req$ in $\rearth$ & $1.7317$ &  $1.7751$\\
         \quad total mass $M$ in $\mearth$ & $5.5932$ & $5.6020$ \\
        \quad relative mass of the core $M_1/M$ & $0.9656$ & $0.9567$\\
        \quad mean mass-density $\langle \rho \rangle$  (g/cm$^3$) & $5.9413$ & $5.8137$\\
   \quad  $\rhoc/\langle \rho \rangle$ & $1.4584$ & $  1.4748$\\
   \quad  $\rho_1^{\rm top}/\rhoc$ & $0.5355$ & $  0.5478$\\
   \quad rotation period $T$ (h & $-$ & $6.3523$\\
   \quad relative size of the core (equator) $\qeq$  & $0.9605$ & $  0.9514$\\
   \quad $\rho_1^{\rm top}/\rho_2^{\rm bottom} \equiv \alpha$ & $2.3200$ & $  2.2868$\\
   \quad $\rhoc/\rho_{s,2}$ & $5.9347$ & $  5.8725$\\
   \quad central pressure $\pc$ ($10^9$ dyn/cm$^2)$ & $8147.64$ & $8147.61$\\
   \quad ocean depth (pole, km) & $30$ & $30$\,\quad\quad\\
   \quad ocean depth (equator, km) & $30$ & $33$\,\quad\quad\\\hline
  \end{tabular}
   $^\star$static case
\end{table}

\subsection{Three-layers and more}

A natural continuation of this work is the introduction of new layers. A three-layer rotating planet made of an iron core, a silicate intermediate layer/mantle and a water envelope, i.e. an IPW planet, would be of particular interest for LHS~1140~b as it is located very close to a pure perovskite composition\cite{val07}. According to our analysis, the internal structure of LHS~1140~b is compatible with i) a small iron core surrounded by a large silicate mantle, and ii) a thin water surface layer on top of a big silicate core, as depicted in Figs. \ref{fig:graph1_lhs1140b.pdf} and \ref{fig:graph1_lhs1140b_oblate005.pdf}. Would a mixture of these two extreme configurations, in the form of a three-layer body, be plausible ? As shown in \cite{dal24}, the silicate mantle should not occupy more $39 \%$ of the planet in size, to the benefit of the water and iron layers. This question merits indeed an extension of our approach to configurations with ${\cal L}=3$. Considering an additional atmosphere, made of hydrogen and helium or water vapor for instance, and studying the response of a gas phase to rotation is another challenge to face \citep{ase08}.

Fitting the mass-radius relationship should become very complicated for ${\cal L}>2$, because the number of parameters increases. Actually, each new layer requires two new parameters (the mass-density and the relative position with respect to the others). This addresses the question of the computing time. Accounting for more layers means is costly. With the modified polytropic EOS running the {\tt DROP} code is already longer, in particular due to the special function (see Sect. \ref{sec:hyperg}). It would be interesting to have both $H(\rho)$ and $\rho(H)$ available in the form of an approximation, for each index $m$ of the EOS. A preliminary inspection shows that $1/{}_2F_1(u)$ can be very well fitted by a low-degree polynomial in $u \in [0,\frac{1}{2}]$. With a second-order polynomial, easily reversible, the relative precision is already as low as few $10^{-4}$ in relative. With such an option, some computing time is saved, and this would enable to increase the number of layers by maintaining the actual resolution. An alternative is to use the method of dimension reduction as proposed in \cite{sh24}, which is currently under examination.

\begin{acknowledgements}
  
  This research has made use of the NASA Exoplanet Archive, which is operated by the California Institute of Technology, under contract with the National Aeronautics and Space Administration under the Exoplanet Exploration Program (retrieved on Dec 13, 2024). This research has made use of data obtained from or tools provided by the portal {\tt exoplanet.eu} of The Extrasolar Planets Encyclopaedia. This work was supported by the Programme National de Planétologie (PNP) of CNRS-INSU co-funded by CNES. Computer time was provided by the computing facilities MCIA (Mésocentre de Calcul Intensif Aquitain) of the Universit\'e de Bordeaux.

    It is a pleasure to thank M. Turbet, F. Chambat and Y. Ricard for very interesting inputs and advices on the article prior to submission,  and J. Leconte.  We are grateful to the referee, Dr. A. Aguichine, for his very detailed report and the many criticisms, questions and suggestions to improve the article.
\end{acknowledgements}

\bibliographystyle{aa}

\begin{appendix}
  
\onecolumn

\section{Mass-radius-oblateness relationships for single-layer and two-layer planets}
\label{app:mrf_table}

\begin{sidewaystable*}[h]
 \caption{Example of data set$^\star$ for a pure water planet with $f=0.1$ ({\it top}) and for IW-planet with $f=0.1$ and $q=0.5$ ({\it bottom}).}
 \label{tab:mr_water_IW}
 \tiny
  \begin{tabular}{rrr|rrrrrrrrrrrrrrrr} \hline \hline
    $f$ & $\biga$ & $\qpol$ & $\log \req/\rearth$ & $\log M/\mearth$ & $M_1/M$ & $\langle \rho \rangle$ & $\rhoc/\langle \rho \rangle$ & $\rho_1^{\rm top}/\rhoc$ & $T$ &  $\qeq$ & $\alpha$  & $\rhoc/a_{\cal L}$ & $\log \pc$ & $\log \interp$ & $C$ & $m$ \\ \hline
    $0.100$ & $0.05$ & $1.0$ & $+0.67404$ & $+2.29083$ &  $  1.0000$ &  $ 11.3857$ &  $  2.5646$ &  $  0.0500$ &  $  2.8976$ &  $  1.0000$ &  $  0.0500$ &  $ 20.0000$ &  $ 14.5484$ &  $ 14.5484$ &  $  0.2967$ &  $  0.1268$\\
$0.100$ & $0.10$ & $1.0$ & $+0.62909$ & $+1.92197$ &  $  1.0000$ &  $  6.6400$ &  $  2.1988$ &  $  0.1000$ &  $  3.9229$ &  $  1.0000$ &  $  0.1000$ &  $ 10.0000$ &  $ 13.9159$ &  $ 13.9159$ &  $  0.3159$ &  $  0.1186$\\
$0.100$ & $0.15$ & $1.0$ & $+0.58704$ & $+1.66724$ &  $  1.0000$ &  $  4.9369$ &  $  1.9715$ &  $  0.1500$ &  $  4.6556$ &  $  1.0000$ &  $  0.1500$ &  $  6.6667$ &  $ 13.5242$ &  $ 13.5242$ &  $  0.3288$ &  $  0.1132$\\
$0.100$ & $0.20$ & $1.0$ & $+0.54624$ & $+1.45730$ &  $  1.0000$ &  $  4.0346$ &  $  1.8094$ &  $  0.2000$ &  $  5.2421$ &  $  1.0000$ &  $  0.2000$ &  $  5.0000$ &  $ 13.2293$ &  $ 13.2293$ &  $  0.3387$ &  $  0.1092$ \\
$0.100$ & $0.25$ & $1.0$ & $+0.50571$ & $+1.26978$ &  $  1.0000$ &  $  3.4659$ &  $  1.6850$ &  $  0.2500$ &  $  5.7385$ &  $  1.0000$ &  $  0.2500$ &  $  4.0000$ &  $ 12.9858$ &  $ 12.9858$ &  $  0.3467$ &  $  0.1061$\\
$0.100$ & $0.30$ & $1.0$ & $+0.46471$ & $+1.09421$ &  $  1.0000$ &  $  3.0702$ &  $  1.5851$ &  $  0.3000$ &  $  6.1718$ &  $  1.0000$ &  $  0.3000$ &  $  3.3333$ &  $ 12.7730$ &  $ 12.7730$ &  $  0.3534$ &  $  0.1035$\\
$0.100$ & $0.35$ & $1.0$ & $+0.42265$ & $+0.92444$ &  $  1.0000$ &  $  2.7766$ &  $  1.5024$ &  $  0.3500$ &  $  6.5598$ &  $  1.0000$ &  $  0.3500$ &  $  2.8571$ &  $ 12.5798$ &  $ 12.5798$ &  $  0.3592$ &  $  0.1013$\\
$0.100$ & $0.40$ & $1.0$ & $+0.37898$ & $+0.75626$ &  $  1.0000$ &  $  2.5485$ &  $  1.4322$ &  $  0.4000$ &  $  6.9115$ &  $  1.0000$ &  $  0.4000$ &  $  2.5000$ &  $ 12.3990$ &  $ 12.3990$ &  $  0.3643$ &  $  0.0994$\\
$0.100$ & $0.45$ & $1.0$ & $+0.33310$ & $+0.58629$ &  $  1.0000$ &  $  2.3654$ &  $  1.3716$ &  $  0.4500$ &  $  7.2334$ &  $  1.0000$ &  $  0.4500$ &  $  2.2222$ &  $ 12.2256$ &  $ 12.2256$ &  $  0.3688$ &  $  0.0978$\\
$0.100$ & $0.50$ & $1.0$ & $+0.28433$ & $+0.41140$ &  $  1.0000$ &  $  2.2146$ &  $  1.3185$ &  $  0.5000$ &  $  7.5325$ &  $  1.0000$ &  $  0.5000$ &  $  2.0000$ &  $ 12.0557$ &  $ 12.0557$ &  $  0.3729$ &  $  0.0963$\\
$0.100$ & $0.55$ & $1.0$ & $+0.23184$ & $+0.22834$ &  $  1.0000$ &  $  2.0877$ &  $  1.2715$ &  $  0.5500$ &  $  7.8117$ &  $  1.0000$ &  $  0.5500$ &  $  1.8182$ &  $ 11.8859$ &  $ 11.8859$ &  $  0.3766$ &  $  0.0950$\\
$0.100$ & $0.60$ & $1.0$ & $+0.17456$ & $+0.03334$ &  $  1.0000$ &  $  1.9793$ &  $  1.2294$ &  $  0.6000$ &  $  8.0740$ &  $  1.0000$ &  $  0.6000$ &  $  1.6667$ &  $ 11.7125$ &  $ 11.7125$ &  $  0.3800$ &  $  0.0938$\\
$0.100$ & $0.65$ & $1.0$ & $+0.11101$ & $-0.17841$ &  $  1.0000$ &  $  1.8853$ &  $  1.1914$ &  $  0.6500$ &  $  8.3215$ &  $  1.0000$ &  $  0.6500$ &  $  1.5385$ &  $ 11.5317$ &  $ 11.5317$ &  $  0.3832$ &  $  0.0927$\\
$0.100$ & $0.70$ & $1.0$ & $+0.03909$ & $-0.41356$ &  $  1.0000$ &  $  1.8028$ &  $  1.1569$ &  $  0.7000$ &  $  8.5562$ &  $  1.0000$ &  $  0.7000$ &  $  1.4286$ &  $ 11.3384$ &  $ 11.3384$ &  $  0.3861$ &  $  0.0917$\\
$0.100$ & $0.75$ & $1.0$ & $-0.04443$ & $-0.68206$ &  $  1.0000$ &  $  1.7298$ &  $  1.1254$ &  $  0.7500$ &  $  8.7795$ &  $  1.0000$ &  $  0.7500$ &  $  1.3333$ &  $ 11.1257$ &  $ 11.1257$ &  $  0.3889$ &  $  0.0907$\\
$0.100$ & $0.80$ & $1.0$ & $-0.14495$ & $-1.00029$ &  $  1.0000$ &  $  1.6646$ &  $  1.0964$ &  $  0.8000$ &  $  8.9952$ &  $  1.0000$ &  $  0.8000$ &  $  1.2500$ &  $ 10.8821$ &  $ 10.8821$ &  $  0.3914$ &  $  0.0898$\\
$0.100$ & $0.85$ & $1.0$ & $-0.27252$ & $-1.39857$ &  $  1.0000$ &  $  1.6059$ &  $  1.0696$ &  $  0.8500$ &  $  9.1994$ &  $  1.0000$ &  $  0.8500$ &  $  1.1765$ &  $ 10.5873$ &  $ 10.5873$ &  $  0.3938$ &  $  0.0890$\\
$0.100$ & $0.90$ & $1.0$ & $-0.44971$ & $-1.94474$ &  $  1.0000$ &  $  1.5528$ &  $  1.0447$ &  $  0.9000$ &  $  9.3954$ &  $  1.0000$ &  $  0.9000$ &  $  1.1111$ &  $ 10.1956$ &  $ 10.1956$ &  $  0.3961$ &  $  0.0883$\\
$0.100$ & $0.95$ & $1.0$ & $-0.74846$ & $-2.85473$ &  $  1.0000$ &  $  1.5043$ &  $  1.0216$ &  $  0.9500$ &  $  9.5839$ &  $  1.0000$ &  $  0.9500$ &  $  1.0526$ &  $  9.5630$ &  $  9.5630$ &  $  0.3983$ &  $  0.0875$\\ \hline \\
   \\\hline \hline
  $f$ & $\biga$ & $\qpol$ & $\log \req/\rearth$ & $\log M/\mearth$ & $M_1/M$ & $\langle \rho \rangle$ & $\rhoc/\langle \rho \rangle$ & $\rho_1^{\rm top}/\rhoc$ & $T$ &  $\qeq$ & $\alpha$  & $\rhoc/a_{\cal L}$ & $\log \pc$ & $\log \interp$ & $C$ & $m$ \\ \hline
$0.100$ & $0.05$ & $0.5$ & $+0.56893$ & $+2.21668$ &  $  0.4838$ &  $ 19.8835$ &  $  6.8295$ &  $  0.4272$ &  $  1.9507$ &  $  0.4721$ &  $  1.9865$ &  $ 93.0093$ &  $ 15.3232$ &  $ 14.5484$ &  $  0.2191$ &  $  0.1606$\\
$0.100$ & $0.10$ & $0.5$ & $+0.53122$ & $+1.83751$ &  $  0.4537$ &  $ 10.7733$ &  $  5.9289$ &  $  0.4906$ &  $  2.7212$ &  $  0.4734$ &  $  2.1463$ &  $ 43.7488$ &  $ 14.6402$ &  $ 13.9159$ &  $  0.2369$ &  $  0.1523$\\
$0.100$ & $0.15$ & $0.5$ & $+0.49057$ & $+1.57024$ &  $  0.4378$ &  $  7.7085$ &  $  5.4176$ &  $  0.5414$ &  $  3.2731$ &  $  0.4739$ &  $  2.3228$ &  $ 28.6037$ &  $ 14.2230$ &  $ 13.5242$ &  $  0.2482$ &  $  0.1471$\\
$0.100$ & $0.20$ & $0.5$ & $+0.44874$ & $+1.34740$ &  $  0.4285$ &  $  6.1594$ &  $  5.0773$ &  $  0.5851$ &  $  3.7094$ &  $  0.4742$ &  $  2.5066$ &  $ 21.4199$ &  $ 13.9136$ &  $ 13.2293$ &  $  0.2562$ &  $  0.1433$\\
$0.100$ & $0.25$ & $0.5$ & $+0.40589$ & $+1.14697$ &  $  0.4228$ &  $  5.2192$ &  $  4.8301$ &  $  0.6242$ &  $  4.0713$ &  $  0.4746$ &  $  2.6945$ &  $ 17.2669$ &  $ 13.6616$ &  $ 12.9858$ &  $  0.2623$ &  $  0.1403$\\
$0.100$ & $0.30$ & $0.5$ & $+0.36173$ & $+0.95849$ &  $  0.4197$ &  $  4.5871$ &  $  4.6387$ &  $  0.6599$ &  $  4.3811$ &  $  0.4758$ &  $  2.8851$ &  $ 14.5742$ &  $ 13.4439$ &  $ 12.7730$ &  $  0.2671$ &  $  0.1379$\\
$0.100$ & $0.35$ & $0.5$ & $+0.31640$ & $+0.77662$ &  $  0.4173$ &  $  4.1270$ &  $  4.4892$ &  $  0.6930$ &  $  4.6502$ &  $  0.4759$ &  $  3.0778$ &  $ 12.6896$ &  $ 13.2479$ &  $ 12.5798$ &  $  0.2711$ &  $  0.1360$\\
$0.100$ & $0.40$ & $0.5$ & $+0.26908$ & $+0.59648$ &  $  0.4161$ &  $  3.7791$ &  $  4.3652$ &  $  0.7240$ &  $  4.8886$ &  $  0.4762$ &  $  3.2721$ &  $ 11.2990$ &  $ 13.0659$ &  $ 12.3990$ &  $  0.2743$ &  $  0.1344$\\
$0.100$ & $0.45$ & $0.5$ & $+0.21954$ & $+0.41504$ &  $  0.4153$ &  $  3.5039$ &  $  4.2627$ &  $  0.7533$ &  $  5.1018$ &  $  0.4762$ &  $  3.4677$ &  $ 10.2303$ &  $ 12.8922$ &  $ 12.2256$ &  $  0.2771$ &  $  0.1331$\\
$0.100$ & $0.50$ & $0.5$ & $+0.16710$ & $+0.22924$ &  $  0.4149$ &  $  3.2814$ &  $  4.1754$ &  $  0.7810$ &  $  5.2944$ &  $  0.4763$ &  $  3.6645$ &  $  9.3845$ &  $ 12.7229$ &  $ 12.0557$ &  $  0.2795$ &  $  0.1319$\\
$0.100$ & $0.55$ & $0.5$ & $+0.11080$ & $+0.03546$ &  $  0.4150$ &  $  3.0980$ &  $  4.0992$ &  $  0.8073$ &  $  5.4690$ &  $  0.4765$ &  $  3.8622$ &  $  8.6979$ &  $ 12.5540$ &  $ 11.8859$ &  $  0.2816$ &  $  0.1310$\\
$0.100$ & $0.60$ & $0.5$ & $+0.04990$ & $-0.16963$ &  $  0.4151$ &  $  2.9424$ &  $  4.0338$ &  $  0.8325$ &  $  5.6301$ &  $  0.4763$ &  $  4.0608$ &  $  8.1296$ &  $ 12.3821$ &  $ 11.7125$ &  $  0.2834$ &  $  0.1301$\\
$0.100$ & $0.65$ & $0.5$ & $-0.01730$ & $-0.39117$ &  $  0.4156$ &  $  2.8102$ &  $  3.9752$ &  $  0.8566$ &  $  5.7797$ &  $  0.4765$ &  $  4.2603$ &  $  7.6515$ &  $ 12.2031$ &  $ 11.5317$ &  $  0.2850$ &  $  0.1292$\\
$0.100$ & $0.70$ & $0.5$ & $-0.09283$ & $-0.63585$ &  $  0.4161$ &  $  2.6955$ &  $  3.9234$ &  $  0.8797$ &  $  5.9188$ &  $  0.4765$ &  $  4.4606$ &  $  7.2435$ &  $ 12.0118$ &  $ 11.3384$ &  $  0.2865$ &  $  0.1285$\\
$0.100$ & $0.75$ & $0.5$ & $-0.17991$ & $-0.91358$ &  $  0.4168$ &  $  2.5952$ &  $  3.8770$ &  $  0.9019$ &  $  6.0474$ &  $  0.4765$ &  $  4.6617$ &  $  6.8914$ &  $ 11.8012$ &  $ 11.1257$ &  $  0.2877$ &  $  0.1278$\\
$0.100$ & $0.80$ & $0.5$ & $-0.28395$ & $-1.24074$ &  $  0.4176$ &  $  2.5066$ &  $  3.8353$ &  $  0.9233$ &  $  6.1665$ &  $  0.4765$ &  $  4.8637$ &  $  6.5847$ &  $ 11.5600$ &  $ 10.8821$ &  $  0.2889$ &  $  0.1273$\\
$0.100$ & $0.85$ & $0.5$ & $-0.41500$ & $-1.64776$ &  $  0.4184$ &  $  2.4279$ &  $  3.7978$ &  $  0.9438$ &  $  6.2789$ &  $  0.4765$ &  $  5.0666$ &  $  6.3155$ &  $ 11.2675$ &  $ 10.5873$ &  $  0.2899$ &  $  0.1268$\\
$0.100$ & $0.90$ & $0.5$ & $-0.59564$ & $-2.20246$ &  $  0.4193$ &  $  2.3574$ &  $  3.7641$ &  $  0.9635$ &  $  6.3854$ &  $  0.4765$ &  $  5.2706$ &  $  6.0778$ &  $ 10.8785$ &  $ 10.1956$ &  $  0.2908$ &  $  0.1262$ \\
$0.100$ & $0.95$ & $0.5$ & $-0.89781$ & $-3.12080$ &  $  0.4204$ &  $  2.2941$ &  $  3.7343$ &  $  0.9824$ &  $  6.4830$ &  $  0.4765$ &  $  5.4761$ &  $  5.8676$ &  $ 10.2486$ &  $  9.5630$ &  $  0.2917$ &  $  0.1258$ \\ \hline \\
  \end{tabular}
  \\
  $^\star$From left to right : the flattening parameter $f=1- \rpol/\req \ge0$, where $\rpol$ is the polar radius, and $\req$ the equatorial radius, the $\biga$-parameter, where ${\cal L} \in \{1,2\}$ is the number of layer, the relative size $\qpol=Z_1/\rpol$ of the core along the axis of rotation, the decimal log. of the equatorial radius $\req$ of the planet in units of the Earth's radius, with $\rearth=6.371 \times 10^8$ cm, the decimal log. of the total mass $M$ of the planet in units of the Earth's mass, with $\mearth=5.972 \times 10^{27}$ g, the relative mass of the core $M_1/M$, the mean mass-density $\langle \rho \rangle = M/V$ in g/cm$^3$, where $V$ is the volume of the planet, the central-to-mean mass-density ratio $\rhoc/\langle \rho \rangle$, the top-to-central mass density ratio in the core $\rho_1^{\rm top}/\rhoc$, the rotation period $T$ in h, the relative size $\qeq=R_1/\req$ of the core at the equator, the interface mass-density jump $\alpha \equiv \rho_1^{\rm top}/\rho_2^{\rm bottom}$, the central-to-surface mass density ratio $\rhoc/a_{\cal L}$, which is identical to $1/\biga$ in the single-layer case, the decimal log. of the central pressure $\pc$ in dyn/cm$^2$, the decimal log. of the interface pressure $\interp \equiv P(Z_1)$ in dyn/cm$^2$, the normalized moment of inertia $C=I_z \big/ M \req^2$, and the $m$-parameter defined by $m= \Omega^2 \req^3 \big/ GM$. We take $G=6.67259 \times 10^{-8}$ for the constant of gravitation.
\end{sidewaystable*}

\newpage

\twocolumn
\section{Note on the hypergeometric function}
\label{sec:hyperg} 

\begin{figure}[h]
       \centering
       \includegraphics[trim={1.5cm 0.1cm 1.5cm 0.1cm},clip,width=0.96\linewidth]{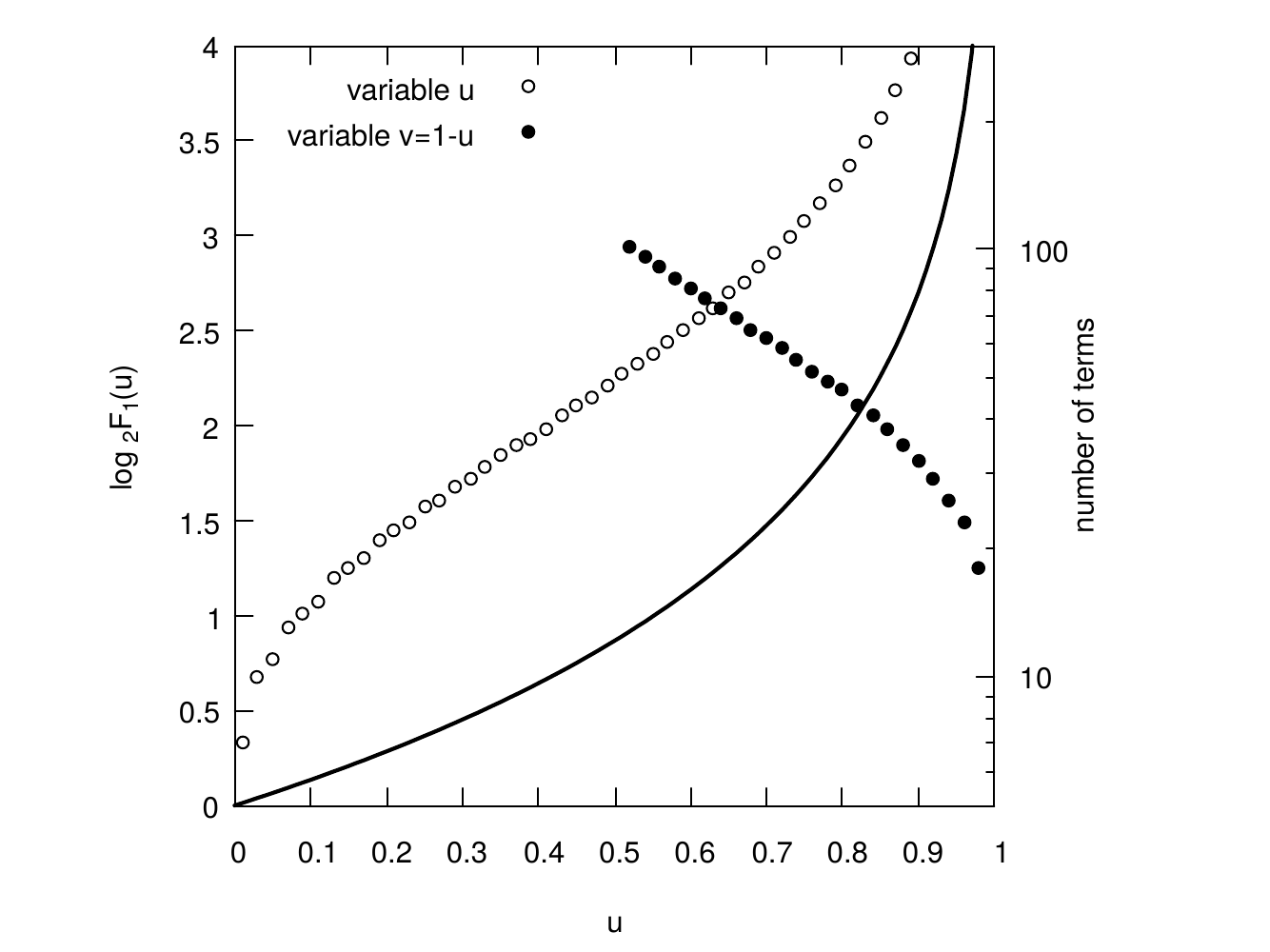}
       \caption{Decimal logarithm of the hypergeometric function ${}_2F_1(u)$ ({\it black line}) appearing in Eq.\eqref{eq:h2F1}, and total number of terms in the Taylor series required to reach the computer precision, i.e. $16$ digits ({\it right axis}). This number can be reduced by a factor two for a $8$-digit precision; see Sec. \ref{sec:hyperg}. The parameter is $\gamma \approx 1.89$ corresponding to iron (see Tab. \ref{tab:eosdata}).}
       \label{fig:sf_2F1}
\end{figure}

As $0 \le u \le 1-\biga < 1$ and $\gamma \approx 2$ in Eq.\eqref{eq:h2F1} for all the materials considered in this study (see Tab. \ref{tab:eosdata}), we can compute the special function ${}_2F_1(u)\equiv {}_2F_1(\gamma,\gamma;\gamma+1;u)$ by direct summation of the terms of the Taylor series \citep{ppm14}. The function ${}_2F_1(u)$ associated with an iron layer is plotted in Fig. \ref{fig:sf_2F1} as an illustration. Convergence at the machine capability ($16$ digits in double precision) is easy and it requires no more than about a hundred terms for $u \le \frac{1}{2}$, as indicated in Fig. \ref{fig:sf_2F1}. Above this value, acceleration is necessary, especially when $\biga$ is very close to zero. Convergence acceleration is obtained by performing a linear transformation $v=1-u$, leaving two hypergeometric functions to estimate. In practical, the switch can therefore be set a little bit beyond $u=\frac{1}{2}$ (see again Fig. \ref{fig:sf_2F1}). So, whatever the argument $u \in [0,1]$, the number of terms in the series remains relatively small (this number is divided by two typically for 8 digits). The second point concerns the Poisson equation. As a matter of fact, the mass density $\rho(H)$ is required in Eq.\eqref{eq:poisson}, which means that the function $H(\rho)$ defined by Eq.\eqref{eq:h2F1} must be reversed at each node of the computational grid (there are $N \times N$ effective nodes in total). This inversion is performed from a standard root-finding method, which involves iterations. We estimate that the computational cost is increased by a factor $3$ at the actual resolution, depending also on $\biga$, and therefore on $u$; see Sec. \ref{sec:conclusion}.

\newpage
\section{Coefficients $c_i$ fitting the mass-radius relationship for single-layer planets (static case)}
\label{sec:coef_singlelayer}

\begin{table}[h]
   \caption{Coefficients in Eq.\eqref{eq:quadfit} for iron.}
   \label{tab:ci_iron}
  \begin{tabular}{rccc} \hline \hline
  $f$         & $c_0$       & $c_1$        & $c_2$ \\\hline
$0.000$ & $-0.1010692$ & $+0.2933718$ & $-0.0183969$ \\
$0.025$ & $-0.0970511$ & $+0.2937060$ & $-0.0183842$ \\
$0.050$ & $-0.0929222$ & $+0.2940590$ & $-0.0183700$ \\
$0.075$ & $-0.0886769$ & $+0.2944307$ & $-0.0183515$ \\
$0.100$ & $-0.0843095$ & $+0.2948240$ & $-0.0183267$ \\
$0.125$ & $-0.0798110$ & $+0.2952435$ & $-0.0182961$ \\
$0.150$ & $-0.0751747$ & $+0.2956862$ & $-0.0182632$ \\
$0.175$ & $-0.0703849$ & $+0.2961575$ & $-0.0182384$ \\
$0.200$ & $-0.0654405$ & $+0.2966548$ & $-0.0182092$ \\\hline
  \end{tabular}
\end{table}

\begin{table}[h]
   \caption{Coefficients in Eq.\eqref{eq:quadfit} for perovskite.}
   \label{tab:ci_perovskite}
   \begin{tabular}{rccc} \hline \hline
  $f$         & $c_0$       & $c_1$        & $c_2$ \\\hline
$0.000$ & $+0.0105525$ & $+0.3010068$ & $-0.0152134$ \\
$0.025$ & $+0.0145022$ & $+0.3012828$ & $-0.0152036$ \\
$0.050$ & $+0.0185591$ & $+0.3015741$ & $-0.0151921$ \\
$0.075$ & $+0.0227294$ & $+0.3018809$ & $-0.0151778$ \\
$0.100$ & $+0.0270190$ & $+0.3022045$ & $-0.0151612$ \\
$0.125$ & $+0.0314328$ & $+0.3025494$ & $-0.0151343$ \\
$0.150$ & $+0.0359814$ & $+0.3029146$ & $-0.0151063$ \\
$0.175$ & $+0.0406767$ & $+0.3033038$ & $-0.0150817$ \\
$0.200$ & $+0.0455257$ & $+0.3037160$ & $-0.0150580$ \\
\hline
  \end{tabular}
\end{table}

\begin{table}[h]
   \caption{Coefficients in Eq.\eqref{eq:quadfit} for water.}
   \label{tab:ci_water}
  \begin{tabular}{rccc} \hline \hline
  $f$         & $c_0$       & $c_1$        & $c_2$ \\\hline
$0.000$ & $+0.1483488$ & $+0.2929427$ & $-0.0172455$ \\
$0.025$ & $+0.1523712$ & $+0.2932570$ & $-0.0172342$ \\
$0.050$ & $+0.1565050$ & $+0.2935890$ & $-0.0172216$ \\
$0.075$ & $+0.1607556$ & $+0.2939389$ & $-0.0172051$ \\
$0.100$ & $+0.1651294$ & $+0.2943090$ & $-0.0171844$ \\
$0.125$ & $+0.1696342$ & $+0.2947054$ & $-0.0171555$ \\
$0.150$ & $+0.1742802$ & $+0.2951243$ & $-0.0171287$ \\
$0.175$ & $+0.1790761$ & $+0.2955672$ & $-0.0171043$ \\
$0.200$ & $+0.1840293$ & $+0.2960352$ & $-0.0170789$ \\\hline
  \end{tabular}
\end{table}

\newpage

\section{Coefficients $c_{ij}$ fitting the mass-radius relationship for single-layer planets}
\label{sec:coef_singlelayer_general} 

\begin{table}[h]
  \caption{Coefficients $c_{ij}$ for the mass-radius relationship fitted with Eq.\eqref{eq:mro_relationship} for single-layer planet, with rotation$^\star$.}
  \label{tab:cij}
  \begin{tabular}{lccc} \hline \hline
             &  $j=0$ &  $j=1$ &  $j=2$  \\\hline
  \multicolumn{4}{l}{$\blacktriangledown$ iron}\\ 
$c_{0j}$ & $-0.1010472$ & $\textcolor{red}{+0.1568129}$ & $+0.1055179$\\
$c_{1j}$ & ${\bf +0.2933768}$ & $+0.0125997$ & $+0.0188474$\\
$c_{2j}$ & $-0.0183993$ & $+0.0005275$ & $+0.0022197$\\\hline
  \multicolumn{4}{l}{$\blacktriangledown$ perovskite}\\ 
$c_{0j}$ & $+0.0105736$ & $\textcolor{red}{+0.1542161}$ & $+0.1021337$\\
$c_{1j}$ & ${\bf +0.3010116}$ & $+0.0103719$ & $+0.0156375$\\
$c_{2j}$ & $-0.0152144$ & $+0.0003446$ & $+0.0022846$\\
\hline
   \multicolumn{4}{l}{$\blacktriangledown$ water} \\
$c_{0j}$ & $+0.1483707$ & $\textcolor{red}{+0.1569754}$ & $+0.1060086$\\
$c_{1j}$ & ${\bf +0.2929471}$ & $+0.0118541$ & $+0.0178539$\\
$c_{2j}$ & $-0.0172474$ & $+0.0004563$ & $+0.0020234$\\\hline 
  \end{tabular}
\\  $^\star$The exponent of the power-law for a static, one Earth's mass body is indicated ({\it bold}), as well as the linear contribution of rotation ({\it red}).
\end{table}

\section{Coefficients $t_{ij}$ fitting the period for single-layer planets}
\label{sec:coef_singlelayer_period}

\begin{table}[h]
  \caption{Coefficients $t_{ij}$ for the mass-period relationship fitted with Eq.\eqref{eq:periodfit} for single-layer planet, with rotation.}
  \label{tab:tij}
  \begin{tabular}{lcc} \hline \hline
             &  $j=0$ &  $j=1$   \\\hline
  \multicolumn{3}{l}{$\blacktriangledown$ iron}\\ 
$t_{0j}$ & $+0.3594785$ & $-0.3414377$ \\
$t_{1j}$ & $-0.0277489$ & $-0.0021638$ \\
$t_{2j}$ & $-0.0117303$ & $-0.0045039$ \\\hline
  \multicolumn{3}{l}{$\blacktriangledown$ perovskite}\\ 
$t_{0j}$ & $+0.3661880$ & $-0.3409544$ \\
  $t_{1j}$ & $-0.0220200$ & $+0.0008478$ \\
$t_{2j}$ & $-0.0128245$ & $-0.0049271$ \\
\hline
   \multicolumn{3}{l}{$\blacktriangledown$ water} \\
$t_{0j}$ & $+0.3577173$ & $-0.3370310$ \\
$t_{1j}$ & $-0.0273110$ & $-0.0012598$ \\
$t_{2j}$ & $-0.0102134$ & $-0.0054784$ \\\hline
  \end{tabular}
  \end{table}

\newpage

\onecolumn
\section{Mass-radius relationship for two-layer, IP planets}
\label{sec:mr_twolayer_IP}

\begin{figure*}[h]
  \centering
  \includegraphics[trim={1.5cm 0.5cm 1.cm 0.5cm},width=17cm]{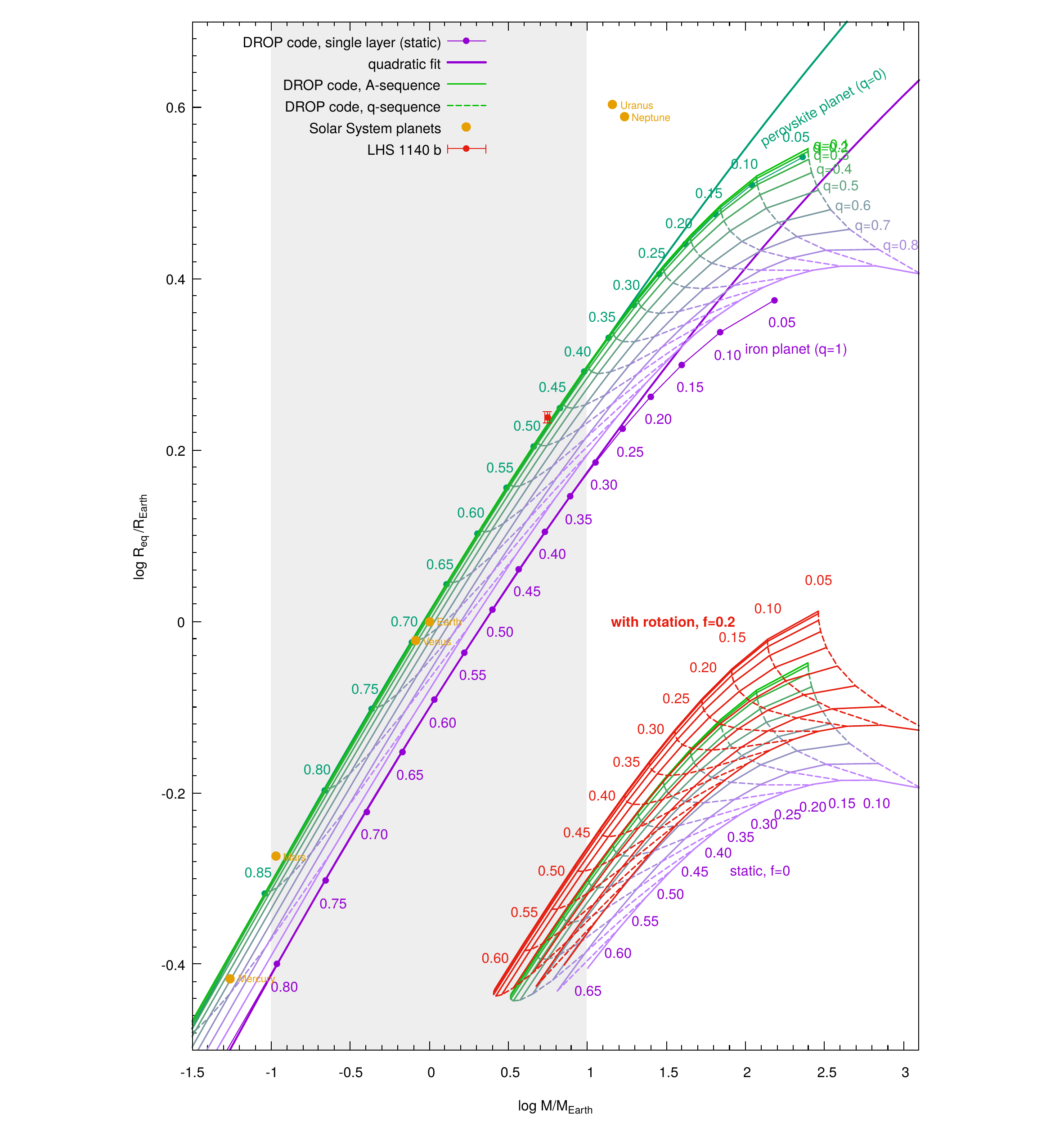}
       \caption{Same caption as for Fig. \ref{fig:mr_static_IW_full} but in a two-layer IP-planet.}
       \label{fig:mr_static_IP_full}
\end{figure*}

\newpage

\onecolumn
\section{Mass-radius relationship for two-layer, PW planets}
\label{sec:mr_twolayer_PW}

\begin{figure*}[h]
  \centering
       \includegraphics[trim={1.5cm 0.5cm 1.cm 0.5cm},width=17cm]{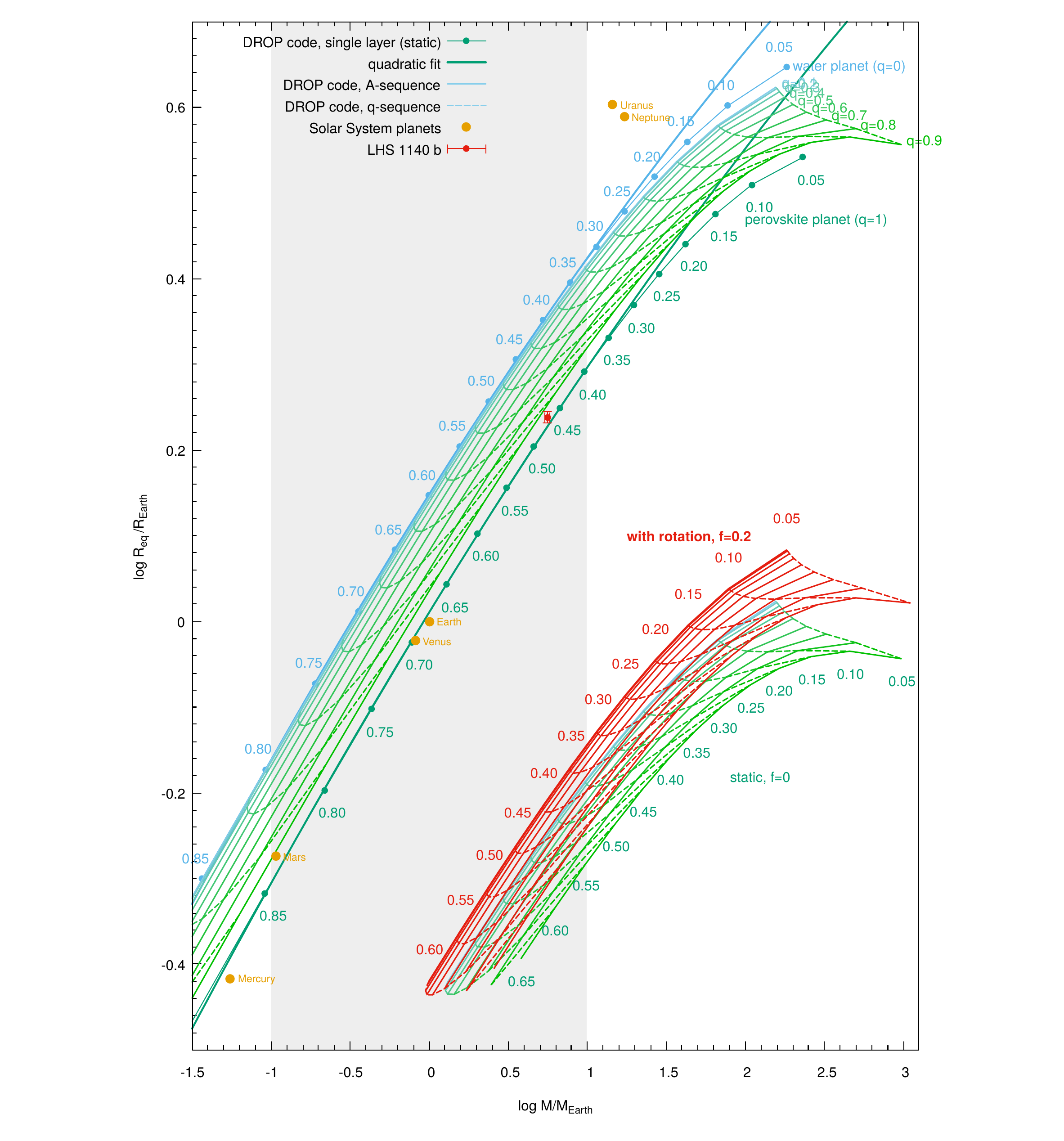}
       \caption{Same caption as for Fig. \ref{fig:mr_static_IW_full} but in a two-layer PW-planet.}
       \label{fig:mr_static_PW_full}
\end{figure*}

\newpage

\twocolumn
\section{Coefficients fitting the mass-radius relationship for two-layer, IW planets}
\label{sec:coef_twolayer_IW}

\begin{table}[h]
  \caption{Coefficients $c'_i$ for the mass-radius relationship fitted with a quadratic law for a two-layer, IW planet for different values of $\qpol$, in the static case$^\star$.}
  \label{tab:ciprim_static_ofqpol}
  \begin{tabular}{rccc} \hline \hline
  $\qpol$  & $c_0'$ & $c_1'$ & $c_2'$ \\\hline
$0.000$ & $+0.1483488$ & $+0.2929427$ & $-0.0172455$ \\
$0.100$ & $+0.1461391$ & $+0.2931076$ & $-0.0174334$ \\
$0.200$ & $+0.1415536$ & $+0.2923524$ & $-0.0182351$ \\
$0.300$ & $+0.1290441$ & $+0.2917313$ & $-0.0176743$ \\
$0.400$ & $+0.1076532$ & $+0.2908377$ & $-0.0178216$ \\
$0.500$ & $+0.0786889$ & $+0.2900709$ & $-0.0187472$ \\
$0.600$ & $+0.0444794$ & $+0.2906534$ & $-0.0184779$ \\
$0.700$ & $+0.0079287$ & $+0.2901063$ & $-0.0183598$ \\
$0.800$ & $-0.0291855$ & $+0.2924474$ & $-0.0182362$ \\
$0.900$ & $-0.0656371$ & $+0.2935405$ & $-0.0181483$ \\    \hline
$\langle c'_i\rangle$  & $+0.0622961$ & $+0.2916497$ & $-0.0181260$\\
$\Delta_i$  & $+0.1058881$ & $+0.0017348$ & $+0.0006569$\\
    \hline
  \end{tabular} \\
  $^\star$Values obtained for the pure-water planet ({\it first row}) have been copied from Tabs. \ref{tab:ci_iron} to \ref{tab:ci_water}. Mean values $\langle c'_i\rangle$ and half-amplitude $\Delta_i$ defined by Eq.\eqref{eq:meancprim} are also given.
\end{table}

\begin{table}[h]
  \caption{Coefficients $d_0$, $d_1$, and $d_2$ for the sigmoidal representation of $c_0'(\qpol)$ for the IW planet, assuming $\sigma$ is a constant.}
  \label{tab:d1d2sigmoid_IW}
  \begin{tabular}{rccc} \hline \hline
$f$      & $d_0$        & $d_1$          & $d_2$ \\\hline
$0.000$ & $+0.1476811$ & $-0.3621996$ & $+1.1962805$  \\
$0.025$ & $+0.1517282$ & $-0.3544016$ & $+1.1517880$  \\
$0.050$ & $+0.1557795$ & $-0.3452736$ & $+1.0779073$  \\
$0.075$ & $+0.1600821$ & $-0.3371772$ & $+1.0164078$  \\
$0.100$ & $+0.1646047$ & $-0.3298797$ & $+0.9572348$ \\
$0.125$ & $+0.1690640$ & $-0.3221491$ & $+0.8844918$  \\
$0.150$ & $+0.1738964$ & $-0.3152371$ & $+0.8130963$ \\
$0.175$ & $+0.1787665$ & $-0.3087314$ & $+0.7462931$  \\
$0.200$ & $+0.1838491$ & $-0.3025681$ & $+0.6676232$  \\
\hline
  \end{tabular}
\end{table}

\newpage

\section{Coefficients fitting the mass-radius relationship for two-layer, IW-, IP- and PW-planets}
\label{sec:coef_twolayer_IWIPPW}

\begin{table}[h]
  \caption{Coefficients$^\star$ $c_{ij}'$,  $d_{ij}$ and $\sigma$\label{sec:mr}
.}
  \label{tab:cprimij_dijsigmoid_IWIPPW}
  \begin{tabular}{lccc} \hline \hline
             &  $j=0$ &  $j=1$ &  $j=2$  \\\hline
  \multicolumn{4}{l}{$\blacktriangledown$ IW-planet}\\ 
$c_{1j}'$ & ${\bf +0.2916497}$ & $+0.0116867$ & $+0.0252880$\\
$c_{2j}'$ & $-0.0181260$ & $+0.0055022$ & -\\
    $d_{0j}$ & $+0.1476811$ & $\textcolor{red}{+0.1563921}$ & $+0.1218398$ \\
    $d_{1j}$ & $-0.3621996$ & $+0.3509368$ & $-0.2597847$  \\
    $d_{2j}$ & $+1.1962807$ & $-2.1953340$ & $-2.2445208$ \\
    $\sigma$ & $+2.7354873$\\\hline

    \multicolumn{4}{l}{$\blacktriangledown$ IP-planet}\\ 
  $c_{1j}'$ & ${\bf +0.2990242}$ & $+0.0141426$ & $+0.0102229$\\
  $c_{2j}'$ & $-0.0153046$ & $+0.0052181$ & -\\
  $d_{0j}$ & $+0.0109148$ & $\textcolor{red}{+0.1547714}$ & $+0.1031154$\\
   $d_{1j}$ & $-0.1653664$ & $+0.1488079$ & $+0.0098217$  \\
    $d_{2j}$& $+0.5304047$ & $-0.6630659$ & $-2.1148875$  \\
  $\sigma$ & $+2.9477349$\\\hline
  \multicolumn{4}{l}{$\blacktriangledown$ PW-planet}\\
$c_{1j}'$ & ${\bf +0.2944043}$ & $+0.0120350$ & $+0.0067155$\\
$c_{2j}'$ & $-0.0176780$ & $+0.0144473$ & -\\
  $d_{0j}$ & $+0.1460727$ & $\textcolor{red}{+0.1501963}$ & $+0.1318709$\\
  $d_{1j}$ & $-0.1261723$ & $+0.0475825$ & $-0.2442127$ \\
  $d_{2j}$ & $+0.5988696$ & $-1.4232417$ & $-0.3751630$ \\
 $\sigma$ & $+2.8544981$ \\ \hline
  \end{tabular}
\\  $^\star$The exponent of the power-law for a static, one Earth's mass body is indicated ({\it bold}), as well as the linear contribution of rotation ({\it red}).
\end{table}

\begin{figure}
       \centering
       \includegraphics[trim={1.3cm 0.2cm 1.2cm 0},clip,width=0.95\linewidth]{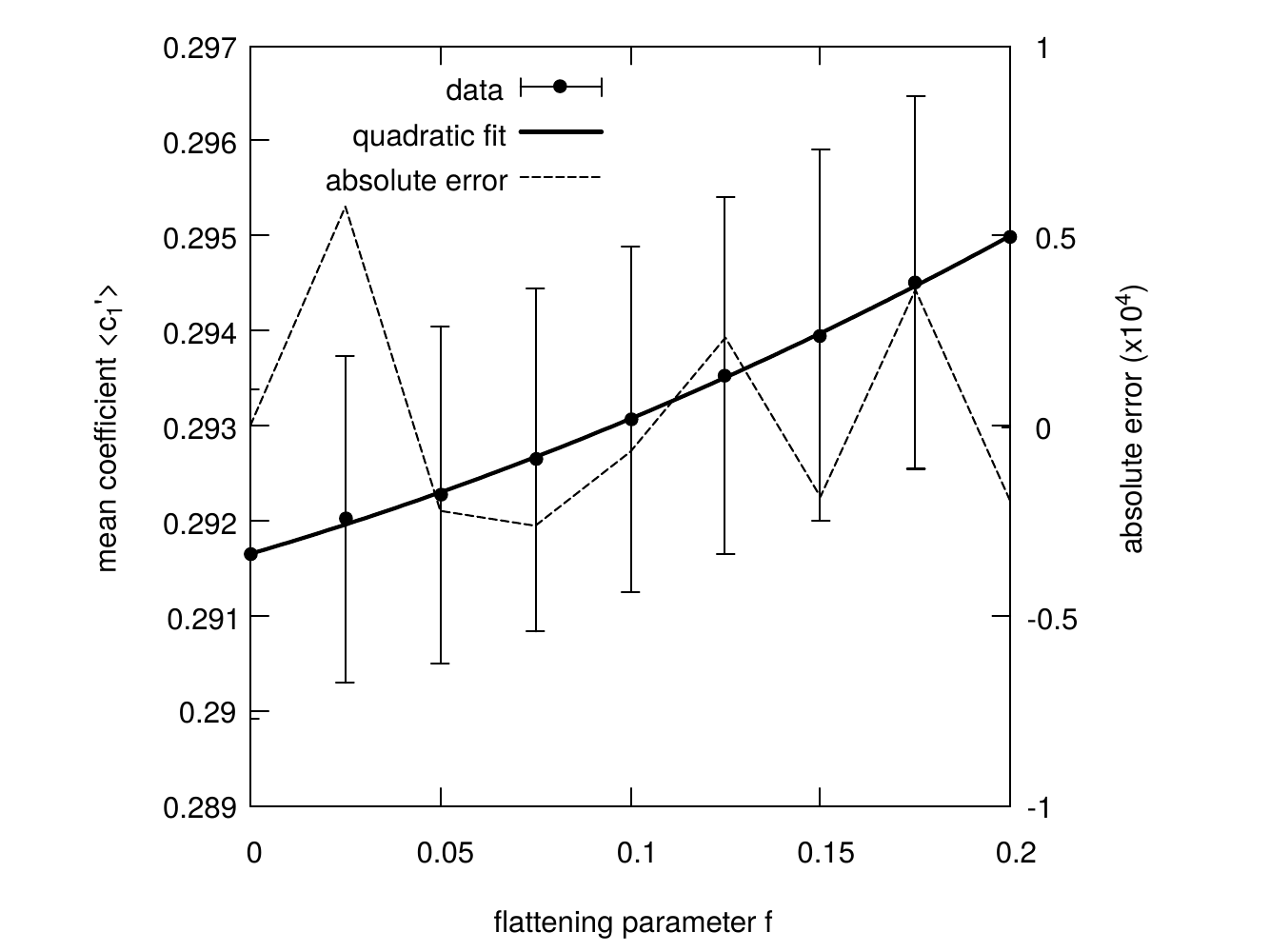}\\
       \includegraphics[trim={1.3cm 0.2cm 1.2cm 0},clip,width=0.95\linewidth]{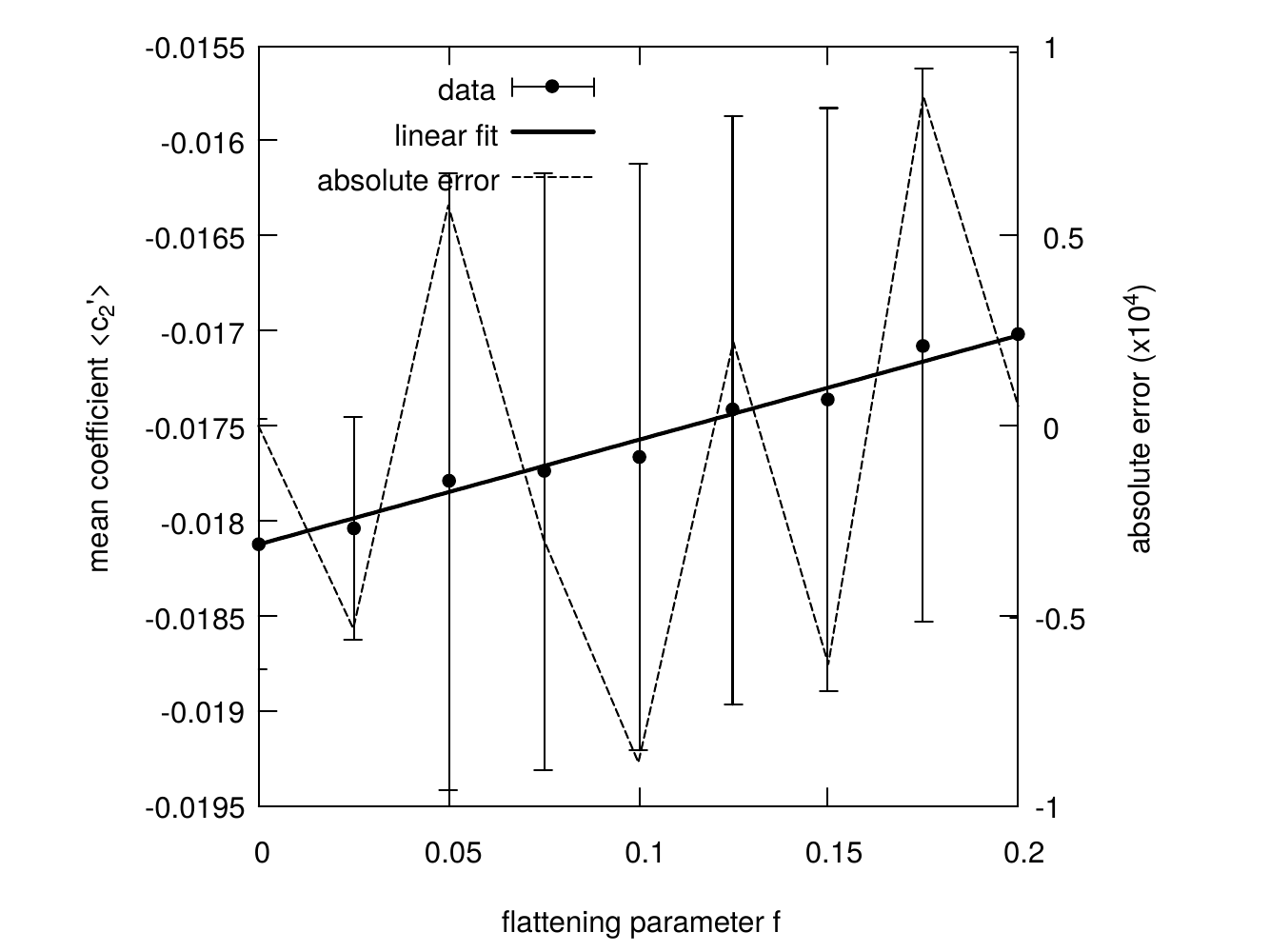}
       \caption{Mean coefficients $\langle c_1' \rangle \pm \Delta_1$ ({\it top}) and $\langle c_2' \rangle \pm \Delta_2$ ({\it bottom}) of the mass-radius relationship for a two-layer IW planet versus the flattening parameter $f$, the fits ({\it plain lines}) and the absolute errors in fitting the means ({\it dashed-lines, right axis}).}
       \label{fig:mr_cprim12mean}
\end{figure}

\newpage
\onecolumn

\section{Central pressure and rotation periods for the IP and PW-planets}
\label{app:mr_cmfIPPW}

\begin{figure*}[h]
       \centering
       \includegraphics[trim={1.8cm 0.5cm 2.9cm 0.5cm},clip,width=0.48\linewidth]{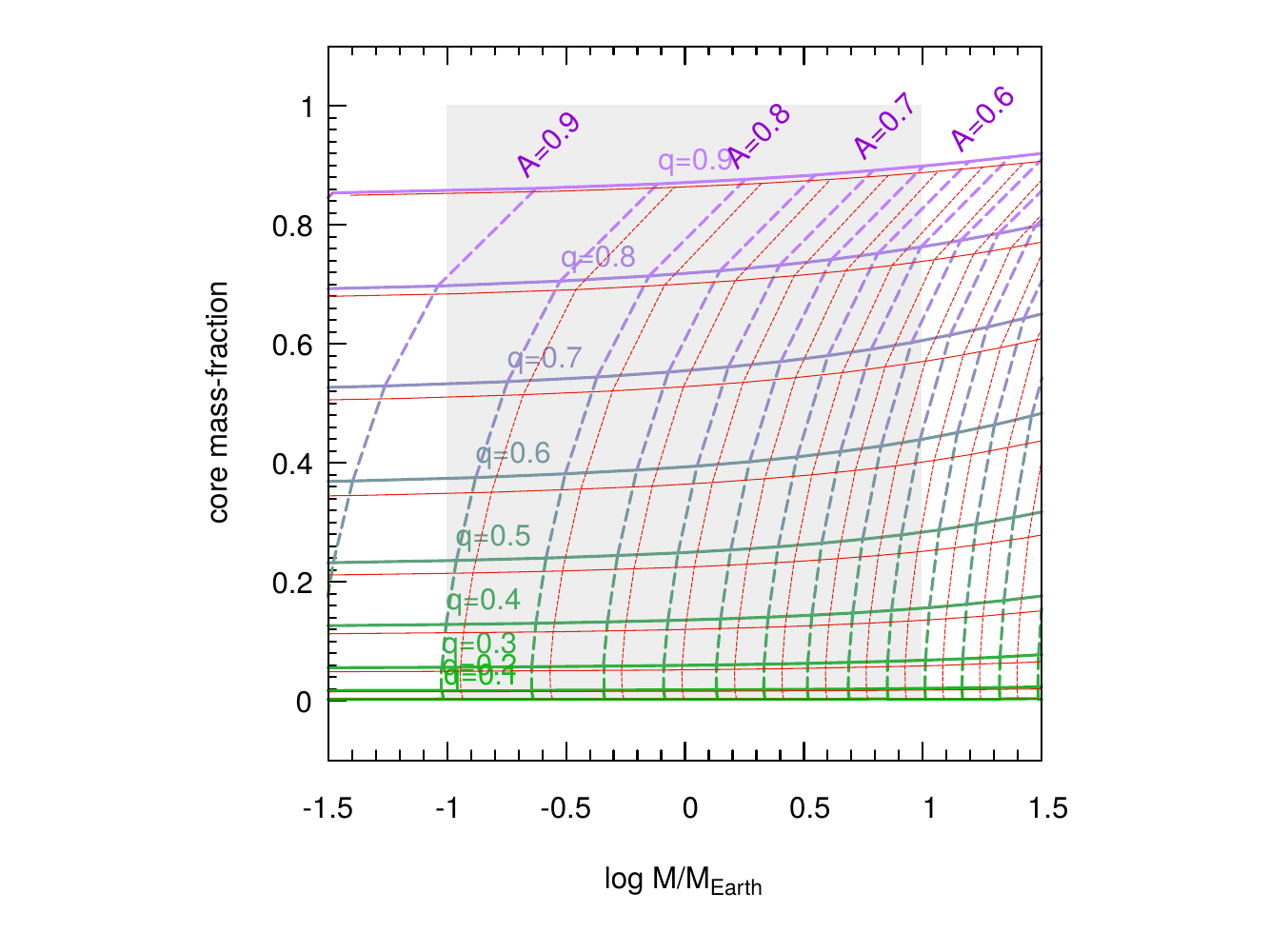}\includegraphics[trim={1.8cm 0.5cm 2.9cm 0.5cm},clip,width=0.48\linewidth]{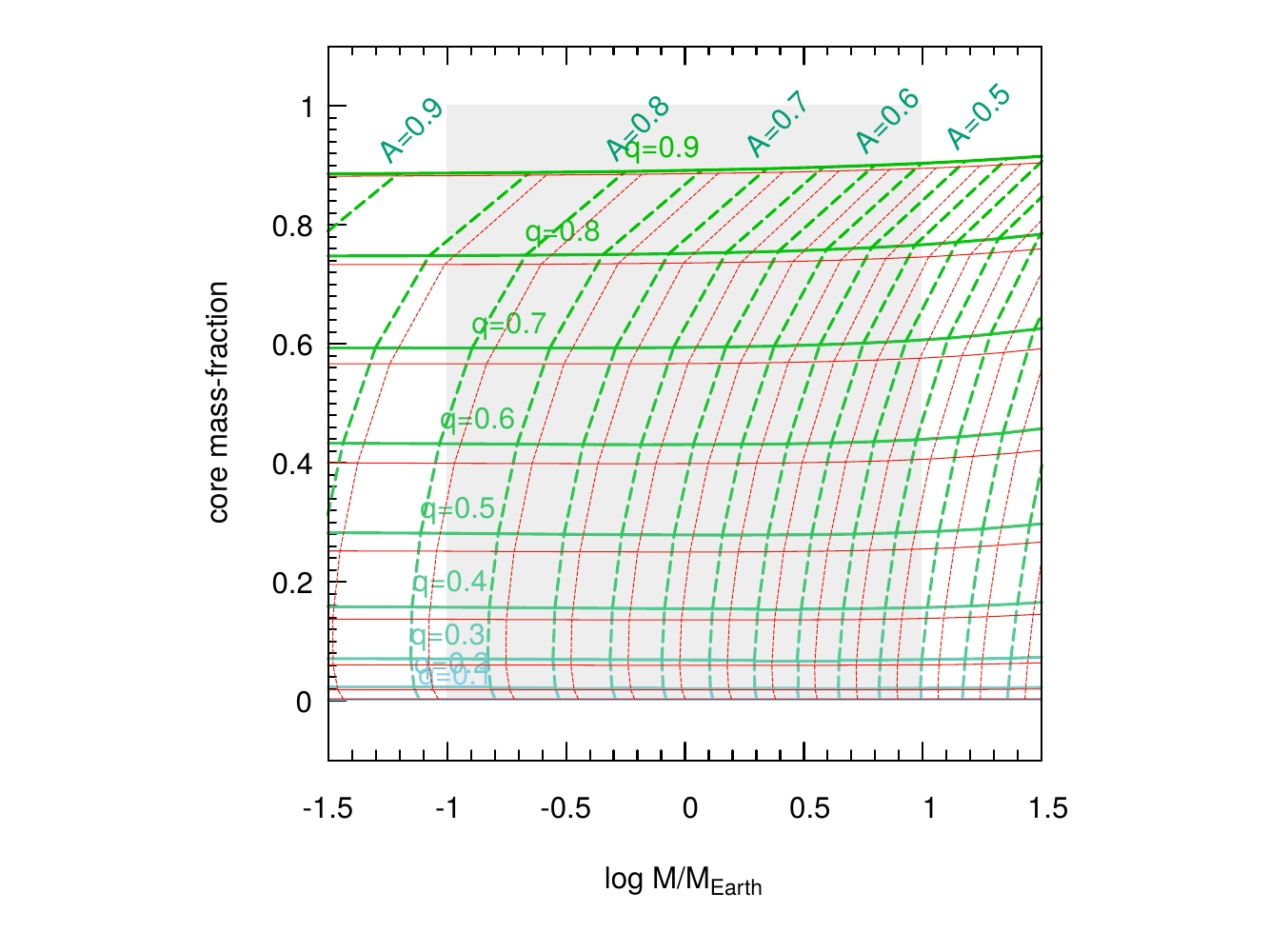}
        \caption{Same caption as for Fig. \ref{fig:mr_cmf}, but for the IP planet ({\it left}) and for the PW planet ({\it right}).}
       \label{fig:mr_cmfIPPW}
\end{figure*}

\begin{figure*}
       \centering
       \includegraphics[trim={1.8cm 0.5cm 2.9cm 0.5cm},clip,width=0.48\linewidth]{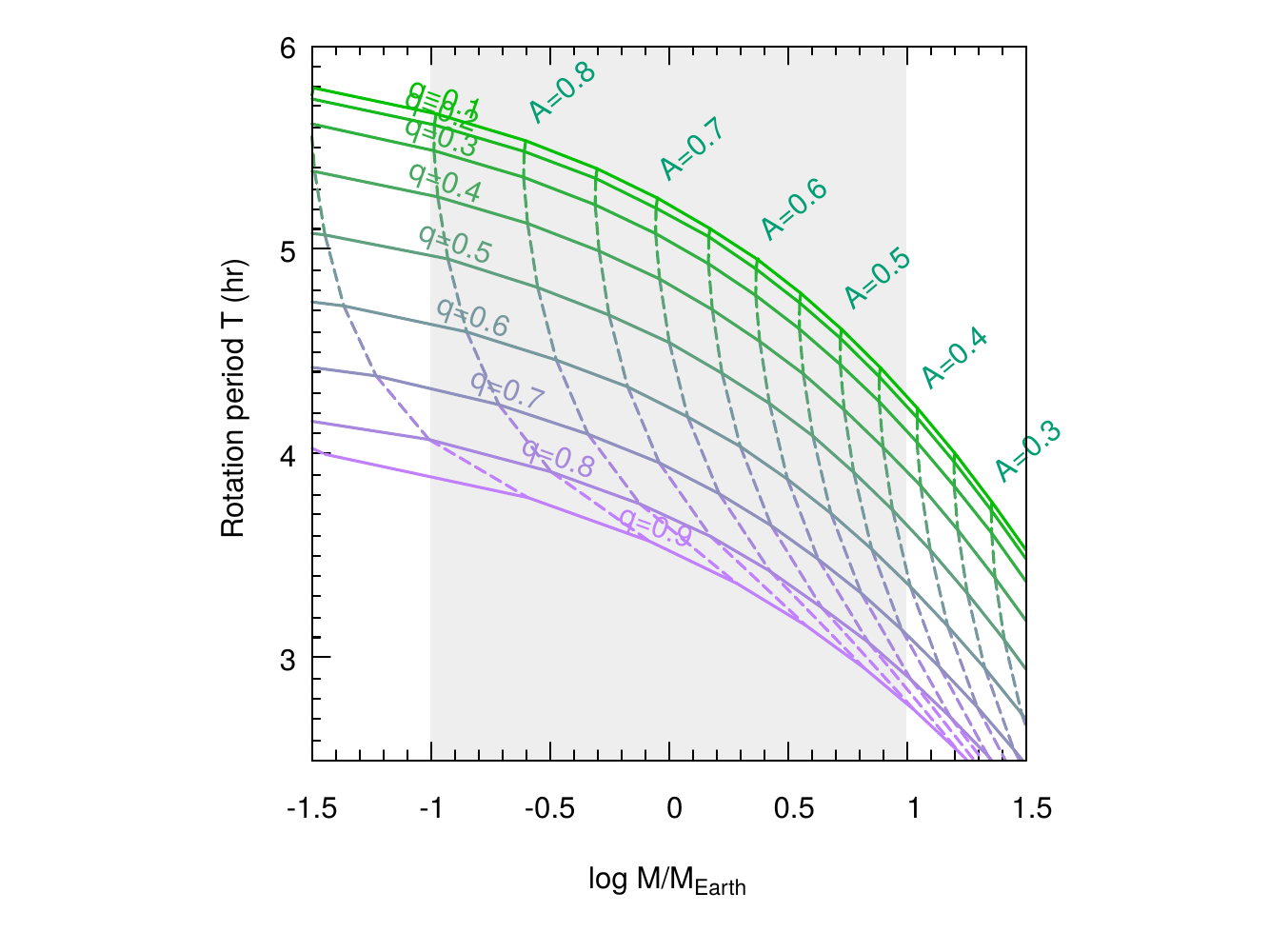}\includegraphics[trim={1.8cm 0.5cm 2.9cm 0.5cm},clip,width=0.48\linewidth]{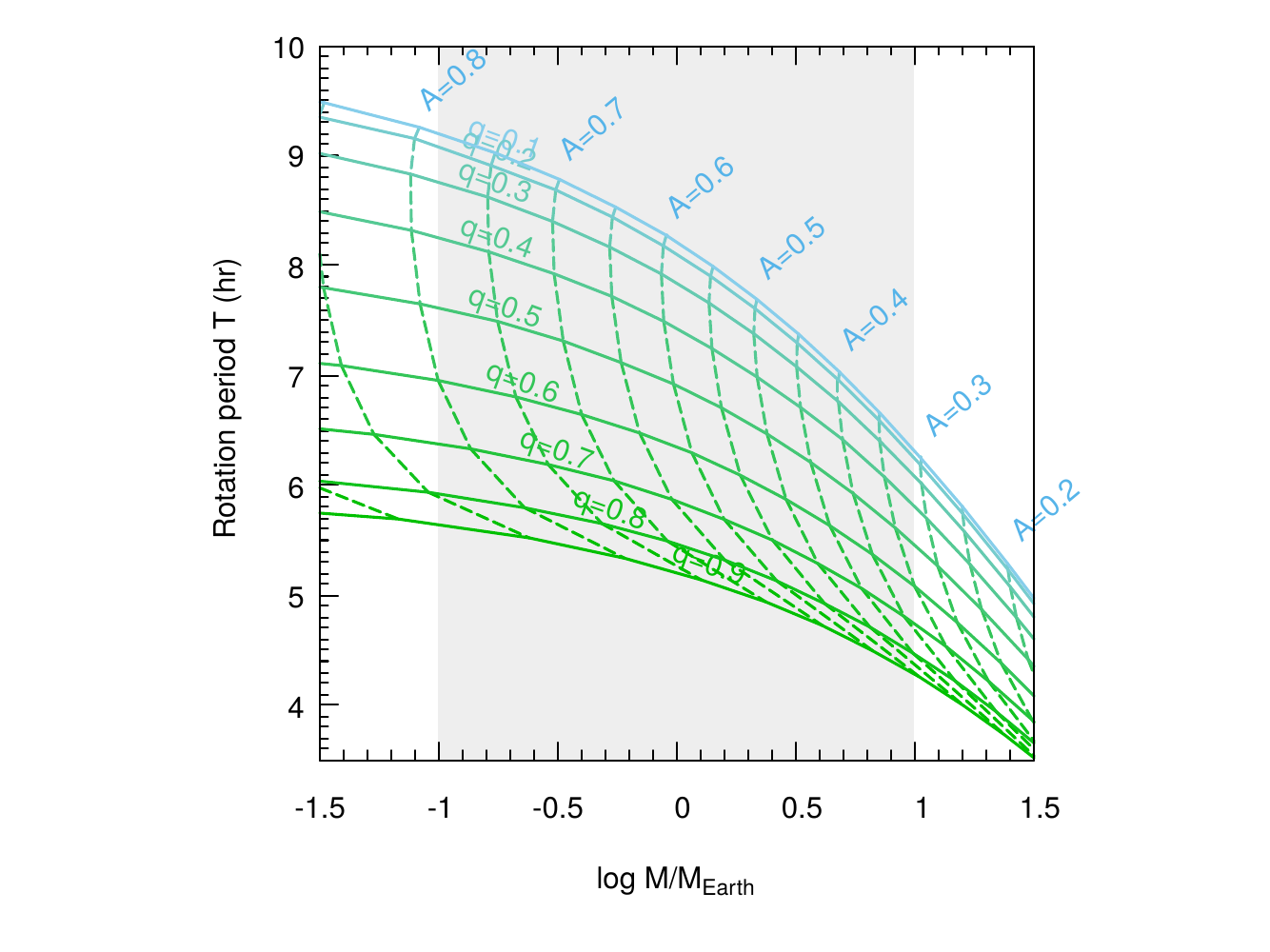}
       \caption{Same caption as for Fig. \ref{fig:mr_period}, but for the IP planet ({\it left}) and for the PW planet ({\it right}).}
       \label{fig:mr_periodIPPW}
\end{figure*}

\end{appendix}

\end{document}